\documentstyle[12pt]{report}

\begin{document}
\begin{center}
\pagenumbering{roman}
\thispagestyle{empty}
{\Huge \bf Studies on Boundary Conditions
\vskip 0.3cm
and  Noncommutativity}
\vskip 0.3cm
{\Huge \bf in String Theory}
\vskip 2.0 true cm
Thesis Submitted for the degree of\\
Doctor of Philosophy (Science)\\
of \\
JADAVPUR UNIVERSITY
\vskip 3.0 true cm
 2008 \\
{\large \bf ARINDAM GHOSH HAZRA}\\
Satyendra Nath Bose National Centre for Basic Sciences\\
JD Block, Sector III \\
Salt Lake City \\
Kolkata  700 098\\
India
\end{center}
\newpage
\centerline{\large \bf CERTIFICATE FROM THE SUPERVISOR} This is to
certify that the thesis entitled {\bf $``$Studies on Boundary
Conditions and  Noncommutativity in String Theory"} submitted by
{\bf Sri Arindam Ghosh Hazra} who got his name registered on {\bf
20.03.2006} for the award of Ph.D.(Science) degree of Jadavpur
University, absolutely based upon his own work under the
supervision of {\bf Dr. Biswajit Chakraborty} and that neither
this thesis nor any part of it has been submitted for any
degree/diploma or any other academic
award anywhere before.\\

\noindent{\bf  Biswajit Chakraborty}\\
    Associate Professor\\
    S.N.Bose National Centre for Basic Sciences, \\
    Salt Lake, Kolkata, India.

\newpage
\thispagestyle{empty}
.
\vskip 6.0cm
\begin{center}
{\Huge \bf  Dedicated}
\vskip 0.5 cm
{\Huge \bf to }
\vskip 0.5cm
{ \Huge \bf   My Parents}
\vskip 0.5 cm
{\Huge \bf and }
\vskip 0.5cm
{\Huge \bf   Grand Parents}
\end{center}
\newpage
\vskip 2.0cm
\begin{center}
{\large \bf ACKNOWLEDGEMENTS}
\end{center}

From the depth of my heart, I express my deep sense of gratitude
to my thesis advisor, Dr. Biswajit Chakraborty for his expert
caring guidance and constant encouragement. I convey my sincere
regards to Dr. Biswajit Chakraborty for helping me throughout the
course of this work. The timely completion of this thesis is a
result of his unflinching support and I am indebted to him. It was
my good luck that I got guidance from Dr. Biswajit Chakraborty who
not only guided me in my thesis but also supported me in various
manners.

\noindent I am thankful to Prof. F. G. Scholtz for the academic
help rendered to me. I also thank Prof. Rabin Banerjee for
encouraging me all through.

I am grateful to Prof. S. Dattagupta, ex. Director of Satyendra
Nath Bose National Centre for Basic Sciences (SNBNCBS), for giving
me the opportunity to do research here. I thank Prof. A.K.
Roychowdhury, Director, SNBNCBS for his support in my work.

\noindent I thank all the academic and administrative staff of
SNBNCBS for helping me in many ways. In particular, I am thankful
to the Library staff for the excellent assistance provided to me.

Special thanks to my friend Mr. Sunandan Gangopadhyay for his
continuous support during my work. I really enjoyed a lot working
with him. Also my thanks to Mr. Saurav Samanta, Mr. Chandrasekhar
Chatterjee, Mr. Anirban Saha and Mr. Saikat Chatterjee for their
help and support in academic matter which made my research an
experience I cherish much.

\noindent Finally and most importantly, I express my heartiest
gratitude to my family members. It is the  love and support of my
parents and grand parents that enabled me to pursue the studies
which finally culminated in this thesis. I am really grateful to
my wife for her encouragement  and support. I dedicate this thesis
to them.
\newpage
\vskip 2.0cm
{\large \bf List of publications}
\begin{enumerate}
\item  Dual families of noncommutativequantum systems\\
F. G. Scholtz, B. Chakraborty, S. Gangopadhyay, A. Ghosh Hazra \\
{\it  Phys. Rev. D}{\bf 71} (2005) 085005.
\item Non(anti) commutativity for open superstrings \\
B. Chakraborty, S. Gangopadhyay, A. Ghosh Hazra, F. G. Scholtz\\
{\it  Phys. Lett. B} {625} (2005) 302-312.
\item  Twisted Galilean symmetry and the Pauli principle at low energies\\
B. Chakraborty, S. Gangopadhyay, A. Ghosh Hazra, F. G. Scholtz \\
{\it  J Phys. A}{\bf 39} (2006) 9557-9572.
\item  Normal ordering and noncommutativity in open bosonic strings\\
B. Chakraborty, S. Gangopadhyay, A. Ghosh Hazra \\
{\it  Phys. Rev. D}{\bf 74} (2006) 105011.
\item  Noncommutativity  in interpolating string: A study of gauge
symmetries in a noncommutative framework\\
S. Gangopadhyay, A. Ghosh Hazra, A. Saha \\
{\it Phys. Rev. D} {\bf 74} (2006) 125023.
\item Normal ordering and non(anti)commutativity  in open super strings\\
S. Gangopadhyay, A. Ghosh Hazra \\
{\it Phys. Rev. D} {\bf 75} (2007) 065026.
\item String non(anti)commutativity for Neveu-Schwarz boundary conditions\\
C. Chatterjee, S. Gangopadhyay, A. Ghosh Hazra, S. Samanta  \\
Accepted in {\it Int J Theor Phys.} (2008), [hep-th:0801.4189]
\end{enumerate}

\newpage
\tableofcontents
\chapter{Introduction and Overview}
\pagenumbering{arabic} Noncommutative theories has a long history
in physics \cite{sny} and has kindled a lot of interest in the
past few years owing to the inspiration from string theory
\cite{sw, dn}. Recent progress in string theory \cite{maldacena,
randall, sundrum} indicates scenarios where our four dimensional
space-time with standard model fields corresponds to a D3-brane
embedded in a larger manifold. Now, since D-branes correspond, in
type II string theories, to the space where the open string
endpoints are attached, our space-time would be affected by string
boundary conditions. One important consequence is the possible
noncommutativity of space-time coordinates at very small length
scales since commuting coordinates are incompatible  with open
string boundary conditions in the presence of anti-symmetric
tensor backgrounds \cite{chu, chu1, ard, rb, schomerus}. This is
one of the main reasons of increasing interest in several aspects
of noncommutative quantum mechanics and quantum 
field theories \cite{sw},\cite{sz}-\cite{sunfgs}.
Furthermore, this illustrates the fact that the string boundary
conditions may play a crucial role in the phenomenology of
four-dimensional physics.

Since the discovery of the role of branes in string theory \cite{polc}
they have frequently shown unexpected properties. They were first
identified as the carriers of R-R charges and very soon after, it
was realised that when $N$ of them merge the space-time coordinates
normal to them become noncommutative \cite{wit} and the $U(N)$
super Yang-Mills theory emerges. Another type of noncommutativity
appears in the bound states of branes with fundamental strings and
with other branes. It has been shown that such brane bound states
correspond to branes with non-zero background internal gauge fields
\cite{wit, has, guk, lu, lu1, hash, arf}. The noncommutativity arising
in the internal structure of these brane bound states is a consequence
of the properties of open strings ending on them. Such open strings
satisfy boundary conditions which are neither Neumann nor Dirichlet,
but a combination of the two, sometimes reffered to as mixed boundary
condition \cite{arf, she, ard1, ard2}. The mixed boundary condition makes the
canonical quantisation of the theory non trivial. Imposing the
standard commutation relation leads to inconsistency. It has been
proposed to remove the inconsistency by relaxing the commutativity
of the space coordinates of the open strings along the tangential
direction of the brane described by mixed boundary conditions
\cite{chu, ard1, ard2, ard3}. The procedure of relaxing the commutativity
of space coordintes adopted in \cite{she, ard3}, was to keep the
standard algebra of the Fourier modes in the mode expansion.

The noncommutativity observed in the above brane system seemed very
similar to that observed by Connes, Douglas and Schwarz \cite{conn}
in the problem of Matrix Model with non-trivial background three
form. They studied compactification of Matrix theory on a noncommutative
torus and realised that it corresponds to the Matrix theory in
such backgrounds. Motivated by this observation, Ardalan
{\it{et al}}. showed that \cite{ard1, ard3}, the noncommutativity
can be derived within the string theory by wrapping branes with
non zero $B_{\mu \nu}$ background field on the compactification
torus.


Different approaches have been adopted to obtain this
noncommutativity. A Hamiltonian operator treatment was provided in
\cite{chu} and a world sheet approach in \cite{som}. Also, an
alternative Hamiltonian (Dirac \cite{di}) approach based on
regarding the Boundary Conditions as constraints was given in
\cite{ar, arkim}; the corresponding Lagrangian (symplectic)
version being done in \cite{br}. The interpretation of the
boundary condition as primary constraints usually led to an
infinite tower of second class constraints \cite{tez}, in contrast
to the usual Dirac formulation of constrained systems \cite{di,
hrt}. Some other approaches to this problem have been discussed in
\cite{zab, and, zwl, whl, ideg, ilor, sgs12, gdr13}.

On the other hand, it has also been shown that non-commutativity
can be obtained in a more transparent manner by modifying the
cannonical Poisson bracket structure, so that it is compatible
with the boundary condition \cite{rb, rbbckk}. In this approach,
the boundary conditions are not treated as constraints. This is
similar in spirit to the treatment of Hanson, Regge and Teitelboim
\cite{hrt}, where modified Poisson brackets were obtained for the
free Nambu-Goto string, in the orthonormal gauge, which is
 the counterpart of the conformal gauge in the free Polyakov string.
Those studies were, however, restricted to the case of the bosonic
string and membrane only. Proceeding further Jing and Long
\cite{jing} obtained the Poisson brackets among the Fourier
components using the Faddeev-Jackiw symplectic formalism
\cite{fj}, so that they are compatible with these boundary
conditions. Using this they obtained the Poisson brackets among
the open string coordinates revealing the noncommutative structure
in the string end points.

\section{Structure of the thesis}
The central theme of this thesis is noncommutativity in string
theory. We explore in detail how noncommutative structures can
emerge in case of the interacting bosonic string and even in the
fermionic sector of superstring theory. We have shown in various
approaches that string coordinates must be noncommutative in order
to be compatible with boundary conditions. These noncommutative
structures lead to new involutive algebra of constraints but
generate same Virasoro algebra, indicating the internal
consistency of our analysis.

On the other hand the action for a string can be chosen, in
analogy with the relativistic particle as the proper area of
the world sheet swept out by the dynamical string.
This gives the Nambu-Goto formalism which, however, poses problems
in quantisation. A redundant description, where the world sheet
metric coefficients are considered as independent fields, has been shown
by Polyakov \cite{pol} to be particularly suitable in this context.
The ensuing action is known as the Polyakov action. The equivalance
between the two approaches can be established on shell by solving the
independent metric in the Polyakov action. The classical correspondence
is assumed to lead to equivalent result at the quantum level. Understanding
this correspondence from different view points will, naturally, be useful.

\noindent $\bullet$ We start with the gauge independent analysis
of Polyakov string and give a review, based on \cite{rb}, of the
emergence of noncommutativity in the context of an open string.
Here the authors, do not treat the boundary conditions as
constraints, but show that they can be systematically implemented
by modifying the canonical Poisson bracket structure. We follow
the same methodology to obtain the noncommutativity among the
string coordinates for both interpolating and super strings in the
following chapters. This is the subject matter of chapter 2.

A deeper connection between Polyakov action and Nambu-Goto string
action has been demonstrated in \cite{rb} by constructing a
Lagrangian description which interpolates between the Nambu-Goto
and Polyakov forms in the free case. The interpolating theory thus
offers a unified pictures for understanding different features of
the basic structures including their various symmetry properties.
In this sense, therefore, it is more general than either the
Nambu-Goto or Polyakov formulation. An added advantage is that it
illuminates the passage from the Nambu-Goto form to the Polyakov
form, which is otherwise lacking. In this context it may be noted
that the Polyakov action has the additional Weyl invariance which
Nambu-Goto action does not have. The interpolating action, which
does not presuppose Weyl invariance, thus offers a proper platform
of discussing the equivalance of the two actions. It also explains
the emergence of the Weyl invariance in a natural way. In this
work we study the interpolating formalism both in free and
interacting case.

\noindent $\bullet$ In chapter 3 we derive a master action for
interacting bosonic strings, interpolating between the Nambu-Goto
and Polyakov formalism. Modification of the basic poisson bracket
structure compatible with boundary conditions followed by the
emergence of the noncommutativity is shown in this formalism (in
case of both free and interacting strings) following the approach
discussed in chapter 2 and \cite{rb, hrt, rbbckk}. Our results go
over smoothly to the Polyakov version once the proper
identifications are made. This noncommutativity leads to a new
involutive constraint algebra which is markedly different from
that obtained in second chapter \cite{rb}. With the above results
at our disposal, we go over to the study of gauge symmetry in the
noncommutative framework. Owing to the new constraint algebra we
find surprising changes in the structure constants of the theory.
Finally, we compute the gauge variations of the fields and show
the underlying unity of diffeomorphism with the gauge symmetry in
the noncommutative framework \cite{agh3}.

So far our attention was basically confined to the classical level.
We now extend parts of the foregoing analysis to the quantum level.
To that end, recall that in quantum field theory, products of
quantum fields at the
same space-time points are in general singular objects. The
same thing happens in string theory if one multiplies position
operators, that can be taken as conformal fields on the world sheet.
This situation is well known and one can remove the singular part of the
operator products by defining normal ordered well behaved objects \cite{pol}.
Normal ordered products of operators are usually defined so as to satisfy
the classical equations of motion at quantum level.

Recently Braga {\it{et al.}} \cite{4cbr} defined normal ordered products
for open string position operators that additionally satisfy the
boundary conditions. This way one can define a normal ordering that will
be valid also at string end-points.

\noindent $\bullet$ In the 4th chapter, noncommutativity in an
open bosonic string moving in the presence of a background
Neveu-Schwarz two-form field $B_{\mu \nu}$ is investigated in a
conformal field theory approach. The mode algebra is first
obtained using the newly proposed normal ordering, which satisfies
both equations of motion and boundary conditions. Using these the
commutator among the string coordinates is obtained.
Interestingly, this new normal ordering  yields the same algebra
between the modes as the one satisfying only  the equations of
motion. In this approach, we find that noncommutativity originates
more transparently and our results match with the existing results
in the literature \cite{agh2}.

Compared to the bosonic string theory, the supersymmetric case as
well as superstring theory has received less attention. In Ref
\cite{chu1}, the authors discussed the fermionic part without
resorting to the dynamical properties. They start from the bosonic
results by supersymmetric transformations. In \cite{haggi}, the
authors find that in order to keep the supersymmetry unbroken in
an open string's end points, it is necessary to add a proper
boundary term to the supersymmetric action. Authors of
\cite{godinho} work in the discrete version. They take the
boundary conditions as constraints and then Fadeev Jackiw method
is employed to get the anti-poisson brackets among the string
coordinates. Jing also obtained the anti-poisson brackets in
\cite{jing1} by following the same methdology of their bosonic
paper \cite{jing}.

\noindent $\bullet$ In chapter 5 we start with the Ramond Neveu
Schwarz superstring action in the conformal gauge and discuss the
super constraint structure of the theory. The
non(anti)commutativity of the theory is then revealed in the
conventional Hamiltonian framework by following \cite{rb, hrt,
rbbckk}. We have obtained this expressions of non(anti)commutative
structure for an open superstring by modifying the canonical
bracket, so that it is compatible with the boundary conditions. We
find that the non(anti)commutative structures not only appear for
an open string moving in the antisymmetric background field but
also in the free case. This is indeed a new result. We have also
shown that this symplectic structure leads to a new involutive
structure for the super constraint algebra at the classical level
\cite{agh1}.

\noindent $\bullet$ In Chapter 6 we extend our methodology
discussed in chapter 4 to analyse an open super string propagating
freely and one moving in a constant antisymmetric background field
\cite{agh4}. We start by reviewing the recent results involving
new normal ordered products of fermionic operators \cite{6cbr}.
The mode algebra is then obtained using the newly proposed normal
ordering, which satisfies both equations of motion and boundary
conditions. Finally we obtain same anti-commutators among the
string coordinates by using the mode algebra.

\noindent $\bullet$ It is surprising that all the studies in the
context of superstring theory is based on Ramond boundary
conditions. But as is well known in the context of fermionic
string there is a choice between Ramond boundary conditions and
Neveu Schwarz boundary conditions. In chapter 5 and chapter 6 we
have a detailed discussion on the problem of
non(anti)commutativity on the basis of Ramond boundary conditions.
In chapter 7, we extend our methodology to the superstring
satisfying the Neveu Schwarz boundary conditions \cite{agh5}.

Finally, in Chapter 8 we summarise the important results.


\newpage

\noindent This thesis is based on the following publications.
\begin{enumerate}
\item Non(anti) commutativity for open superstrings
 \cite{agh1}\\
B. Chakraborty, S. Gangopadhyay, {\bf{A. Ghosh Hazra}}, F. G. Scholtz\\
{\it  Phys. Lett. B} {625} (2005) 302-312.
\item  Normal ordering and noncommutativity in open bosonic strings
\cite{agh2}\\
B. Chakraborty, S. Gangopadhyay, {\bf{A. Ghosh Hazra}} \\
{\it  Phys. Rev. D}{\bf 74} (2006) 105011. \item  Noncommutativity
in interpolating string: A study of gauge symmetries in a
noncommutative framework
\cite{agh3}\\
S. Gangopadhyay, {\bf{A. Ghosh Hazra}}, A. Saha \\
{\it Phys. Rev. D} {\bf 74} (2006) 125023. \item Normal ordering
and non(anti)commutativity  in open super strings
\cite{agh4}\\
S. Gangopadhyay, {\bf{A. Ghosh Hazra}} \\
{\it Phys. Rev. D} {\bf 75} (2007) 065026. \item String
non(anti)commutativity for Neveu-Schwarz boundary conditions
\cite{agh5}\\
C. Chatterjee, S. Gangopadhyay, {\bf{A. Ghosh Hazra}}, S. Samanta  \\
Accepted in {\it Int J Theor Phys.} (2008), [hep-th:0801.4189].
\end{enumerate}


\chapter{Review of Bosonic String}
An intriguing connection between string theory, noncommutative
geometry and noncommutative Yang-Mills theory was revealed in
\cite{sw}. The study of open string, in the presence of a
background Neveu-Schwarz two-form field $B_{\mu \nu}$, leads to a
noncommutative(NC) structure which manifests in the
noncommutativity at the end points of the string which are
attached to D-branes. Different approaches have been adopted to
obtain this result.

In this present chapter, we discuss the Polyakov action and also
the essential result of \cite{rb} in which the authors provide an
 exhaustive analysis of the noncommutativity in open string theory
moving in the presence of a constant Neveu-Schwarz field, in the
conventional Hamiltonian framework. In contrast to the usual studies,
this model of string theory is very general in the sense that no gauge
is fixed at the beginning. Let us recall that all computations of
noncommutativity, mentioned before, were done in the
conformal gauge. This gauge independent analysis
yields a new noncommutative structure, which correctly
reduces to the usual one in conformal gauge. This shows
the compatibility of the present analysis with the existing literature.
In the general case, the noncommutativity is
manifested at all points of the string, in contrast
to conformal gauge results where it appears only at the
boundaries. Indeed, in this gauge independent scheme,
one finds a noncommutative algebra among the coordinates,
even for a free string, a fact that was not observed before. Expectedly,
this noncommutativity vanishes in the conformal gauge.
Note however, that there is no gauge for which
noncommutativity vanishes in the interacting theory.

At the outset, let us point out the crucial difference between
existing Hamiltonian analysis \cite{ard} and this approach. This
is precisely in the interpretation of the boundary conditions(BC)
arising in the string theory. The general consensus has been to
consider the boundary conditions as primary constraints of the
theory and attempt a conventional Dirac constraint analysis
\cite{di}. The aim is to induce the noncommutativity in the form
of Dirac Brackets between coordinates. The subsequent analysis
turns out to be ambiguous since it involves the presence of
$\delta (0)$-like factors, (see Chu and Ho in \cite{ard}).
Different results are obtained depending on the interpretation of
these factors.

Here on the other hand we do not treat the BCs as constraints, but
show that they can be systematically implemented by modifying the
canonical Poisson Bracket(PB)  structure. In this sense this
approach is quite similar in spirit to that of Hanson, Regge and
Teitelboim \cite{hrt}, where modified PBs were obtained for the
free Nambu-Goto string, in the orthonormal gauge, which is the
counterpart of the conformal gauge in the free Polyakov string.

\section{The free string in Polyakov formalism}
In this section, we analyze the Polyakov formulation of the free
string. The Polyakov action for a free bosonic string reads,
\begin{eqnarray}
S_P= -{1\over 2} \int_{-\infty}^{+\infty} d\tau\int_0^\pi
d\sigma\,\sqrt{-g} g^{ab}{\partial}_aX^{\mu}\,{\partial}_bX_{\mu}
\label{paction}
\end{eqnarray}
where $\tau $ and $\sigma$ are the usual world-sheet parameters and
$g_{ab}$, up to a Weyl factor, is the induced metric on the
world-sheet. $X^{\mu}(\sigma)$ are the string coordinates in the
D-dimensional Minkowskian target space with metric $\eta_{\mu
\nu} =$ diag $(-1,1,1..,1)$. \\
\noindent
This action has the following symmetries:
\begin{itemize}
\item 1. D-dimensional Poincar\'{e} invariance:
\begin{eqnarray}
X^{'\mu}(\tau, \sigma) &=& {\Lambda^{\mu}}_{\nu} X^{\nu}(\tau ,
\sigma)+ a^{\mu}\nonumber\\
g^{'}_{ab}(\tau, \sigma) &=& g_{ab}(\tau, \sigma)
\end{eqnarray}
\item 2. Diffeomorphism Invariance:
\begin{eqnarray}
X^{'\mu}(\tau^{'}, \sigma^{'}) &=& X^{\mu}(\tau ,\sigma)\nonumber \\
\frac{\partial \sigma^{'c}}{\partial \sigma^{a}}\,
\frac{\partial \sigma^{'d}}{\partial \sigma^{b}}\,
 g^{'}_{cd}(\tau^{'}, \sigma^{'}) &=& g_{ab}(\tau, \sigma)
\end{eqnarray}
for new coordinates $ \sigma^{'a}(\tau, \sigma)$.
\item 3. Two-dimensional Weyl invariance:
\begin{eqnarray}
X^{\prime \mu}(\tau,\sigma) &=& X^{\mu}(\tau,\sigma)\nonumber\\
 g^{\prime}_{ab}(\tau, \sigma) &=& {\mathrm{exp}}(2\omega(\tau,
\sigma))\,g_{ab}(\tau, \sigma)
\end{eqnarray}
for arbitrary $\omega(\tau, \sigma)$.\\
\end{itemize}
Here we carry out our analysis in the
complete space by regarding both $X^{\mu}$ and $g_{ab}$ as
independent dynamical variables \cite{holt}. The canonical momenta
are,
\begin{eqnarray}
\Pi_{\mu} &=& {\delta {\cal L}_P \over \delta \left(\partial_\tau\,
X^{\mu}\right)}=-{\sqrt {-g}}\, \partial_\tau X_{\mu}\nonumber \\
\pi_{ab} &=& {\delta {\cal L}_P \over \delta (\partial_\tau\,g^{ab})}=0 .
\label{mom}
\end{eqnarray}
It is clear that while $\Pi_{\mu}$ is a genuine momenta, $\pi_{ab}\approx 0$
are the primary constraints of the theory.
The conservation of the above primary constraints leads to the secondary
constraints $\Omega_1(\sigma)$ and  $\Omega_2(\sigma)$.
These secondary constraints also follow from the
equation obtained by varying $g_{ab}$ since this is basically a
Lagrange multiplier. This imposes the vanishing of the symmetric
energy-momentum tensor,
\begin{eqnarray}
T_{ab}={2\over {\sqrt {-g}}}{\delta S_P\over \delta g^{ab}}
=-\partial_a X^{\mu}\,\partial_b{X_{\mu}} + {1\over
2}g_{ab}g^{cd}\partial_cX^{\mu}\, \partial_dX_{\mu} = 0.
\label{emt}
\end{eqnarray}
Because of the Weyl invariance, the energy-momentum tensor is traceless,
$${T^a}_a=g^{ab}T_{ab}=0 $$
so that only two components of $T_{ab}$ are independent. These components, which
are the constraints of the theory, are given by,
\begin{eqnarray}
\Omega_1(\sigma) &=& g\,T^{00} = - T_{11} = \left(
\Pi^2(\sigma) + X^{\prime\,2}(\sigma)\right) = 0 \nonumber \\
\Omega_2(\sigma) &=& {\sqrt {-g}}{T^0}_1 =
{\Pi}(\sigma) \cdot X^{\prime}(\sigma) = 0
\label{constraints}
\end{eqnarray}
The canonical Hamiltonian obtained from (\ref{paction}) by a
Legendre transformation is given by,
\begin{eqnarray}
H = \int d\sigma\, {\sqrt {-g}}{T^0}_0 = \int d\sigma {\sqrt
{-g}}\left({1\over 2 g_{11}}\, \Omega_1(\sigma) + {g_{01} \over
{\sqrt {-g}}\, g_{11}}\,\Omega_2(\sigma)\right) \label{ham}
\end{eqnarray}
expectedly, the Hamiltonian turns out to be a linear combination of the constraints.

Just as variation of $g_{ab}$ yields the constraints, variation of $X^{\mu}$
gives the equation of motion,
\begin{eqnarray}
\partial_a\,({\sqrt {-g}}g^{ab}\, \partial_bX^{\mu}) = 0
\label{eomb}
\end{eqnarray}
Finally, there is a mixed BC,
\begin{eqnarray}
\partial^{\sigma}\, X^{\mu}(\tau ,\sigma)\vert_{\sigma = 0,\pi} = 0
\label{bcb}
\end{eqnarray}
In the covariant form involving phase space variables, this is given by
\begin{eqnarray}
\left(\partial_\sigma\,X^{\mu} + {\sqrt {-g}}g^{01}\,\Pi^{\mu}\right)
\vert_{\sigma = 0,\pi}= 0.
\label{bcb1}
\end{eqnarray}
The non trivial basic PBs of the theory are:
\begin{eqnarray}
\left\{X^{\mu}(\tau ,\sigma ),\Pi_{\nu}(\tau ,\sigma^\prime)\right\} &=&
\delta^{\mu}_{\nu}\, \delta (\sigma - \sigma^{\prime}) \nonumber \\
\left\{g_{ab}(\tau ,\sigma ),\pi^{cd}(\tau ,\sigma^{\prime})\right\}
&=& {1\over 2}({\delta^c}_a
{\delta^d}_b+ {\delta^d}_a{\delta^c}_b)\delta (\sigma -\sigma^{\prime})
\label{bpb}
\end{eqnarray}
where $\delta (\sigma -\sigma^{\prime})$ is the usual one-dimensional
Dirac delta function.  From the basic PB, it is easy
to generate the following first class (involutive) algebra,
\begin{eqnarray}
\left\{\Omega _1(\sigma ),\Omega _1(\sigma^{\prime})\right\} &=&
4\, (\Omega _2(\sigma ) + \Omega _2(\sigma^{\prime}))
\partial_{\sigma}\delta(\sigma - \sigma^{\prime}), \nonumber \\
\{\Omega _2(\sigma ),\Omega _1(\sigma^{\prime})\} &=& (\Omega _1(\sigma ) +
\Omega_1(\sigma^{\prime}))\partial_{\sigma}\delta(\sigma -\sigma^{\prime}),
\nonumber \\
\{\Omega_2(\sigma ),\Omega _2(\sigma^{\prime})\} &=& (\Omega _2(\sigma )+\Omega _2
(\sigma^{\prime}))\partial_{\sigma }\delta(\sigma -\sigma^{\prime}).
\label{alg}
\end{eqnarray}

\section{Modified brackets for the Polyakov string}
Let us again consider the BCs for the Polyakov string,
\begin{eqnarray}
\left(\partial_\sigma\,X^{\mu} + {\sqrt {-g}}g^{01}\,\Pi^{\mu}\right)
\vert_{\sigma = 0,\pi}= 0.
\label{bcb1r}
\end{eqnarray}
It is easily seen that the above BCs are incompatible with the
first of the basic PBs (\ref{bpb}). Hence the brackets should be
modified suitably. The modification of PBs can be done in spirit
to the treatment of Hanson {\it{et al.}} \cite{hrt}, where
modified PBs were obtained for the free Nambu-Goto string.

We would also like to mention that there is an apparent
contradiction of the constraint $\pi_{ab}\approx 0$ with the
second PB (\ref{bpb}). However this equality is valid in
Dirac's ``weak" sense only, so that it can be set equal to zero
only after the relevant brackets have been computed. These weak equalities will be designated by $\approx$,
rather than an equality, which is reserved only for a strong equality. In this
sense, therefore, there is no clash between this constraint and the relevant PB.
Indeed, we can even ignore the canonical pair $(g_{ab}, \pi^{cd})$ from the
basic PB.

The situation is quite similar to usual electrodynamics. There the
Lagrange multiplier is $A_0$, which corresponds to $g_{ab}$ in the
string theory. The multiplier $A_0$ enforces the
Gauss constraint just as $g_{ab}$ enforces the constraints $\Omega_1$
and $\Omega_2$. Furthermore, the Gauss constraint generates
the time independent gauge transformations, while
$\Omega_1$, $\Omega_2$ generate the diffeomorphism transformations.

The BC (\ref{bcb1r}), on the other hand, is not a constraint in
the Dirac sense \cite{di}, since it is applicable only at the
boundary. Thus, there has to be an appropriate modification in the
PB, to incorporate this condition. This is not unexpected and
occurs, for instance, in the example of a free scalar field $\phi
(x) $ in $(1+1)$ dimension, subjected to periodic BC of period,
say, $2\pi$ ($\phi (t,x+2 \pi )= \phi (t,x)$). There the PB
between the field $\phi (t,x)$ and its conjugate momentum $\pi
(t,x)$ are given by,
\begin{eqnarray}
\left\{\phi (t,x),\pi (t,y)\right\}= \delta_P (x-y)
\label{sca}
\end{eqnarray}
where,
\begin{eqnarray}
\delta_P(x-y) = \delta_P(x - y + 2\pi) = {1\over {2\pi}}\sum_{n\in
{\cal Z}}e^{in(x-y)} \label{pdf}
\end{eqnarray}
is the periodic delta function of period $2\pi$ \cite{schwinger}
and occurs in the closure properties of the basis functions
$e^{inx}$ for the space of square integrable functions, defined on
the unit circle $S^1$. This periodic delta function is related to
the usual Dirac delta function as $\delta_P(x-y) = \sum_{n\in
{\cal Z}} \delta(x - y + 2\pi n)$


Before discussing the mixed type condition (\ref{bcb1r}), that emerged in a
completely gauge independent formulation of the Polyakov action,
consider the simpler Neumann type condition
$(\partial_\sigma X^{\mu})\vert_{\sigma
 =0,\pi}=0$ in an orthonormal (conformal) gauge.
It is easy to find the solutions to the equations of motion (\ref{eomb})
which are compatible with the Neumann BCs
\begin{eqnarray}
X^{\mu}(\tau, \sigma) = x^{\mu} + p^{\mu}\,\tau
+ i \sum_{n \neq 0 }\frac{\alpha^{\mu}_{n}}{n}\,e^{-in\tau}\,
\mathrm{cos}(n\sigma)
\label{mode}
\end{eqnarray}
Reality of $X^{\mu}(\tau, \sigma)$ implies that $x$, $p$ are real and
\begin{eqnarray}
\alpha^{\mu\, \star}_{n} = \alpha^{\mu}_{-n} \quad \mathrm{for}\ n\neq 0.
\label{mode1}
\end{eqnarray}
We enlarge the domain of definition of the bosonic field $X^{\mu}$
from $[0,\pi]$ to $[-\pi,\pi]$ by observing the fact
\begin{eqnarray}
X^{\mu}(\tau , -\sigma) = X^{\mu}(\tau , \sigma) \ \mathrm{under}\ \sigma \to -\sigma
\label{even}
\end{eqnarray}
which further yields
\begin{eqnarray}
X^{\mu}(-\pi) = X^{\mu}(\pi).
\label{per}
\end{eqnarray}
Now we start by noting that the usual properties of a delta
function is also satisfied by $\delta_P (x)$ (\ref{pdf}),
\begin{eqnarray}
\int_{-\pi}^{\pi} dx'\, \delta_{P}(x' - x)\,f(x') = f(x)
\label{dpop}
\end{eqnarray}
for any periodic function $f(x) = f(x + 2\pi)$ defined in the interval
$[-\pi , \pi]$. Then by using (\ref{even}), the above integral (\ref{dpop})
reduces to the following:
\begin{eqnarray}
\int_{0}^{\pi} d\sigma'\, \Delta_{+}(\sigma', \sigma)\,
X^\mu (\sigma') = X^\mu (\sigma)
\label{dpop1}
\end{eqnarray}
where
\begin{eqnarray}
\Delta_{+}(\sigma', \sigma)\, = \delta_P (\sigma' - \sigma) +
\delta_P (\sigma' + \sigma).
\label{Del}
\end{eqnarray}
Using (\ref{pdf}), the explicit form of $\Delta_{+}(\sigma', \sigma)$
can be given as,
\begin{eqnarray}
\Delta_{+}(\sigma', \sigma)\, = {1\over \pi } + {1\over \pi }
\sum_{n \neq 0}\mathrm{cos}(n\sigma' )\mathrm{cos}(n\sigma ) .
\label{Del1}
\end{eqnarray}
It thus follows that the appropriate PB is given by,
\begin{eqnarray}
\left\{ X^{\mu}(\tau, \sigma), \Pi_{\nu}(\tau, \sigma')\right\}
= \delta^{\mu}_{\nu} \Delta_{+}(\sigma', \sigma).
\label{mpb1}
\end{eqnarray}
It is clearly consistent with Neumann BC as $\partial_{\sigma}\,
\Delta_{+}(\sigma ,\sigma')\vert_{\sigma = 0,\pi } =
\partial_{\sigma'}\,\Delta_{+}(\sigma ,\sigma')\vert_{\sigma = 0,
\pi } = 0$ and is automatically satisfied. Observe also that the
other brackets
\begin{eqnarray}
\left\{X^{\mu}(\tau ,\sigma ),X^{\nu }(\tau ,\sigma')\right\} &=& 0
\label{mpb2} \\
\left\{\Pi^{\mu}(\tau ,\sigma ), \Pi^{\nu }(\tau ,\sigma )\right\} &=& 0
\label{mpb3}
\end{eqnarray}
are already consistent with the Neumann BCs and hence remain
unchanged.

For a gauge independent analysis, we take recourse to the mixed condition
(\ref{bcb1r}). A simple inspection
shows that this is also compatible with the modified brackets (\ref{mpb1}
, \ref{mpb3}), but not with (\ref{mpb2}). Hence the bracket
among the coordinates should be altered suitably.
We therefore make an ansatz,
\begin{eqnarray}
\{X^{\mu}(\tau ,\sigma ),X^{\nu }(\tau ,\sigma')\} =
C^{\mu \nu}(\sigma ,\sigma')
\label{ans}
\end{eqnarray}
where,
$$C^{\mu \nu}(\sigma ,\sigma') = - C^{\nu \mu}(\sigma' ,\sigma). $$
Imposing the BC (\ref{bcb1r}) on this algebra, we get,
\begin{eqnarray}
\partial_{\sigma'}\,C^{\mu\nu}(\sigma ,\sigma')\vert_{\sigma'=0,\pi}
&=& \partial_{\sigma }\,C^{\mu\nu}(\sigma ,\sigma')
\vert_{\sigma =0,\pi} \nonumber\\
&=& - {\sqrt {-g}}g^{01}\{\Pi^{\mu}(\tau ,\sigma),
X^{\nu}(\tau,\sigma')\}\nonumber\\
&=&{\sqrt {-g}}g^{01}\eta^{\mu \nu}\Delta_{+}(\sigma ,\sigma')
\label{ans1}
\end{eqnarray}
For an arbitrary form of the metric tensor, it might be
technically problematic to find a solution for $C^{\mu\nu}(\sigma
,\sigma')$. However, for a restricted class of metric{\footnote
{Such conditions also follow from a standard treatment of the
light-cone gauge \cite{pol}}} that satisfy
$$\partial_\sigma g_{ab}=0$$
it is possible to give a quick solution of $C^{\mu\nu}(\sigma ,\sigma')$ as,
\begin{eqnarray}
C^{\mu\nu}(\sigma ,\sigma') = {\sqrt {-g}}g^{01}\eta^{\mu\nu}
\left[\Theta(\sigma ,\sigma') - \Theta(\sigma',\sigma )\right]
\label{csol}
\end{eqnarray}
where the generalised step function $\Theta (\sigma ,\sigma')$ satisfies,
\begin{eqnarray}
\partial_{\sigma }\Theta (\sigma ,\sigma') =
\Delta_{+}(\sigma ,\sigma')
\label{theta}
\end{eqnarray}
An explicit form of $\Theta $ is given by \cite{hrt},
\begin{eqnarray}
\Theta (\sigma ,\sigma') = {\sigma \over \pi} + {1\over \pi }
\sum_{n\neq 0}{1\over n}\mathrm{sin}(n\sigma )\mathrm{cos}
(n\sigma' )~,
\label{csol1}
\end{eqnarray}
having the properties,
\begin{eqnarray}
\Theta (\sigma ,\sigma') &=& 1~~~ \mathrm{for} ~~\sigma >\sigma'~, \nonumber \\
\Theta (\sigma ,\sigma') &=& 0 ~~~\mathrm{for}~~ \sigma <\sigma'.
\label{the2}
\end{eqnarray}
Using these relations, the simplified structure of
noncommutative algebra follows,
\begin{eqnarray}
\{X^\mu (\tau,\sigma ),X^{\nu}(\tau, \sigma' )\} &=& 0~~~
\mathrm{for} ~~\sigma =\sigma' \nonumber \\
\{X^\mu (\tau,\sigma ),X^{\nu}(\tau, \sigma' )\} &=& \pm {\sqrt
{-g}}\,g^{01}\, \eta^{\mu \nu}~~~\mathrm{for} ~~\sigma  \neq
\sigma' \label{fnon}
\end{eqnarray}
respectively. Thus a noncommutative algebra for distinct
coordinates $\sigma \neq \sigma' $ of the string emerges
automatically in a free string theory if a gauge independent
analysis is carried out like this. But this non-commutativity can
be made to vanish in gauges like conformal gauge, where
$g^{01}=0$, thereby restoring the usual commutative structure.

Now using the modified basic brackets we obtain the following
involutive constraint algebra\footnote{Note that there were some
errors in \cite{rb}}
\begin{eqnarray}
\{\Omega_1(\sigma) , \Omega_1(\sigma^{\prime})\} &=&
\Omega_1(\sigma^{\prime})\partial_{\sigma}\Delta_{+} \left(\sigma
, \sigma^{\prime}\right) + \Omega_1(\sigma)
\partial_{\sigma}\Delta_{-}\left(\sigma , \sigma^{\prime}\right) \,
\nonumber \\
\{\Omega_1(\sigma) , \Omega_2(\sigma^{\prime})\} &=&  \left(
\Omega_2(\sigma) + \Omega_2(\sigma^{\prime})\right)
\partial_{\sigma}\Delta_{+}\left(\sigma , \sigma^{\prime}\right)\,
\nonumber \\
\{\Omega_2(\sigma) , \Omega_2(\sigma^{\prime})\} &=& 4 \left(
\Omega_1(\sigma)\partial_{\sigma}\Delta_{+} \left(\sigma ,
\sigma^{\prime}\right) + \Omega_1(\sigma^{\prime})
\partial_{\sigma}\Delta_{-}\left(\sigma , \sigma^{\prime}\right)\right).
\label{modbr}
\end{eqnarray}
A crucial intermediate step in the above derivation is to use the
relation,
\begin{eqnarray}
\{X^{\prime\mu}(\sigma), X^{\prime\nu}(\sigma^{\prime})\} = 0
\label{importantstep}
\end{eqnarray}
which follows from the basic bracket (\ref{fnon}) \cite{rb}.

\section{Interacting Polyakov string}
The Polyakov action for a bosonic string moving in the presence of
a constant background Neveu-Schwarz two-form field $B_{\mu \nu}$
is given by,
\begin{eqnarray}
S_P = -{1\over 2}\int d\tau d\sigma \left( {\sqrt{-g}}g^{ab}\,
{\partial}_a X^{\mu}\,{\partial}_bX_{\mu}
+ e \epsilon^{ab}\,B_{\mu \nu}\,\partial_aX^{\mu}\,\partial_bX^{\nu}\right)
\label{ipoly}
\end{eqnarray}
where $\epsilon^{01}=-\epsilon^{10}=+1$.
A usual canonical analysis leads to the following set of
primary first class constraints,
\begin{eqnarray}
gT^{00} &=& {1\over 2}\left[(\Pi_{\mu} + eB_{\mu \nu}\partial_\sigma X^{\nu})^2
+ (\partial_\sigma X)^2\right] \approx 0 \nonumber \\
{\sqrt {-g}}\, T^{0}_{\ 1} &=& {\Pi} .{\partial}_\sigma X \approx 0
\label{icstr}
\end{eqnarray}
where
\begin{eqnarray}
\Pi_{\mu} = -{\sqrt {-g}}\partial^\tau X_{\mu}
+ e B_{\mu \nu}\,\partial_\sigma X^{\nu}
\label{1i1}
\end{eqnarray}
is the momentum conjugate to $X^{\mu}$. The boundary condition
written in terms of phase-space variables is
\begin{eqnarray}
\left[\partial_\sigma X_\mu + \Pi^\rho \left(N M^{-1}\right)_{
\rho \mu}\right]_{\sigma = 0, \pi} = 0
\label{1i2}
\end{eqnarray}
where,
\begin{eqnarray}
{M^{\rho}}_{\mu}&=& {1\over g_{11}}[{\delta^{\rho}}_{\mu}
-{{2e}\over {\sqrt{-g}}}\,g_{01}{B^{\rho}}_{\mu}
+ e^2\,B^{\rho \nu}B_{\nu \mu}] \nonumber \\
N_{\nu \mu} &=& -{g^{01}\over g^{00}{\sqrt{-g}}}\,\eta_{\nu \mu}-
{1\over g_{11}}\,e\,B_{\nu \mu}
\label{1i3}
\end{eqnarray}
are two matrices.

The $\{X^\mu , \Pi_\nu\}$ Poisson bracket is the same as that of the
free string whereas considering the general structure (\ref{ans})
and exploiting the above boundary condition, one obtains
\begin{eqnarray}
\partial_\sigma C_{\mu\nu}(\sigma ,\sigma ')\mid _{\sigma =0,\pi}
= \left(NM^{-1}\right)_{\nu \mu}\Delta_+(\sigma, \sigma')
\mid_{\sigma = 0,\pi}.
\label{1i4}
\end{eqnarray}
As in the free case, we restrict to the class of metrics defined
satisfying $\partial_\sigma g_{ab} = 0$, the above equation has a solution
\begin{eqnarray}
C_{\mu\nu}(\sigma ,\sigma ') &=&
{1\over 2}(NM^{-1})_{(\nu \mu )}\left[\Theta (\sigma, \sigma')
-\Theta (\sigma', \sigma )\right]\nonumber \\
&& +{1\over 2}(NM^{-1})_{[\nu \mu]}\left[\Theta (\sigma, \sigma')
+ \Theta (\sigma', \sigma ) - 1\right].
\label{1i5}
\end{eqnarray}
where $(NM^{-1})_{(\nu \mu )}$ the symmetric and $(NM^{-1})_{[\nu
\mu]}$ the antisymmetric part of $(NM^{-1})_{\nu \mu}$. The
modified algebra is gauge dependent; it depends on the choice of
the metric. However, there is no choice, for which the
non-commutativity vanishes. To show this, note that the origin of
the non-commutativity is the presence of non-vanishing
$N^{\nu\mu}$ in the BC (\ref{1i2}). Vanishing $N^{\nu\mu}$ would
make $B_{\mu\nu}$ and $\eta_{\mu\nu}$ proportional to each other
which obviously cannot happen, as the former is an antisymmetric
and the latter is a symmetric tensor. Hence non-commutativity will
persist for any choice of world-sheet metric $g_{ab}$.

\section{Summary}
In this chapter we have discussed the Polyakov string and derived
the expressions for a noncommutative algebra \cite{rb}, that are
more general than the standard results found in the conformal
gauge. The origin of any modification in the usual Poisson algebra
is the presence of boundary conditions. This phenomenon is quite
well known for a free scalar field subjected to periodic boundary
conditions. We showed that its exact analogue is the conformal
gauge fixed free string, where the boundary condition is of
Neuman-type. This led to a modification only in the $\{X^{\mu
}(\sigma ),\Pi_{\nu }(\sigma' )\}$ algebra, where the usual Dirac
delta function got replaced by $\Delta_+ (\sigma ,\sigma' )$.
Using certain algebraic consistency requirements, we showed that
the boundary conditions in the free theory naturally led to a
noncommutative structure among the coordinates. This
non-commutativity, however, vanishes in the conformal gauge, as
expected. The same technique was adopted for the interacting
string. Here, on contrast, we find that there is a genuine
noncommutativity at the string end points and can not be made to
vanish in any gauge.

\chapter{Interpolating string: A study of gauge symmetries in a noncommutative framework}
The dynamics of a bosonic string is described either by
Nambu--Goto(NG)  or Polyakov action. Both these actions, though
very well-known in the literature, poses certain degrees of
difficulty. NG formalism is inconvenient for path integral
quantisation whereas Polyakov action involves many redundant
degrees of freedom. However, yet another formulation,
interpolating between these two versions of string action, has
also been put forward in the literature \cite{rb, agh3}. This
interpolating Lagrangian, in a certain sense, is a better
description of the theory in the sense that it neither objects to
quantization nor has as many redundancies as in the Polyakov
version. Further, it gives a perfect platform to study the gauge
symmetries vis-\`{a}-vis reparametrisation symmetries of the
various free string actions by a constrained Hamiltonian approach
\cite{rbpmas1, rbpmas2}.

In the present chapter, acknowledging the above facts, we derive a
master action for interacting bosonic strings, interpolating
between the NG and Polyakov formalism. Modification of the basic
PB structure compatible with BC(s) followed by the emergence of
the noncommutativity is shown in this formalism (in case of both
free and interacting strings) following the approach discussed in
previous chapter \cite{rb, rbbckk, agh1}. Our results go over
smoothly to the Polyakov version once proper identifications are
made. These modified PBs lead to a new involutive constraint
algebra which is markedly different from that given in \cite{rb}.
With the above results at our disposal, we go over to the study of
gauge symmetry in the NC framework. Owing to the new constraint
algebra we find surprising changes in the structure constants of
the theory. Finally, we compute the gauge variations of the fields
and show the underlying unity of diffeomorphism with the gauge
symmetry in the NC framework.

%
\section{The interacting theory: Nambu-Goto Formulation}
Although the Polyakov and NG formulations for free strings are
regarded to be classically equivalent, there are are some subtle
issues. Indeed, the structures of BCs in the two formulations are
different. Also more complications are expected in the presence of
interactions. In this section, we analyse the NG formulation of
the interacting bosonic string. As we shall see in the next
section, this is essential in the construction of the
Interpolating Lagrangian for the interacting string. The NG action
for a bosonic string moving in the presence of a constant
background Neveu-Schwarz two-form field $B_{\mu\nu}$ reads:
\begin{eqnarray}
S_{NG}=\int^{\infty}_{- \infty} d\tau \int^{\pi}_{0}d\sigma
\left[{\mathcal L}_{0} + eB_{\mu\nu}\dot{X}^\mu
X^{\prime \nu}\right]
\label{NGaction}
\end{eqnarray}
where ${\mathcal L}_{0}$ is the free NG Lagrangian density
given by
\begin{eqnarray}
{\mathcal L}_{0} = -\left[(\dot{X}.X^{\prime})^2-\dot{X}^{2}X^{\prime 2}\right]
^\frac{1}{2} .
\label{freeNGaction}
\end{eqnarray}
The Euler-Lagrange equations and BC obtained by varying the action
read:
\begin{eqnarray}
\dot{\Pi}^{\mu} + K^{\prime \mu} &=& 0 \nonumber \\
K^{\mu}\vert_{\sigma = 0, \pi} &=& 0
\label{EL}
\end{eqnarray}
where,
\begin{eqnarray}
\Pi_{\mu} &=& \frac{\partial {{\mathcal L}}}{\partial \dot{X}^{\mu}}
= {\mathcal L}_{0}^{-1}\left(-X^{\prime 2}\dot{X}_{\mu}
+(\dot{X}.X^{\prime})X^{\prime}_{\mu}\right) + eB_{\mu\nu}X^{\prime \nu}
\nonumber \\
K_{\mu} &=& \frac{\partial {{\mathcal L}}}{\partial X^{\prime \mu}}
= {\mathcal L}_{0}^{-1}\left(- \dot{X}^2 X^{\prime}_{\mu}
+ (\dot{X} \cdot X^{\prime})\dot{X}_{\mu}\right) -e B_{\mu\nu}\dot{X}^{\nu}.
\label{NGmomentum}
\end{eqnarray}
Note that $\Pi_{\mu}$ is the canonically conjugate momentum
to $X^{\mu}$.
The first of (\ref{bpb}) is the only nontrivial PB of NG theory.
The primary constraints of the theory are:
\begin{eqnarray}
\Omega_{1} &=& \Pi_{\mu}X^{\prime \mu} = 0\nonumber\\
\Omega_{2} &=& \left(\Pi_{\mu} - e B_{\mu \nu} X^{\prime
\nu}\right)^2 + X^{\prime 2} = 0
\label{NGconstraints}
\end{eqnarray}
and they generate the same first class (involutive) algebra as that of
Polyakov string (\ref{alg})

Now as happens for a reparametrisation invariant theory,
 the canonical Hamiltonian density defined by a Legendre transform vanishes
\begin{eqnarray}
{\mathcal H}_{c} = \Pi_{\mu}\dot{X}^{\mu} - {\mathcal L} = 0.
\label{CanonicalH1}
\end{eqnarray}
This can be easily seen by substituting
(\ref{NGmomentum}) in (\ref{CanonicalH1}).
The total Hamiltonian density is thus given by
a linear combination of the first class
constraints (\ref{NGconstraints}):
\begin{eqnarray}
{\mathcal H}_{T} = - \rho \,\Omega_{1} - \frac{\lambda}{2}\,\Omega_{2}
\label{totalH}
\end{eqnarray}
where $\rho$ and $\lambda$ are Lagrange multipliers. It is easy
to check that time preserving the primary constraints yields no
new secondary constraints. Hence the total set of constraints
 of the NG theory is given by the first-class system (\ref{NGconstraints}).

As in the case of Polyakov string, here also we enlarge the domain
of definition of the bosonic field $X^{\mu}$ from $[0,\pi]$ to
$[-\pi,\pi]$ in order to write down the generators of $\tau$ and
$\sigma$ reparametrisation in a compact form. We define
\begin{eqnarray}
X^{\mu}(\tau , -\sigma) = X^{\mu}(\tau , \sigma)\ \ ; \ \
B_{\mu \nu} \to - B_{\mu \nu} \ \mathrm{under}\ \sigma \to -\sigma.
\label{37}
\end{eqnarray}
The second condition implies that $B_{\mu \nu}$, albeit a constant,
transforms as a pseudo scalar under $\sigma \to -\sigma$
in the extended interval. This ensures that the interaction term
$eB_{\mu\nu}\dot{X}^{\mu}X^{\prime \nu}$ in (\ref{NGaction})
remains invariant under $\sigma \to - \sigma$ like the
free NG Lagrangian density ${\cal{L}}_{0}$ (\ref{freeNGaction}).
Consistent with this, we have
\begin{eqnarray}
\Pi^{\mu}(\tau , -\sigma) = \Pi^{\mu}(\tau , \sigma),
\quad
X^{\prime \mu}(\tau , -\sigma) = - X^{\prime \mu}(\tau , \sigma).
\label{37h}
\end{eqnarray}
Now, from  (\ref{NGconstraints}), (\ref{37}) we note that the
constraints $\Omega_1(\sigma) = 0$ and $\Omega_2(\sigma) = 0$ are
odd and even respectively under $\sigma \rightarrow -\sigma$. Now
demanding the total Hamiltonian density $\mathcal{H_T}$
(\ref{totalH}) also remains invariant under $\sigma \to - \sigma$,
one finds that $\rho$ and $\lambda$ must be odd and even
respectively under $\sigma \to - \sigma$.

\noindent
We may then write the generator of all $\tau$ and $\sigma$
reparametrisation as the functional \cite{hrt}:
\begin{eqnarray}
L[f] = \frac{1}{2}\int_{0}^{\pi} d\sigma \{f_+(\sigma) \Omega_2(\sigma)
+ 2 f_-(\sigma) \Omega_1(\sigma)\}\,,
\label{32u}
\end{eqnarray}
where, $f_{\pm}(\sigma)=\frac{1}{2}(f(\sigma)\pm f(-\sigma))$ are by
construction even and odd function and  $f(\sigma)$ is an arbitary
differentiable function defined in the extended interval $[-\pi, \pi]$.
The above expression can be simplified to:
\begin{eqnarray}
L[f] &=& \frac{1}{4}\int_{-\pi}^{\pi} d\sigma f(\sigma)
\left[\Omega_2(\sigma) + 2\Omega_1(\sigma)\right] \nonumber \\
&=& \frac{1}{4}\int_{-\pi}^{\pi} d\sigma f(\sigma)
\left[\Pi_{\mu}(\sigma) + X^{\prime}_{\mu}(\sigma)
-e B_{\mu \nu}X^{\prime \nu}(\sigma) \right]^2
\label{33u}.
\end{eqnarray}
It is now easy to verify (using (\ref{alg})) that
the above functional (\ref{33u}) generates the following Virasoro algebra:
\begin{eqnarray}
\{L[f(\sigma)] , L[g(\sigma)]\} &=& L[f(\sigma)g^{\prime}(\sigma)
- f^{\prime}(\sigma)g(\sigma)].
\label{34a}
\end{eqnarray}
Defining
\begin{eqnarray}
L_m = L[e^{-im\sigma}]\,,
\label{34y}
\end{eqnarray}
one can write down an equivalent form of the Virasoro algebra
\begin{eqnarray}
\{L_m , L_n\} &=& i (m-n) L_{m + n}\,.
\label{34b}
\end{eqnarray}
Note that we do not have a central extension here, as
the analysis is entirely classical.

\section{Interacting String in interpolating formalism}
In the previous section we have reviewed the salient
features of the interacting NG string.
We now pass on to the construction of the interpolating action of the
interacting string\footnote{The construction of the interpolating
action for the free string has been discussed in \cite{rb}.}. To
achieve this end, we write down the Lagrangian
 of the interacting NG action in the
first-order form:
\begin{eqnarray}
{\mathcal L}_{I} = \Pi_{\mu}\dot{X}^{\mu} - {\mathcal H}_{T}\,.
\label{1stL}
\end{eqnarray}
Substituting (\ref{totalH}) in (\ref{1stL}), ${\mathcal L}_{I}$ becomes
\begin{eqnarray}
{\mathcal L}_{I} = \Pi_{\mu}\dot{X}^{\mu}+\rho\Pi_{\mu}X^{\prime\mu}
+\frac{\lambda}{2}\left[(\Pi^2 + X^{\prime 2})-2eB_{\mu\nu}
\Pi^{\mu}X^{\prime\nu}+e^2B_{\mu\nu}B^{\mu}_{\ \rho}X^{\prime\nu}
X^{\prime\rho}\right]\,.
\label{subL}
\end{eqnarray}
The advantage of working with the interpolating action is that
it naturally leads to either the NG or the Polyakov formulations
of the string. In the Lagrangian (\ref{subL}), $\lambda$ and  $\rho$
originally introduced as Lagrange multipliers, will be treated as
independent fields, which behave as scalar and pseudo-scalar
fields respectively in the extended world-sheet, as was
discussed in the previous section.
 We will eliminate $\Pi_{\mu}$ from (\ref{subL}) as
it is an auxiliary field. The Euler-Lagrange equation for
 $\Pi_{\mu}$ is:
\begin{eqnarray}
\dot{X}^{\mu}+\rho X^{\prime\mu}+\lambda\Pi^{\mu}-
e\lambda B^{\mu\nu}X^{\prime}_{\nu}=0\,.
\label{momentumeqn}
\end{eqnarray}
Substituting $\Pi_{\mu}$ from (\ref{momentumeqn}) back in
(\ref{subL}) yields:
\begin{eqnarray}
{\mathcal L}_{I} = -\frac{1}{2\lambda}\left[\dot{X}^{2}+
2\rho(\dot{X}.X^{\prime})+(\rho^2-\lambda^2)X^{\prime 2}-2\lambda eB_{\mu\nu}
\dot{X}^{\mu}X^{\prime\nu}\right]\,.
\label{interpolatingL}
\end{eqnarray}
This is the form of the interpolating Lagrangian of the
interacting string.

The reproduction of the NG action (\ref{NGaction}) from the
interpolating action of the interacting string is trivial and can
be done by eliminating $\rho$ and $\lambda$ from their respective
Euler-Lagrange equations of motion following from
(\ref{interpolatingL}),
\begin{eqnarray}
\rho &=& - \frac{\dot{X} \cdot X^{\prime}}{X^{\prime 2}}\nonumber \\
\lambda^{2} &=& \frac{\left(\dot{X} \cdot X^{\prime}\right)^2
- \dot{X}^2 X^{\prime 2}}{X^{\prime 2} X^{\prime 2}}.
\label{rholambdaeqn}
\end{eqnarray}
From this equation $\lambda$ is determined modulo a sign which can be
fixed by demanding the consistency of Eq. (\ref{NGmomentum})
with Eq. (\ref{momentumeqn}). Accordingly,
\begin{eqnarray}
\lambda = - \frac{\left[\left(\dot{X} \cdot X^{\prime}\right)^2
- \dot{X}^2 X^{\prime 2}\right]^{1 \over 2}}{X^{\prime 2} X^{\prime 2}}.
\label{lambdaeqn}
\end{eqnarray}
If, on the other hand, we identify $\rho$ and $\lambda$
with the following contravariant
components of the world-sheet metric,
\begin{eqnarray}
g^{ab}=(-g)^{-{1\over 2}}\pmatrix{{1\over \lambda} & {\rho \over
\lambda}\cr {\rho \over \lambda} &
{(\rho^2-\lambda^2)\over {\lambda}}}
\label{idm}
\end{eqnarray}
then the above Lagrangian (\ref{interpolatingL})
reduces to the Polyakov form,
\begin{eqnarray}
{\mathcal L}_P= -{1\over 2}\left( {\sqrt {-g}}g^{ab}{\partial}_a
X^{\mu}{\partial}_bX_{\mu}
-e\epsilon^{ab}B_{\mu \nu}\partial_{a} X^{\mu}\partial_{b} X^{\nu}\right)
\ \ ; \  \ \left(a,b = \tau,\sigma\right).
\label{Polyaction}
\end{eqnarray}
\noindent In this sense, therefore, the Lagrangian in (\ref{interpolatingL})
is referred to as an interpolating Lagrangian.
It should be noted that the interpolating action has only
two additional degrees of freedom, $\lambda$ and $\rho$,
 which does not fully match the three degrees of freedom of the
worldsheet metric of the Polyakov action. However, due to Weyl
invariance of the Polyakov action, only two of the three different
metric coefficients $g_{ab}$ are really independent. This Weyl
invariance is special to the Polyakov string, the higher branes do
not share it.

We can now, likewise construct the interpolating BC
from the interpolating Lagrangian (\ref{interpolatingL}),
\begin{eqnarray}
K^{\mu} = \left[\partial {\cal{L}}_{I}\over \partial
X^{\prime}_{\mu}\right]_{\sigma =0,\pi} = \left({\rho \over \lambda}
{\dot X}^{\mu}+{\rho^2 -
\lambda^2 \over \lambda}X'^{\mu} + eB^{\mu}_{\ \nu}\dot{X}^{\nu}
\right)_{\sigma = 0, \pi} = 0.
\label{16}
\end{eqnarray}
The fact that this can be easily interpreted as interpolating
BC, can be easily seen by using the expressions
(\ref{rholambdaeqn}) for $\rho$ and $\lambda$ in  (\ref{16})
to yield:
\begin{eqnarray}
\left[{\mathcal L}^{- 1}_{0}\left( - \dot{X}^{2} X^{\prime \mu} +
\left(\dot{X} X^{\prime}\right)\dot{X}^{\mu}\right) - e B^{\mu
\nu}\dot{X}_{\nu}\right]_{\sigma = 0, \pi} = 0\, ,
 \label{17}
\end{eqnarray}
which is the BC of the interacting NG string (\ref{NGmomentum}).

\noindent On the other hand, we can identify $\rho$ and $\lambda$
with the metric components as in (\ref{idm}) to recast (\ref{16})
as:
\begin{eqnarray}
\left(g^{1a}\partial_{a}X^{\mu}(\sigma) +
\frac{1}{\sqrt{-g}}e B^{\mu}_{\ \nu}\partial_{0}
X^{\nu}(\sigma)
\right)_{\sigma = 0, \pi} = 0.
\label{16a}
\end{eqnarray}
which is easily identifiable with Polyakov form of BC \cite{rb}
following from the action (\ref{Polyaction}).

\noindent
Using phase space variables $X^{\mu}$ and
$\Pi_{\mu}$, (\ref{16}) can be rewritten as
\begin{eqnarray}
K^{\mu} = \left[\left(\rho \Pi^{\mu} + \lambda X^{\prime \mu}\right)
+ e B^{\mu}_{\ \nu}\left(\Pi^{\nu} - e B^{\nu}_{\ \rho}
X^{\prime \rho}\right)\right]_{\sigma = 0, \pi} = 0.
\label{18}
\end{eqnarray}
 Hence it is possible to interpret either of
(\ref{16}) or (\ref{18}) as an interpolating BC.

\noindent
Now we come to the discussion of the constraint structure
of the interpolating interacting string.
Note that the independent fields in
(\ref{interpolatingL}) are $X^{\mu}$, $\rho$ and $\lambda$.
The corresponding momenta denoted by $\Pi_{\mu}$,
$\pi_{\rho}$ and $\pi_{\lambda}$, are given as:
\begin{eqnarray}
\Pi_{\mu} &=&  -\frac{1}{\lambda}\left(\dot{X}_{\mu} +
\rho X^{\prime}_{\mu}\right) + eB_{\mu\nu}X^{\prime \nu} \nonumber\\
\pi{\rho}& = & 0 \nonumber\\
\pi_{\lambda}& = & 0\,.
\label{211}
\end{eqnarray}
In addition to the PB(s) similar to (\ref{bpb}), we now have:
\begin{eqnarray}
 \{\rho\left(\tau,\sigma\right),
 \pi_{\rho}\left(\tau,\sigma^{\prime}\right)\} =
  \delta\left(\sigma - \sigma^{\prime}\right) \nonumber\\
 \{\lambda\left(\tau,\sigma\right),
 \pi_{\lambda}\left(\tau,\sigma^{\prime}\right)\} =
  \delta\left(\sigma - \sigma^{\prime}\right).
\label{interpolatingPB}
\end{eqnarray}
The canonical Hamiltonian following from (\ref{interpolatingL}) reads:
\begin{equation}
{\cal{H}}_c = -\rho \Pi_{\mu}X^{\prime\mu}
- \frac{\lambda}{2}\left\{\left(\Pi_{\mu} - e B_{\mu \nu}
X^{\prime \nu}\right)^2 + X^{\prime 2}\right\}
\label{canonicalH2}
\end{equation}
which reproduces the total Hamiltonian (\ref{totalH}) of the NG action.
From the definition of the canonical momenta we can
easily identify the primary constraints:
\begin{eqnarray}
\Omega_{3} = \pi{\rho}& = & 0 \nonumber\\
\Omega_{4} = \pi_{\lambda}& = & 0\,.
\label{int_primary}
\end{eqnarray}
The conservation of the above primary constraints leads to the
secondary constraints $\Omega_1$ and $\Omega_2$ of
(\ref{NGconstraints}). The primary constraints of the NG action
appear as secondary constraints in this formalism. The system of
constraints for the Interpolating Lagrangian thus comprises of the
set (\ref{int_primary}) and (\ref{NGconstraints}). The PB(s) of
the constraints of (\ref{int_primary}) vanish within themselves.
Also the PB of these with (\ref{NGconstraints}) vanish.

\section{Modified brackets for Interpolating String}
\subsection{Free Interpolating String:}
Let us consider boundary condition for free interpolating string
which can be obtained by setting $B_{\mu \nu} = 0$ in (\ref{18}):
\begin{eqnarray}
K^{\mu} = \left[\left(\rho \Pi^{\mu} + \lambda X^{\prime \mu}\right)
 \right]_{\sigma = 0, \pi} = 0.
\label{20}
\end{eqnarray}
It is now easy to note that the above BC is not compatible with
the basic PB (\ref{bpb}). To incorporate this, an appropriate
modification in the PB is in order. In the previous chapter, the
equal time brackets were given in terms of certain combinations
($\Delta_{+}(\sigma , \sigma^{\prime})$) of periodic delta
function \cite{rb, hrt, rbbckk, agh1},
\begin{eqnarray}
\{X^{\mu}(\tau , \sigma) ,  \Pi_{\nu}(\tau , \sigma^{\prime})\}
= \delta^{\mu}_{\nu} \Delta_{+}(\sigma, \sigma^{\prime})
\label{23}
\end{eqnarray}
where,
\begin{eqnarray}
\Delta_{+}\left(\sigma , \sigma^{\prime}\right)
&=& \delta_{P}(\sigma - \sigma^{\prime})
 + \delta_{P}(\sigma + \sigma^{\prime})
= \frac{1}{\pi} + \frac{1}{\pi}\sum_{n\neq 0}
\mathrm{cos}(n\sigma^{\prime})\mathrm{cos}(n\sigma)
\nonumber \\
\Delta_{-}\left(\sigma , \sigma^{\prime}\right)
&=& \delta_{P}(\sigma - \sigma^{\prime})
 - \delta_{P}(\sigma + \sigma^{\prime})
=  \frac{1}{\pi}\sum_{n\neq 0}
\mathrm{sin}(n\sigma^{\prime})\mathrm{sin}(n\sigma)
\label{24}
\end{eqnarray}
 rather than an ordinary delta function to ensure
compatibility with Neumann BC
\begin{eqnarray}
\partial_{\sigma}X^{\mu}(\sigma)\vert_{\sigma = 0, \pi} = 0\,,
\label{16b}
\end{eqnarray}
in the bosonic sector. Remember that the other brackets
\begin{eqnarray}
\{X^{\mu}\left(\sigma\right), X^{\nu}\left(\sigma^{\prime}\right)\}
 & = & 0
\label{24q}\\
\{\Pi^{\mu}\left(\sigma\right), \Pi^{\nu}
\left(\sigma^{\prime}\right)\} & = & 0
\label{24p}
\end{eqnarray}
are consistent with the Neumann boundary condition (\ref{16b}).

\noindent Now a simple inspection shows that the BC (\ref{20}) is
also compatible with (\ref{23})\footnote{Note that there is no
inconsistency in (\ref{16b}) as
$\partial_{\sigma}\Delta_{+}\left(\sigma , \sigma^{\prime}\right)
\vert_{\sigma = 0, \pi}= 0$.} and (\ref{24p}), but not with
(\ref{interpolatingPB}) and (\ref{24q}). Hence the brackets
(\ref{interpolatingPB}) and (\ref{24q}) should be altered
suitably.

\noindent Now, since $\rho$ and $\lambda$ are odd and even
functions of $\sigma$
respectively, we propose:
\begin{eqnarray}
\{\rho(\tau,\sigma ) , \pi_{\rho}(\tau, \sigma^{\prime} )\} &=&
\Delta_{-}(\sigma , \sigma^{\prime}) \nonumber \\
\{\lambda(\tau,\sigma ) , \pi_{\lambda}(\tau, \sigma^{\prime} )\} &=&
\Delta_{+}(\sigma , \sigma^{\prime}).
\label{32j}
\end{eqnarray}
and also make the following ansatz for the bracket among
the coordinates (\ref{24q}):
\begin{eqnarray}
\{X^{\mu}(\tau , \sigma) ,  X^{\nu}(\tau , \sigma^{\prime})\}
= C^{\mu \nu}(\sigma, \sigma^{\prime})\ \ ;\ \
\mathrm{where} \ \ \ C^{\mu \nu}(\sigma, \sigma^{\prime}) =
-\  C^{\nu \mu}(\sigma^{\prime}, \sigma)\,.
\label{25}
\end{eqnarray}
One can easily check that the brackets (\ref{32j}) are
indeed compatible with the BC (\ref{20}).
Now imposing the BC (\ref{20}) on the above equation (\ref{25}), we obtain
the following condition:
\begin{eqnarray}
\partial_{\sigma}C^{\mu \nu}\left(\sigma,
\sigma^{\prime}\right)|_{\sigma = 0, \pi}
= \frac{\rho}{\lambda}\eta^{\mu \nu} \Delta_{+}\left(\sigma,
\sigma^{\prime}\right)|_{\sigma = 0, \pi}\,.
\label{26}
\end{eqnarray}
Now to find a solution for
$C^{\mu\nu}(\sigma ,\sigma^{\prime})$,
we choose{\footnote {The condition (\ref{27}) reduces to a restricted
class of metric for Polyakov
formalism that satisfy $\partial_{\sigma}g_{01}=0$.
Such conditions also follow from a standard
treatment of the light-cone gauge \cite{pol}.}}:
\begin{eqnarray}
\partial_{\sigma}\left(\frac{\rho}{\lambda}\right) = 0
\label{27}
\end{eqnarray}
which gives a solution of
$C^{\mu\nu}(\sigma ,\sigma^{\prime})$ as:
\begin{eqnarray}
C^{\mu\nu}(\sigma ,\sigma^{\prime}) = \eta^{\mu \nu}\left[
\kappa(\sigma)\Theta(\sigma ,\sigma^{\prime}) -
\kappa(\sigma^{\prime})\Theta(\sigma^{\prime} ,\sigma )\right]
\label{28}
\end{eqnarray}
where the generalised step function $\Theta
(\sigma ,\sigma^{\prime})$ satisfies,
\begin{eqnarray}
\partial_{\sigma }\Theta (\sigma ,\sigma^{\prime}) =
\Delta_{+}(\sigma ,\sigma^{\prime})\,.
\label{29}
\end{eqnarray}
Here, $\kappa(\sigma) = \frac{\rho}{\lambda}(\sigma)$ is a pseudo-scalar.
The $\sigma$ in the parenthesis has been included deliberately
to remind the reader that it transforms as a pseudo-scalar under
$\sigma \to -\sigma$ and should not be read as a functional dependence.
The pseudo-scalar property of $\kappa(\sigma)$ is necessary for
$C^{\mu\nu}(\sigma ,\sigma^{\prime})$ to be an even function of
$\sigma$ as $X(\sigma)$ is also an even function of $\sigma$
in the extended interval $[-\pi, \pi]$ of the string (\ref{37}).
\noindent
An explicit form of $\Theta(\sigma ,\sigma^{\prime})$ is given by \cite{hrt}:
\begin{eqnarray}
\Theta (\sigma ,\sigma^{\prime})={\sigma \over \pi} + {1 \over \pi }
\sum_{n\neq 0}{1\over n}\mathrm{sin}(n\sigma)\mathrm{cos}(n\sigma^{\prime})
\label{30}
\end{eqnarray}
having the properties,
\begin{eqnarray}
\Theta (\sigma ,\sigma^{\prime}) &=& 1~~~ \mathrm{for}
~~\sigma >\sigma^{\prime}
\nonumber \\
\mathrm{and} \quad
\Theta (\sigma ,\sigma^{\prime}) &=& 0 ~~~\mathrm{for}
~~ \sigma <\sigma^{\prime}.
\label{31jj}
\end{eqnarray}
Using the above relations, the simplified structure of
(\ref{28}) reads,
\begin{eqnarray}
\{X^\mu (\tau,\sigma ),X^{\nu}(\tau, \sigma^{\prime} )\} &=& 0
~~~\mathrm{for} ~~\sigma = \sigma^{\prime} \nonumber \\
\{X^\mu (\tau,\sigma ),X^{\nu}(\tau, \sigma^{\prime} )\} &=&
\kappa(\sigma)\,\eta^{\mu \nu}~~~\mathrm{for} ~~\sigma >\sigma^{\prime}
\nonumber \\
&=& -\kappa(\sigma^\prime)\,\eta^{\mu \nu}~~~
\mathrm{for} ~~\sigma <\sigma^{\prime}.
\label{32}
\end{eqnarray}

\noindent
We therefore propose the brackets (\ref{23}) and (\ref{32})
as the basic PB(s) of the theory and using these
one can easily obtain the following involutive
algebra between the constraints:
\begin{eqnarray}
\{\Omega_1(\sigma) , \Omega_1(\sigma^{\prime})\} &=&
\Omega_1(\sigma^{\prime})\partial_{\sigma}\Delta_{+}
\left(\sigma , \sigma^{\prime}\right) + \Omega_1(\sigma)
\partial_{\sigma}\Delta_{-}\left(\sigma , \sigma^{\prime}\right) \,
\nonumber \\
\{\Omega_1(\sigma) , \Omega_2(\sigma^{\prime})\} &=&  \left(
\Omega_2(\sigma) + \Omega_2(\sigma^{\prime})\right)
\partial_{\sigma}\Delta_{+}\left(\sigma , \sigma^{\prime}\right)\,
\nonumber \\
\{\Omega_2(\sigma) , \Omega_2(\sigma^{\prime})\} &=& 4 \left(
\Omega_1(\sigma)\partial_{\sigma}\Delta_{+}
\left(\sigma , \sigma^{\prime}\right) + \Omega_1(\sigma^{\prime})
\partial_{\sigma}\Delta_{-}\left(\sigma , \sigma^{\prime}\right)\right).
\label{33}
\end{eqnarray}
The above algebra is exactly similar with the modified involutive
algebra (\ref{modbr}), between the constraints of the Polyakov
theory.

We now compute the algebra between the Virasoro functionals
using the modified constraint algebra (\ref{33}),
\begin{eqnarray}
\{L[f(\sigma)] , L[g(\sigma)]\} &=& L[f(\sigma)g^{\prime}(\sigma)
- f^{\prime}(\sigma)g(\sigma)].
\label{0d}
\end{eqnarray}
Interestingly, the Virasoro algebra has the same form
as that of (\ref{34a}) at the classical level.
Consequently, the alternative forms of Virasoro algebra
(\ref{34b}) is also reproduced here.

We shall study the consequences of the above algebra (\ref{33}) in later
section where we make an exhaustive analysis of gauge symmetry.
\subsection{Interacting Interpolating String:}
The Interpolating action for a bosonic string moving in
the presence of a constant background Neveu-Schwarz two-form
field $B_{\mu \nu}$ is given by,
\begin{eqnarray}
S_I= \int d\tau d\sigma \left\{ -\frac{1}{2\lambda}\left[\dot{X}^{2}+
2\rho(\dot{X}.X^{\prime})+(\rho^2-\lambda^2)X^{\prime 2}
- \lambda e\epsilon^{ab}B_{\mu\nu}
\partial_{a}X^{\mu}\partial_{b}X^{\nu}\right]\right\}
\label{40}
\end{eqnarray}
where $\epsilon^{01}=-\epsilon^{10}=+1$.
The constraint structure has already been discussed in the section 3.\\
\noindent The BC (\ref{18}) can be written in a completely
covariant form as:
\begin{eqnarray}
\left[M^{\mu}_{\ \nu}\left(\partial_{\sigma}X^{\nu}\right)
+ N^{\mu \nu}\Pi_{\nu}\right]\vert_{\sigma = 0, \pi} = 0
\label{41}
\end{eqnarray}
where,
\begin{eqnarray}
M^{\mu}_{\ \nu} &=& \left(\lambda\, \delta^{\mu}_{\nu}
- e^2  B^{\mu \rho}B_{\rho \nu}\right) \nonumber \\
N^{\mu \nu} &=& \left(\rho\, \eta^{\mu \nu} + eB^{\mu \nu}\right).
\label{42}
\end{eqnarray}
This nontrivial BC leads to a modification in the original
(naive) PBs (\ref{bpb}).\\
\noindent
The BC (\ref{41}) can be recast as:
\begin{eqnarray}
\left(\partial_{\sigma}X^{\mu} + \Pi_{\rho}
\left(N M^{-1}\right)^{\rho \mu}\right)\vert_{\sigma = 0, \pi} = 0.
\label{43}
\end{eqnarray}
The $\{X^{\mu}(\sigma) , \Pi^{\nu}(\sigma^{\prime})\}_{
\mathrm{PB}}$ is the same as that of the free string (\ref{23}).
We therefore make similar ansatz like (\ref{25}) and
using the BC (\ref{43}), we get:
\begin{eqnarray}
\partial_\sigma C_{\mu\nu}(\sigma ,\sigma^{\prime})
\mid_{\sigma =0,\pi} = (NM^{-1})_{\nu \mu}\Delta_+ (\sigma, \sigma^{\prime})
\mid_{\sigma =0,\pi}.
\label{44}
\end{eqnarray}
As in the free case, we restrict to the class defined by
$\partial_\sigma (NM^{-1})_{\nu \mu} = 0$ which reduces to a
restricted class of metric for Polyakov formalism. This reproduces
the corresponding equation in interacting Polyakov string theory
(see second chapter, in particular Eq \ref{1i4}). We therefore,
obtain the following solution:
\begin{eqnarray}
C_{\mu\nu}(\sigma ,\sigma^{\prime}) &=&
{1\over 2}(NM^{-1})_{(\nu \mu )}(\sigma)\Theta (\sigma, \sigma^{\prime})
- {1\over 2}(NM^{-1})_{(\nu \mu )}(\sigma^{\prime})
\Theta (\sigma^{\prime}, \sigma )\nonumber\\
&& + {1\over 2}(NM^{-1})_{[\nu \mu]}(\sigma)[\Theta (\sigma, \sigma^{\prime}) - 1]
+ {1\over 2}(NM^{-1})_{[\nu \mu]}(\sigma^{\prime})\Theta (\sigma^{\prime}, \sigma)
\label{45}
\end{eqnarray}
with $(NM^{-1})_{(\nu \mu )}$ the symmetric and $(NM^{-1})_{[\nu \mu]}$
the antisymmetric part of $(NM^{-1})_{\nu \mu}$.
\section{Gauge symmetry}
 In this section we will discuss the gauge symmetries of the
different actions and investigate their correspondence with the
reparametrisation invariances. This has been done earlier for the
free string case \cite{rbpmas1}, however the canonical symplectic
structure for the open string were not compatible with the general
BC(s) of the theory. In the last two sections we have been
discussing the BCs (\ref{16}, \ref{18}) and have shown how the
basic PB structures has to be modified suitably to be compatible
with the BC(s). Now we shall investigate the gauge symmetry with
the new modified PB structures (discussed in the earlier sections)
which correctly takes into account the BC(s) of the theory.
Importantly, the modified PB structure reveals a NC behavior among
the string coordinates (\ref{28}, \ref{32}). For simplicity the
following analysis of the gauge symmetry is done for the case of
the free strings.

All the constraints are first class and therefore generate gauge
transformations on ${\cal{L}}_{I}$ but the number of independent
gauge parameters is equal to the number of independent primary
first class constraints, i.e. two.  In the following analysis we
will apply a systematic procedure of abstracting the most general
local symmetry transformations of the Lagrangian. A brief review
of the procedure of \cite{brr, brr12} will thus be appropriate.

    Consider a theory with first class constraints only. The set of
constraints $\Omega_{a}$ is assumed to be classified as
\begin{equation}
\left[\Omega_{a}\right] = \left[\Omega_{a_1}
                ;\Omega_{a_2}\right]
\label{215}
\end{equation}
where $a_1$ belong to the set of primary and $a_2$ to the set of
secondary constraints. The total Hamiltonian is
\begin{equation}
H_{T} = H_{c} + \Sigma\lambda^{a_1}\Omega_{a_1}
\label{216}
\end{equation}
where $H_c$ is the canonical Hamiltonian and $\lambda^{a_1}$ are Lagrange multipliers enforcing the primary constraints. The most general expression for the generator of gauge transformations is obtained according to the Dirac conjecture as
\begin{equation}
G = \Sigma \epsilon^{a}\Omega_{a}
\label{217}
\end{equation}
where $\epsilon^{a}$ are the gauge parameters, only
$a_1$ of which are independent. By demanding the
commutation of an arbitrary gauge variation with the
total time derivative,(i.e. $\frac{d}{dt}
\left(\delta q \right) = \delta \left(\frac{d}{dt} q \right) $)
 we arrive at the following equations \cite{brr, htz}
\begin{equation}
\delta\lambda^{a_1} = \frac{d\epsilon^{a_1}}{dt}
                 -\epsilon^{a}\left(V_{a}^{a_1}
                 +\lambda^{b_1}C_{b_1a}^{a_1}\right)
                              \label{218}
\end{equation}
\begin{equation}
  0 = \frac{d\epsilon^{a_2}}{dt}
 -\epsilon^{a}\left(V_{a}^{a_2}
+\lambda^{b_1}C_{b_1a}^{a_2}\right)
\label{219}
\end{equation}
Here the coefficients $V_{a}^{a_{1}}$ and $C_{b_1a}^{a_1}$ are the structure
functions of the involutive algebra, defined as
\begin{eqnarray}
\{H_c,\Omega_{a}\} = V_{a}^b\Omega_{b}\nonumber\\
\{\Omega_{a},\Omega_{b}\} = C_{ab}^{c}\Omega_{c}.
\label{2110}
\end{eqnarray}
Solving (\ref{219}) it is possible to choose $a_1$ independent
gauge parameters from the set $\epsilon^{a}$ and express $G$ of
(\ref{217}) entirely in terms of them. The other set (\ref{218})
gives the gauge variations of the Lagrange multipliers.{\footnote{
 It can be shown that
these equations are not independent conditions but appear as
internal consistency conditions. In fact the conditions
(\ref{218}) follow from (\ref{219}) \cite{brr, brr12}.}}

We begin the analysis with the interpolating Lagrangian
(\ref{interpolatingL}). It contains additional fields $\rho$ and
$\lambda$. We shall calculate the gauge variation of these extra
fields and explicitly show that they are connected to the
reparametrization by a mapping between the gauge parameters and
the diffeomorphism parameters. These maps will be obtained later
in this section by demanding the consistency of the variations
$\delta X^{\mu}$ due to gauge transformation and reparametrization

 .

The full constraint structure of the theory comprises of the
constraints (\ref{int_primary}) along with (\ref{NGconstraints}).
We could proceed from these and construct the generator of gauge
transformations. The generator of the gauge transformations of
(\ref{interpolatingL}) is obtained by including the whole set of
first class constraints $\Omega_{i}$ given by (\ref{int_primary})
and (\ref{NGconstraints}) as
\begin{equation}
G =\int d\sigma \alpha_{i}\Omega_{i}
\label{3112}
\end{equation}
where only two of the $\alpha_{i}$'s are the independent gauge
parameters. Using (\ref{219}) the dependent gauge parameters could
be eliminated. After finding the gauge generator in terms of the
independent gauge parameters, the variations of the fields $X^{\mu}$,
 $\rho$ and $\lambda$ can be worked out. But the number of independent
 gauge parameters are same in both NG (\ref{NGaction}) and interpolating
 (\ref{interpolatingL}) version. So the gauge generator\footnote{Note
that the gauge parameters $\alpha_{1}$ and $\alpha_{2}$ are odd and
even respectively under $\sigma \to -\sigma$.}
is the same for both the cases, namely:
\begin{equation}
G =\int d\sigma\left( \alpha_{1}\Omega_{1} +  \alpha_{2}\Omega_{2}\right)
\label{3112aaa}
\end{equation}
Also, looking at the intermediate first order form (\ref{subL}) it
appears that the fields $X^{\mu}$ were already there in the NG
action (\ref{NGaction}). The other two fields of the interpolating
Lagrangian are $\rho$ and $\lambda$ which are nothing but the
Lagrange multipliers enforcing the first class constraints
(\ref{NGconstraints}) of the NG theory. Hence their gauge
variation can be worked out from (\ref{218}). We prefer to take
this alternative route. For convenience we relabel $\rho$ and
$\lambda$ by $\lambda_{1}$ and $\lambda_{2}$
\begin{equation}
\lambda_{1} = \rho \hspace{1cm} \rm{and} \hspace{1cm}
\lambda_{2} = \frac{\lambda}{2}
\label{313}
\end{equation}
and their variations are obtained from (\ref{218})
\begin{equation}
\delta\lambda_{i}\left( \sigma\right) = -  \dot \alpha_{i}
- \int d\sigma^{\prime} d\sigma^{\prime \prime} C_{kj}{}^{i}
\left(\sigma^{\prime}, \sigma^{\prime \prime}, \sigma\right)
\lambda_{k}\left(\sigma^{\prime}\right) \alpha_{j}
\left(\sigma^{\prime \prime}\right)
\label{314}
\end{equation}
where
$ C_{kj}{}^{i}
\left(\sigma^{\prime}, \sigma^{\prime \prime}, \sigma\right)$
are given by
\begin{equation}
\left\{ \Omega_{\alpha}\left(\sigma\right),\Omega_{\beta}
\left(\sigma^{\prime}\right)\right\} =  \int d\sigma^{\prime \prime}
C_{\alpha \beta}{}^{\gamma}\left(\sigma, \sigma^{\prime}, \sigma^{\prime
\prime}\right) \Omega_{\gamma}\left(\sigma^{\prime \prime}\right)
\label{315}
\end{equation}
Observe that the structure function $ V_{a}{}^{b}$ does not
appear in (\ref{314}) since $H_{c} = 0$ for the NG theory.
The nontrivial structure functions $C_{\alpha \beta}{}^{\gamma}
\left(\sigma, \sigma^{\prime}, \sigma^{\prime \prime}\right)$
are obtained from the
constraint algebra (\ref{33}) as:
\begin{eqnarray}
C_{1 1}{}^{1}\left(\sigma ,\sigma^{\prime},\sigma^{\prime \prime}\right) &=&
\left(\partial_{\sigma}\Delta_{+}\left(\sigma , \sigma^{\prime}
\right)\right)\Delta_{-}\left(\sigma^{\prime} , \sigma^{\prime \prime}\right)
+ \left(\partial_{\sigma}\Delta_{-}\left(\sigma , \sigma^{\prime}
\right)\right)\Delta_{-}\left(\sigma , \sigma^{\prime \prime}\right)
\nonumber \\
C_{2 2}{}^{1}\left(\sigma, \sigma^{\prime}, \sigma^{\prime \prime}\right) &=&
4\left(\partial_{\sigma}\Delta_{+}\left(\sigma , \sigma^{\prime}
\right)\right)\Delta_{-}\left(\sigma , \sigma^{\prime \prime}\right)
+ 4\left(\partial_{\sigma}\Delta_{-}\left(\sigma , \sigma^{\prime}
\right)\right)\Delta_{-}\left(\sigma^{\prime} , \sigma^{\prime \prime}\right)
\nonumber \\
C_{1 2}{}^{2}\left(\sigma, \sigma^{\prime}, \sigma^{\prime \prime}\right) &=&
\partial_{\sigma}\Delta_{+}\left(\sigma , \sigma^{\prime}
\right)
\left[\Delta_{+}\left(\sigma , \sigma^{\prime \prime}\right)
+ \Delta_{+}\left(\sigma^{\prime} , \sigma^{\prime \prime}\right)
\right]
\nonumber \\
C_{2 1}{}^{2}\left(\sigma, \sigma^{\prime}, \sigma^{\prime \prime}\right) &=&
\partial_{\sigma}\Delta_{-}\left(\sigma , \sigma^{\prime}
\right)
\left[\Delta_{+}\left(\sigma , \sigma^{\prime \prime}\right)
+ \Delta_{+}\left(\sigma^{\prime} , \sigma^{\prime \prime}\right)
\right]
\label{319}
\end{eqnarray}
all other $ C_{\alpha b}{}^{\gamma}$'s are zero. Note that these
structure functions are potentially different from those appearing
in \cite{rbpmas1, rbpmas2} in the sense that here periodic delta
functions are introduced to make the basic brackets compatible
with the nontrivial BC. Using the expressions of the structure
functions (\ref{319}) in equation (\ref{314}) we can easily
derive:
\begin{eqnarray}
\delta \lambda_{1} &=& - \dot \alpha_{1}
+ \left(\alpha_{1}\partial_{1}\lambda_{1}
 - \lambda_{1}\partial_{1}\alpha_{1} \right)
+ 4 \left(\alpha_{2}\partial_{1}\lambda_{2} - \lambda_{2}\partial_{1}
\alpha_{2}\right)\nonumber\\
\delta \lambda_{2} &=& -\dot \alpha_{2}
+\left(\alpha_{2}\partial_{1}\lambda_{1} - \lambda_{1} \partial_{1}\alpha_{2}
\right)
+\left(\alpha_{1}\partial_{1}\lambda_{2} - \lambda_{2} \partial_{1}\alpha_{1}
\right)
\label{GM41}
\end{eqnarray}
From the correspondence (\ref{313}), we get the variations of $\rho$
and $\lambda$ as:
\begin{eqnarray}
\delta \rho &=& - \dot \alpha_{1}
+ \left(\alpha_{1}\partial_{1}\rho
 - \rho\partial_{1}\alpha_{1} \right) + 2 \left(\alpha_{2}\partial_{1}\lambda - \lambda \partial_{1}\alpha_{2}\right)
\nonumber\\
\delta \lambda &=& -2 \dot \alpha_{2}
+2 \left(\alpha_{2}\partial_{1}\rho - \rho \partial_{1}\alpha_{2}
\right)
+\left(\alpha_{1}\partial_{1}\lambda - \lambda \partial_{1}\alpha_{1}\right)
\label{GM51}
\end{eqnarray}
 In the above we have found out the full set of symmetry transformations of the fields in the
 interpolating Lagrangian (\ref{interpolatingL}). These symmetry transformations (\ref{GM51})
 were earlier given in \cite{kaku, kaku12} for the free string case. But the results were found there
 by inspection\footnote{ For easy comparison identify
$ \alpha_{1} = \eta $ and $ 2 \alpha_{0} = \epsilon $ }.
In our approach \cite{rbpmas1, rbpmas2} the appropriate transformations are
obtained systematically by a general method applicable
to a whole class of string actions.

We are still to investigate to what extent the exact correspondence
between gauge symmetry and reparametrisation holds in our
modified NC framework. This can be done very easily if we stick to the
method discussed in \cite{rbpmas1, rbpmas2}.

To work out the mapping between the gauge parameters and the
diffeomorphism parameters we now take up the
Polyakov action (\ref{Polyaction}) (with $B = 0$).
Here the only dynamic fields are $X^{\mu}$.
The transformations of $X^{\mu}$  under (\ref{3112aaa})
can be worked out resulting in the following:
\begin{equation}
\delta X_{\mu}(\sigma) = \left\{X_{\mu}(\sigma), G \right\}
 = \left( \alpha_{1} X_{\mu}^{\prime}(\sigma) +  2\alpha_{2} \Pi_{\mu}(\sigma)\right)
\label{3114}
\end{equation}
We can now substitute for $\Pi_{\mu}$ (obtained from (\ref{Polyaction}))
to obtain:
\begin{equation}
\delta X_{\mu}
 = \left(\alpha_1 - 2\alpha_2 \, \sqrt{- g}\, g^{01}\right)X^{\prime}_{\mu}
- 2\sqrt{- g}\, g^{00}\,\alpha_2\,\dot{X}_{\mu}
\label{31141}
\end{equation}
This is the gauge variation of $X^{\mu}$ in terms of $X^{\prime}{}_{\mu}$
and $\dot{X}_{\mu}$ where the cofficients appear as arbitrary functions
of $\sigma$ and $\tau$. So we can identify them with the arbitrary
parameters  $\Lambda_{1}$ and $\Lambda_{0}$ characterising the
infinitesimal reparametrization \cite{holt}:
\begin{eqnarray}
\tau^{\prime} & = & \tau- \Lambda_{0}\nonumber\\
\sigma^{\prime} & = & \sigma - \Lambda_{1}\nonumber\\
\delta X^{\mu} & = & \Lambda^{a} \partial_{a} X^{\mu}
= \Lambda_{0} \dot X^{\mu} + \Lambda_{1}  X^{\prime \mu}
\label{3117}
\end{eqnarray}
and that of $g_{ab}$ as:
\begin{eqnarray}
\delta g_{ab} = D_{a}\Lambda_{b} + D_{b}\Lambda_{a}
\label{3117b}
\end{eqnarray}
where
\begin{equation}
D_{a}\Lambda_{b} = \partial_{a} \Lambda_{b} - \Gamma_{ab}{}^{c} \Lambda_{c}
\label{3123}
\end{equation}
$\Gamma_{ab}{}^{c}$ being the usual Christoffel symbols \cite{holt}.
The infinitesimal parameters $\Lambda^{a}$ characterizes reparametrisation.

Comparing (\ref{31141}) and (\ref{3117}), we get the map
connecting the gauge parameters with the diffeomorphism parameters:
\begin{eqnarray}
\Lambda_{0} &=&  - 2\sqrt{- g}\, g^{00}\,\alpha_2 \nonumber \\
\Lambda_{1} &=& \left(\alpha_1 - 2\alpha_2 \, \sqrt{- g}\, g^{01}\right)
\label{the_map}
\end{eqnarray}
Using the definitions (\ref{idm}), this map can be cast in a
better shape:
\begin{eqnarray}
\Lambda_{1} & = &  \left(\alpha_1 - 2\frac{\alpha_{2}\rho}{\lambda}
\right)\nonumber \\
\Lambda_{0} & = & - \frac{ 2 \alpha_{2}}{\lambda}
\label{themap}
\end{eqnarray}
All that remains now is to get the variation of $\rho$ and $\lambda$
induced by the reparametrisation (\ref{3117b}).

\noindent The identification (\ref{idm}) and (\ref{3117b})
reproduces (\ref{GM51}) as the variations of $\rho$ and $\lambda$.
This establishes complete equivalance
of the gauge transformations with the diffeomorphisms of the string.

%
\section{Summary}
In this chapter, we have developed a new action formalism for
interacting bosonic string and demonstrated that it interpolates
between the NG and Polyakov form of interacting bosonic actions.
This is similar to the interpolating action formalism for free
string proposed in \cite{rb}. We have also modified the basic PBs
in order to establish consistency of the BC with the basic PBs. We
stress that contrary to standard approaches, BC(s) are not treated
as primary constraints of the theory. Our approach is similar in
spirit with the previous treatment of string theory \cite{rb, hrt,
rbbckk, agh1}. The NC structures derived in this chapter go over
smoothly to the Polyakov version once suitable identifications are
made. We then set out to study the status of gauge symmetries
vis-\`{a}-vis reparametrisation in this NC set up and establish
the connection between gauge symmetry and diffeomorphism
transformations.

\chapter{Normal ordering and noncommutativity in open bosonic strings}

So far we have discussed bosonic string theory classically now it
would be quite interesting to study whether NC structure can be
obtained from the conformal field theory techniques where the
analysis is carried out in a quantum setting. Indeed, as has been
stressed in \cite{sw} that it is very important to understand this
noncommutativity from different perspectives. More importantly, it
is necessary to check explicitly whether the central charge gets
effected by the modified BCs in this case, as the central charge
can be related to the Casimir energy arising from finite size of
the string and is a purely quantum effect \cite{pol}.

It may be recalled in this context that in quantum field theory,
 products of quantum
fields at the same space-time points are in general singular objects.
 The same thing is true in string theory when one
multiplies position operators, that can be taken as
conformal fields on the world sheet.
This situation is well known and one can remove the singular part of the
operator products by defining normal ordered operators which have
well behaved properties \cite{pol}.
This is important, for example,  when one builds up the generators of conformal
transformations and investigates the realization of classical symmetries
at quantum level.

Usually normal ordered products of operators are defined so as to
satisfy the classical equations of motion at quantum level.
However, in a recent paper \cite{4cbr}, new normal ordered
products have been defined for open string position operators that
additionally satisfy the BCs. This way one obtains a normal
ordering that is also valid at string end points. The mathematical
problem posed by defining the normal ordering is related  to that
of calculating Green's functions \cite{4cAGNY, 4cSchomerus:2002dc,
4cDN, 4cDolan:2002px, 4cpmrp, 4cpmrp1}. The normal ordered product
is defined by subtracting out the corresponding Green's functions.
So we can find normal ordered products satisfying open string
boundary condition using the solutions to open string Green's
functions.

In this chapter, we shall consider the problem of noncommutativity
using this new normal ordering given in \cite{4cbr}. By using the
contour argument and the new $XX$ operator product expansion
(OPE), we shall first find the commutator among the
 Fourier components and then the commutation relations among string's coordinates
 which reproduces the same noncommutative structure obtained in previous chapters
 and also in \cite{rb, jing}, as mentioned earlier.
We would also like to stress that the above commutator computed
using the $XX$ OPE satisfying the equation of motion
only \cite{pol} (where the modifications due to BCs are not taken into account)
 also leads to the same NC structure.

\section{New Normal ordered products}
Let us first consider the free Polyakov string action,
\begin{equation}
\label{4c1}
S_{P} \,= -\,
{ 1 \over 4\pi \alpha^\prime }
\int_{\Sigma} d\tau d\sigma \Big(
G^{ab} \eta_{\mu\nu } \partial_a X^\mu \partial_b X^\nu \,+ \,
\epsilon^{ab} B_{\mu\nu} \partial_a X^\mu \partial_b X^\nu\Big)
\end{equation}
\noindent
where $\tau, \sigma$ are the usual world-sheet parameters,
 $G_{ab}$ is the induced world-sheet metric with
$G_{\tau\tau} = -1, \, G_{\sigma\sigma} = 1$ (upto a Weyl factor)
in conformal gauge and the antisymmetric
tensor is chosen by $\epsilon^{\tau \sigma} = 1$.
$X^\mu (\tau, \sigma)$ are the string coordinates
in the $D$ dimensional Minkowskian target space with metric
$\eta_{\mu\nu} = (-1, 1, ...., 1)$.

We now make a Wick rotation by defining
$\sigma^2 = i\tau$   and obtain the classical action
for a bosonic string taking a world-sheet with Euclidean signature:
\begin{equation}
\label{4c2}
S \,=\, { 1 \over 4\pi \alpha^\prime }
\int_{\Sigma} d^2\sigma \Big(
g^{ab} \eta_{\mu\nu } \partial_a X^\mu \partial_b X^\nu \,+ \,i\,
\varepsilon^{ab} B_{\mu\nu} \partial_a X^\mu \partial_b X^\nu\Big)
\end{equation}
where  $g^{ab}$ can now be taken to be proportional to the
unit matrix and $\varepsilon^{12} = - \varepsilon^{21} = 1$.
Note that the $D$ dimensional target space-time still has the Lorentzian
signature.

\noindent
The variation of the action (\ref{4c2}) gives the equation of motion,
\begin{equation}
(\, \partial^2_1 \, +\, \partial^2_2\,)  X^\mu\,=\,0
\label{4c3}
\end{equation}
and a boundary term that yields the following BCs:
\begin{eqnarray}
\left(\partial_1 X^{\mu}(\sigma^1, \sigma^2)\,+ \,i\, B_{\mu\nu}
\partial_2 X^{\nu}(\sigma^1, \sigma^2)\right)\vert_{\sigma=0, \pi} = 0.
\label{4c4}
\end{eqnarray}
It is convenient, to introduce complex world sheet coordinates \cite{pol}:
$ z \,=\,  \sigma^1 +i \sigma^2 \, ; \,  \bar{z} \,=\, \sigma^1 - i \sigma^2$
and $\partial_z = \frac{1}{2}(\partial_1 - i\partial_2), \, \partial_{\bar{z}} =
\frac{1}{2}(\partial_1 + i\partial_2)$.

In this notation the action is
\begin{eqnarray}
\label{4c5}
S \,=\,{ 1 \over 2\pi \alpha^\prime }
\int d^2 z \Big(\partial_{z} X^\mu \partial_{\bar{z}}X_\mu - B_{\mu \nu}
\partial_{z} X^\mu \partial_{\bar{z}}X^{\nu}\Big)
\end{eqnarray}
while the classical equations of motion and boundary conditions take the form
\begin{eqnarray}
\partial_{\bar z} \partial_z  X^{\mu}(z, \bar{z}) &=& 0
\label{4c6}\\
\left[\eta^{\mu \nu}\Big(\partial_z + \partial_{\bar z}\Big)
- B^{\mu \nu}\Big(\partial_z - \partial_{\bar z}\Big)\right]
X_\nu\vert_{z = -\bar{z},\,  2\pi - \bar{z}} &=& 0
\label{4c7}
\end{eqnarray}
We now study the properties of quantum operators, corresponding to
the classical variables, by considering the expectation values.
Defining the expectation value of an operator ${\cal F}$ as
\cite{pol}:
\begin{eqnarray}
\langle {\cal F} [X] \rangle \,=\,\int [dX] \exp ( - S[X] ) {\cal F} [X]
\label{4c28}
\end{eqnarray}
\noindent
and using the fact that the path integral of a total derivative vanishes one
finds:
\begin{eqnarray}
0 &=& \int [dX] {\delta \over \delta X^\nu (z^\prime , {\bar z}^\prime )}
exp ( - S[X] ) \,=\,
\Big\langle \,{1\over \pi\alpha^\prime } \partial_{\bar z^\prime}
\partial_{z^\prime} X_\nu ( z^\prime, \bar z^\prime ) \Big\rangle \nonumber\\
&+& {1\over 2\pi \alpha^\prime}  \oint_{_{\partial \Sigma}} \delta^2 (z - z^\prime)
\Big\langle \Big( \eta_{\nu\mu} ( \partial_z + \partial_{\bar z} ) +
 B_{\nu\mu} ( \partial_z - \partial_{\bar z} ) \Big) X^\mu (z, {\bar z} )
 \,\Big\rangle dz.
\label{4c29}
\end{eqnarray}
\noindent
The last (singular) term is integrated over the boundary,
where $dz = - d \bar z$. We thus find that this equation
 implies that both string equations of motion
and boundary condition hold as expectation values.
So the corresponding quantum position operators ${\hat X}^\mu$
(in target space) satisfy (as long as they are not
multiplied by other local operators coincident at
the same world-sheet point) the following equations:
\begin{eqnarray}
\label{4c30a}
\partial_{\bar z} \partial_{z} {\hat X}^\nu ( z, \bar z ) &=& \,0  \\
\Big( \eta_{\nu\mu} ( \partial_z + \partial_{\bar z} ) -
 B_{\nu\mu} ( \partial_z - \partial_{\bar z} ) \Big) {\hat X}^{\mu}
\vert_{\mathrm{Bound}.} &=& 0
\label{4c30}
\end{eqnarray}
which are nothing but the quantum version of (\ref{4c6}, \ref{4c7}).
Proceeding in the same way, we can consider a pair of local
operators which may now be coincident to show that their products
 at the quantum level satisfy \cite{4cbr} :
\begin{equation}
\partial_{\bar z^\prime} \partial_{z^\prime} {\hat X}^\mu ( z^\prime, {\bar z}^\prime)
{\hat X} ^\nu (z^{\prime\prime}, {\bar z^{\prime\prime}} ) = \,- \pi \alpha^\prime\,
\eta^{\mu\nu}
\, \delta^2
( z^\prime - z^{\prime\prime} , {\bar z}^\prime - {\bar z}^{\prime\prime} )
\end{equation}
\begin{equation}
\Big( \eta_{\nu\mu} ( \partial_{z^\prime} + \partial_{\bar z^\prime} ) -
 B_{\nu\mu} ( \partial_{z^\prime} - \partial_{\bar z^\prime} ) \Big)
{\hat X}^\mu ( z^\prime, {\bar z}^\prime)
{\hat X}^\rho ( z^{\prime\prime}, {\bar z^{\prime\prime}} )
\vert_{_{\mathrm{Bound}.}} \,=\, 0.
\end{equation}

Now if we introduce the operation of normal ordering in the
standard way \cite{pol},
\begin{eqnarray}
:\,{\hat X}^\mu (z, \bar z )\, :\, &=& \, {\hat X}^\mu (z, \bar z )\nonumber\\
:\,{\hat X}^\mu (z, \bar z )\, {\hat X}^\nu (z^\prime, {\bar z}^\prime)\,:
\, &=& \, {\hat X}^\mu (z , \bar z  ) \,{\hat X}^\nu (z^\prime ,{\bar z}^\prime)\, +
{ \alpha^\prime \over 2} \eta^{\mu\nu} {\mathrm{ln}} \vert z - z^\prime \vert^2
\label{4c9}
\end{eqnarray}
it satisfies the equation of motion (\ref{4c6}) at the quantum level,
but fails to satisfy the boundary conditions (\ref{4c7}). In \cite{4cbr},
the authors have introduced a different kind of normal
ordered product satisfying both equation of motion and boundary conditions.

At this point it is more convenient to choose world sheet
coordinates, related to these $z$ coordinates by conformal
transformation, that simplify the representation of the boundary,
\begin{eqnarray}
\omega\, = \,\mathrm{exp}\left(-iz\right)\, =\, e^{- i \sigma^1 +
\sigma^2} \,\,\,;\,\,\,{\bar \omega} = e^{i \sigma^1 + \sigma^2}.
\label{4c10}
\end{eqnarray}
In this present coordinates the complete boundary corresponds
just to the region $\omega = {\bar \omega}$.
On the other hand, the action (\ref{4c5})  along with equation of motion (\ref{4c30a})
in terms of $\omega, {\bar \omega}$ has still the same form, while
the form of BCs are slightly altered:
\begin{eqnarray}
\label{4c31}
\partial_{\bar{\omega}}\partial_{\omega} {\hat X}^\mu(\omega, \bar{\omega}) &=& 0 \\
\Big( \eta_{\mu\nu} ( \partial_\omega - \partial_{\bar \omega} ) -
 B_{\mu\nu} ( \partial_\omega + \partial_{\bar \omega} ) \Big)
{\hat X}^\nu\vert_{_{\omega = \bar \omega}} \,&=&\, 0\,\,.
\label{4c31q}
\end{eqnarray}
The corresponding new normal ordering is given by \cite{4cbr}:
\begin{eqnarray}
\label{4c11}
\mbox{{\bf :}} \, {\hat X}^{\mu}(\omega, \bar \omega )\, {\hat X}^{\nu}(\omega^\prime, {\bar \omega}^\prime)
\,\mbox{{\bf :}} &=&
 {\hat X}^{\mu}(\omega, \bar \omega ) \, {\hat X}^{\nu}(\omega^\prime, {\bar \omega}^\prime)
+ \frac{\alpha'}{2}
\eta^{\mu \nu} \mathrm{ln} \vert \omega-\omega^\prime \vert^2
+ \frac{\alpha'}{2}
\left( [ \eta - B ]^{-1}\,[ \eta + B ] \right)^{\mu \nu}\nonumber \\
&& \mathrm{ln}\left(\omega - \bar{\omega}^{\prime}\right) +  \frac{\alpha'}{2}
\Big( [ \eta  - B ] \, [ \eta + B ]^{-1} \,\Big)^{\mu \nu} \,\,
\mathrm{ln} (\bar{\omega}- \omega^{\prime} )
\end{eqnarray}
which satisfy both equation of motion and open string
BCs (\ref{4c31}, \ref{4c31q}) at the quantum level. These additional terms
can be understood easily as `image' contribution as in electrostatics.

Now for any arbitary functional ${\cal{F}}[X]$, the new
 normal ordering (in absence of the $B$ field)
can be compactly written as:
\begin{eqnarray}
\label{4c32}
\mbox{{\bf :}}\,{\cal{F}}\,\mbox{{\bf :}} = \exp\left(
{\alpha^\prime \over 4} \int d^2 \omega_1 d^2 \omega_2\, \left[\mathrm{ln}
\vert \omega_1 - \omega_2\vert^2 + \mathrm{ln} \vert
\omega_1 - \bar{\omega}_2 \vert^2\right]\frac{\delta}{\delta X^{\mu}
(\omega_1, \bar{\omega}_1)}\,\frac{\delta}{\delta X_{\mu}
(\omega_2, \bar{\omega}_2)}\right){\cal{F}}.
\end{eqnarray}
For example, this reproduces correctly the expression given in (\ref{4c11}),
as one can easily verify for $B=0$.\\
The OPE for any pair of operators, satisfying the BCs, can be generated from
\begin{eqnarray}
\label{4c33}
\mbox{{\bf :}}\,{\cal{F}}\,\mbox{{\bf :}} \,
\mbox{{\bf :}}\,{\cal{G}}\,\mbox{{\bf :}}= \exp\left(
{\alpha^\prime \over 4} \int d^2 \omega_1 d^2 \omega_2\, \left[\mathrm{ln}
\vert \omega_1 - \omega_2\vert^2 + \mathrm{ln} \vert
\omega_1 - \bar{\omega}_2 \vert^2\right]\frac{\delta}{\delta X^{\mu}
(\omega_1, \bar{\omega}_1)}\,\frac{\delta}{\delta X_{\mu}
(\omega_2, \bar{\omega}_2)}\right)\mbox{{\bf :}}\,{\cal{F}}\, {\cal{G}}\,\mbox{{\bf :}}.
\end{eqnarray}
It is now easy to verify that the $TT$ OPE involving
energy-momentum tensor
\begin{eqnarray}
\label{4c32q}
T(\omega) = - \frac{1}{\alpha^\prime} \mbox{{\bf :}}\,
\partial X^{\mu}(\omega) \partial X_{\mu}(\omega) \,\mbox{{\bf :}}
\end{eqnarray}
undergoes no modification.
Indeed, using the above definition we obtain the following OPE:
\begin{eqnarray}
\label{4c41}
\mbox{{\bf :}}\,\partial X^\mu(\omega)\, \partial X_\mu(\omega)\,\mbox{{\bf :}}\,
\mbox{{\bf :}}\,\partial^{\prime}X^\nu(\omega^{\prime})\,
\partial^{\prime}X_\nu(\omega^{\prime})\,\mbox{{\bf :}}
&=& \mbox{{\bf :}}\,\partial X^\mu(\omega)\, \partial X_\mu(\omega)\,\partial^{\prime}X^\nu(\omega^{\prime})\,
\partial^{\prime}X^\nu(\omega^{\prime})\,\mbox{{\bf :}} \nonumber \\
&& -4 \cdot\frac{\alpha^\prime}{2}\left(\partial\partial^{\prime}\mathrm{ln}\vert
\omega - \omega^{\prime}\vert^2\right)\mbox{{\bf :}}\,\partial X^\mu(\omega)\,\partial^{\prime}X_\nu(\omega^{\prime})\,\mbox{{\bf :}} \nonumber \\
&& +2 \cdot \delta^{\mu}_{\ \mu}\left(\frac{\alpha^{\prime}}{2}
\partial\partial^{\prime}\mathrm{ln}\vert \omega - \omega^{\prime}\vert^2\right)^2 \nonumber \\
&\sim & \frac{D\alpha^{\prime 2}}{2\left(\omega - \omega^{\prime}\right)^4}
- \frac{2\alpha^{\prime}}{\left(\omega - \omega^{\prime}\right)^2}\mbox{{\bf :}}\,
\partial^{\prime}X^\mu(\omega^{\prime})\, \partial^{\prime}X_\mu(\omega^{\prime})
\,\mbox{{\bf :}}\nonumber \\
&& - \frac{2\alpha^{\prime}}{\left(\omega - \omega^{\prime}\right)}
\partial^{\prime 2}X^\mu(\omega^{\prime})\, \partial^{\prime}X_\mu(\omega^{\prime})
\,\mbox{{\bf :}}
\end{eqnarray}
where $\sim$ mean `equal upto nonsingular terms'. The above result
is same as that of \cite{pol} which is obtained by using the usual
normal ordering satisfying the equation of motion only. This also
implies that the Virasoro algebra remains the same as that of
\cite{pol}. So the new normal ordering (\ref{4c11}) (with $B= 0$)
has no impact on the central charge.

We shall make use of the results discussed here in the next section
where we study both free and interacting open bosonic strings.
\section{Mode expansions and Non-Commutativity for bosonic strings}
\subsection{Free open strings}
In this section, we consider the mode expansions of free
($B_{\mu \nu} = 0$) bosonic strings.
We start with the closed string first.
In the $X^{\mu}$ theory (\ref{4c5}), $\partial X^{\mu}$ and $\bar{\partial} X^{\mu}$ are
(anti)holomorphic and so have the following Laurent expansions,
\begin{eqnarray}
\label{4c12}
\partial X^{\mu}(\omega) &=& -i \left(\frac{\alpha^{\prime}}{2}\right)^{\frac{1}{2}}
\sum^{\infty}_{m= -\infty} \frac{\alpha^{\mu}_{m}}{\omega^{m+1}}\nonumber \\
 \bar{\partial} X^{\mu}(\bar{\omega}) &=& -i \left(\frac{\alpha^{\prime}}{2}
\right)^{\frac{1}{2}}
\sum^{\infty}_{m= -\infty} \frac{{\tilde{\alpha}}^{\mu}_{m}}{\bar{\omega}^{m+1}}.
\end{eqnarray}
Now the BC(s) (\ref{4c31q}) in case of free open strings
(i.e. $B_{\mu\nu} = 0$) requires
$\alpha = \tilde{\alpha}$ in the expansions
(\ref{4c12})\footnote{Note that the BC(s) (\ref{4c31q}) in case of free
open strings are the usual Neumann BC(s).}.
The expansion for $X^{\mu}$ is then:
\begin{eqnarray}
\label{4c13}
X^{\mu}(\omega, \bar{\omega}) = x^{\mu} -i\alpha^{\prime}p^{\mu}
\mathrm{ln}\vert \omega \vert^2 + i \left(\frac{\alpha^{\prime}}{2}
\right)^{\frac{1}{2}}\sum_{m \neq 0}\frac{\alpha^{\mu}_{m}}{m}
\left( \omega^{-m} + \bar{\omega}^{-m}\right)
\end{eqnarray}
where, $x^{\mu}$ and $p^{\mu} = \frac{1}{\sqrt{2 \alpha^{\prime}}}
\alpha^{\mu}_{0}$ are the center of mass coordinate and momentum
respectively.

Now the expressions (\ref{4c12}) for open strings can be equivalently written as:
\begin{eqnarray}
\alpha^{\mu}_{m} &=& \left(\frac{2}{\alpha^{\prime}}\right)^{\frac{1}{2}}
\oint \frac{d\omega}{2\pi}\, \omega^m \partial X^{\mu}(\omega)\, =
\,\oint \frac{d\omega}{2\pi i}\, j^{\mu}_{m}(\omega)
\label{4c38}
\end{eqnarray}
where, $j^{\mu}_{m}(\omega) = \sqrt{\frac{2}{\alpha^{\prime}}}
 i\, \omega^m \partial X^{\mu}(\omega)$ is the corresponding holomorphic current.
The commutation relation between $\alpha$'s can be worked out from
the contour argument and the $X\, X$ OPE\footnote{ Note that  here
we have used the new normal ordering (\ref{4c11}) for free open
string, yet the commutation relations (\ref{4c14}) remain same
(see \cite{pol}).} \cite{pol},
\begin{eqnarray}
\label{4c14}
\left[\alpha^{\mu}_{m}, \alpha^{\nu}_{n}\right] &=&
\oint \frac{d\omega_2}{2\pi i}\, \mathrm{Res}_{\omega_1 \to \omega_2}\left(
j^{\mu}_{m}(\omega_1)\,j^{\nu}_{n}(\omega_2)\right) \nonumber \\
&=& m \, \delta_{m, -n}\eta^{\mu \nu}
\end{eqnarray}
At this stage, it should be noted that the above approach
 does not give the algebra
among the zero modes, i.e. the center of mass variables $[x^\mu,
p^\nu]$ and $[x^\mu, x^\nu]$ of the open string. However, the
results can be derived using standard techniques (as has been done
in \cite{jing}) and read,
\begin{eqnarray}
\left[x^{\mu}, p^{\nu}\right] &=& i \eta^{\mu \nu}.
\label{4c14a}
\end{eqnarray}
The conjugate momenta $\Pi_{\mu} = \frac{1}{2\pi\,
\alpha^{\prime}} \dot{X^{\mu}}$
corresponding to $X^{\mu}$
can be calculated from the action (\ref{4c1}) which has a Lorentzian
signature for the world-sheet. In order to make a transition to Euclidean
world-sheet, we make use of Wick rotation as before,
by defining $\sigma^2 = i \tau$, so that $\dot{X^{\mu}} = i
\frac{\partial X^{\mu}}{\partial \sigma^2}$. The $\Pi^{\mu}(\sigma^1, \sigma^2)$ can be recast as
a function of $\omega$ and $\bar{\omega}$ using (\ref{4c10}), so that its mode
expansion becomes:
\begin{eqnarray}
\Pi^{\mu}(\omega, \bar{\omega}) = \frac{1}{2\pi\, \alpha^{\prime}}
\left[2 \alpha^{\prime} p^{\mu} + \left(\frac{\alpha^{\prime}}{2}
\right)^{\frac{1}{2}} \sum_{m \neq 0}\alpha^{\mu}_{m}
\left( \omega^{-m} + \bar{\omega}^{-m}\right)\right]
\label{4c15}
\end{eqnarray}
where we have made use of the mode expansion of
$X^{\mu}(\omega, \bar{\omega})$ (\ref{4c13}).
The commutation relations between $X^{\mu}(\omega, \bar{\omega})$ and
$\Pi^{\nu}(\omega^{\prime}, \bar{\omega^{\prime}})$ are then obtained by using
(\ref{4c14}, \ref{4c14a}) as,
\begin{eqnarray}
\left[X^{\mu}(\omega, \bar{\omega}), \Pi^{\nu}(\omega^{\prime},
\bar{\omega^{\prime}})\right] = i \eta^{\mu \nu}\left(\frac{1}{\pi}
+ \frac{1}{4\pi} \sum_{m \neq 0}\left(\omega^{-m} + \bar{\omega}^{-m}\right)
\left( \omega^{\prime m} + \bar{\omega^{\prime}}^{m}\right)\right).
\label{4c16}
\end{eqnarray}
To obtain the usual equal time (i.e. $\tau = \tau^{\prime}$)
commutation relation we first rewrite (\ref{4c16}) in ``$z$ frame''
using (\ref{4c10}) and then in terms of $\sigma^1, \ \sigma^2$ to find,
\begin{eqnarray}
\left[X^{\mu}(\sigma^1, \sigma^2), \Pi^{\nu}(\sigma^{1\,\prime}, \sigma^{2\,\prime})\right]
= i \eta^{\mu \nu}\left[\frac{1}{\pi}
+ \frac{1}{\pi} \sum_{m \neq 0} \exp^{-m\left(\sigma^2 -
\sigma^{2\,\prime}\right)} \mathrm{cos}\left(m \sigma^1\right)
\mathrm{cos}\left(m \sigma^{\prime\,1}\right)\right].
\label{4c17}
\end{eqnarray}
Finally substituting $\tau = \tau^{\prime}$ i.e. $\sigma^2 = \sigma^{2\,\prime}$
and $\sigma^1 = \sigma$ we get
back the usual equal time commutation relation,
\begin{eqnarray}
\left[X^{\mu}(\tau, \sigma), \Pi^{\nu}(\tau, \sigma^{\prime})\right]
= i \eta^{\mu \nu} \Delta_{+}\left(\sigma , \sigma^{\prime}\right)
\label{4c18}
\end{eqnarray}
where,
\begin{eqnarray}
\Delta_{+}\left(\sigma , \sigma^{\prime}\right) = \left[\frac{1}{\pi}
+ \frac{1}{\pi} \sum_{m \neq 0} \mathrm{cos}\left(m \sigma\right)
\mathrm{cos}\left(m \sigma^{\prime}\right)\right].
\label{4c18aa}
\end{eqnarray}
It is easy to see that (\ref{4c18}) is compatible with Neumann BCs
and reproduces the result in the previous chapters and \cite{rb,
hrt, jing}.
\subsection{Open string in the constant B-field background}
We now analyse the open string moving in presence of a background
Neveu-Schwarz two form field $B_{\mu \nu}$.
To begin with, let us again consider the Laurent expansion of $\partial X^{\mu}(\omega)$
and $\bar{\partial} X^{\mu}(\bar{\omega})$ (\ref{4c12}) for the case of closed
string. But now the corresponding open string Laurent expansion is
obtained by imposing the BCs, given in (\ref{4c31q}) with $B_{\mu \nu} \neq 0$
consequently, the modes $\alpha$ and $\tilde{\alpha}$ now satisfy:
\begin{eqnarray}
\label{4c34}
\alpha^{\mu}_{m} - B^{\mu}_{\ \nu}\alpha^{\nu}_{m} =
\tilde{\alpha}^{\mu}_{m} + B^{\mu}_{\ \nu}\tilde{\alpha}^{\nu}_{m}.
\end{eqnarray}
So there exists only one set of independent modes
$\gamma^{\mu}_{m}$, which can be thought of as the modes of free
strings and is related to $\alpha^{\mu}_{m}$ and $\tilde{\alpha}^{\mu}_{m}$ by:
\begin{eqnarray}
\label{4c35}
\alpha^{\mu}_{m} &=& \left(\delta^{\mu}_{\ \nu} + B^{\mu}_{\ \nu}
\right)\gamma^{\nu}_{m} := \left[({1\!\mbox{l}} + B)\gamma
\right]^{\mu}_{m} \nonumber \\
\tilde{\alpha}^{\mu}_{m} &=& \left(\delta^{\mu}_{\ \nu} - B^{\mu}_{\ \nu}
\right)\gamma^{\nu}_{m} := \left[({1\!\mbox{l}} - B)\gamma\right]^{\mu}_{m} .
\end{eqnarray}
Note that under world-sheet parity transformation
$\alpha^{\mu}_{m} \leftrightarrow \tilde{\alpha}^{\mu}_{m}$, as $B_{\mu \nu}$
is a world-sheet pseudo-scalar.
Substituting (\ref{4c35}) in (\ref{4c12}), we obtain  the
following Laurent expansions for $\partial X^{\mu}$ and $\bar{\partial} X^{\mu}$:
\begin{eqnarray}
\label{4c19}
\partial X^{\mu}(\omega) &=& -i \left(\frac{\alpha^{\prime}}{2}\right)^{\frac{1}{2}}
\sum^{\infty}_{m= -\infty} \frac{\left[({1\!\mbox{l}} + B)
\gamma\right]^{\mu}_{m}}{\omega^{m+1}} \\
\bar{\partial} X^{\mu}(\bar{\omega}) &=& -i \left(\frac{\alpha^{\prime}}{2}
\right)^{\frac{1}{2}}
\sum^{\infty}_{m= -\infty} \frac{\left[({1\!\mbox{l}} - B)
\gamma\right]^{\mu}_{m}}{\bar{\omega}^{m+1}}\nonumber.
\end{eqnarray}
Integrating the expansion (\ref{4c19}) we obtain the mode expansion
of $X^{\mu}(\omega, \bar{\omega})$ for the interacting string:
\begin{eqnarray}
\label{4c20}
X^{\mu}(\omega, \bar{\omega}) &=& x^{\mu} -i\alpha^{\prime}p^{\mu}
\mathrm{ln}\vert \omega \vert^2 - i \alpha^{\prime} B^{\mu}_{\ \nu}p^{\nu}
\left(\mathrm{ln}\omega - \mathrm{ln}\bar{\omega}\right) \nonumber \\
&&+ \, i \left(\frac{\alpha^{\prime}}{2}
\right)^{\frac{1}{2}}\sum_{m \neq 0}\left[\frac{\gamma^{\mu}_{m}}{m}
\left( \omega^{-m} + \bar{\omega}^{-m}\right) + \frac{1}{m}B^{\mu}_{\ \nu}\,
\gamma^{\nu}_{\ m}\left( \omega^{-m} - \bar{\omega}^{-m}\right)\right].
\end{eqnarray}
Now the expressions (\ref{4c19}) for open interacting strings can
also be equivalently written as:
\begin{eqnarray}
\left[({1\!\mbox{l}} + B)\gamma\right]^{\mu}_{m} &=&
\left(\frac{2}{\alpha^{\prime}}\right)^{\frac{1}{2}}
\oint \frac{d\omega}{2\pi}\, \omega^m \partial X^{\mu}(\omega) \nonumber \\
\left[({1\!\mbox{l}} - B)\gamma\right]^{\mu}_{m} &=&
- \left(\frac{2}{\alpha^{\prime}}\right)^{\frac{1}{2}}
\oint \frac{d\bar{\omega}}{2\pi}\, \bar{\omega}^m \partial X^{\mu}(\bar{\omega}).
\label{4c39}
\end{eqnarray}
The commutation relation between $\gamma$'sigma can be obtained
from the contour argument (using (\ref{4c39})) and the $X\, X$ OPE (\ref{4c11}):
\begin{eqnarray}
\label{4c22}
\left[\gamma^{\mu}_{m}, \gamma^{\nu}_{n}\right]
= m\, \delta_{m, -n}\left[\left({1\!\mbox{l}}- B^2\right)^{-1}\right]^{\mu \nu}
=  m\, \delta_{m, -n} \left(M^{-1}\right)^{\mu \nu}
\end{eqnarray}
where, $M = ({1\!\mbox{l}} - B^2)$ ; $(B^2)^{\mu \nu} = B^{\mu}_{\
\rho}B^{\rho \nu}$\footnote{Here we should note that
$\left({1\!\mbox{l}}\right)^{\mu \nu} = \eta^{\mu \nu}$.}. Once
again the algebra among the zero modes i.e. the center of mass
variables $[x^\mu, p^\nu]$ and $[x^\mu, x^\nu]$ of the open string
can not be obtained from the above contour arguments. The results
can be derived using standard techniques (as discussed earlier)
and read \cite{jing},
\begin{eqnarray}
\label{4c23}
\left[x^{\mu}, p^{\nu}\right] &=& i \left(M^{-1}\right)^{\mu \nu} \nonumber \\
\left[x^{\mu}, x^{\nu}\right] &=& -2i\,\alpha^\prime
\pi \left(M^{-1}B\right)^{\mu \nu}.
\end{eqnarray}
Now proceeding as in the free case, the conjugate momentum
$\Pi^{\mu}(\omega. \bar{\omega})$ corresponding to (\ref{4c20}) is:
\begin{eqnarray}
\label{4c27}
\Pi^{\mu}(\omega, \bar{\omega}) &=& \frac{1}{2\pi \alpha^{\prime}}
\left[2 \alpha^{\prime}\, M^{\mu \rho}\,p_{\rho}
+ \left(\frac{\alpha^{\prime}}{2}\right)^{\frac{1}{2}}
\sum_{m \neq 0}M^{\mu}_{\ \rho}\,\gamma^{\rho}_{\ m}
\left( \omega^{-m} - \bar{\omega}^{-m}\right)\right].
\end{eqnarray}
The commutators among the canonical variables $X^{\mu}(\omega, \bar{\omega}),
\Pi^{\nu}(\omega, \bar{\omega})$ can be computed by
using (\ref{4c22}), (\ref{4c23}),
\begin{eqnarray}
\label{4c24}
\left[X^{\mu}(\omega, \bar{\omega}), \Pi^{\nu}(\omega^{\prime},
\bar{\omega^{\prime}})\right] &=& i \eta^{\mu \nu}\left(\frac{1}{\pi}
+ \frac{1}{4\pi} \sum_{m \neq 0}\left(\omega^{-m} + \bar{\omega}^{-m}\right)
\left( \omega^{\prime m} + \bar{\omega^{\prime}}^{m}\right)\right. \nonumber \\
&& \left. + \frac{1}{4\pi}\sum_{m \neq 0}B^{\mu \nu}
\left(\omega^{-m} - \bar{\omega}^{-m}\right)
\left( \omega^{\prime m} + \bar{\omega^{\prime}}^{m}\right)\right)
 \\
\left[X^{\mu}(\omega, \bar{\omega}),
X^{\nu}(\omega^{\prime}, \bar{\omega^{\prime}})\right] &=& \alpha^{\prime}
\left(M^{-1}\right)^{\mu \nu}\left(\mathrm{ln}\vert \omega^{\prime}\vert^2
- \mathrm{ln}\vert \omega \vert^2\right) - \alpha^{\prime}\left(M^{-1} B
\right)^{\mu \nu}\left(\mathrm{ln}\frac{\omega^{\prime}}{\bar{\omega^{\prime}}}
+ \mathrm{ln}\frac{\omega}{\bar{\omega}} + 2 i\, \pi\right) \nonumber \\
&& + \frac{\alpha^{\prime}}{2}\left(\sum_{m \neq 0}\frac{1}{m}\left[
\left(M^{-1}\right)^{\mu \nu}\left(\omega^{-m} + \bar{\omega}^{-m}\right)
\left( \omega^{\prime m} + \bar{\omega^{\prime}}^{m}\right)\right.\right. \nonumber \\
&&\left.\left.+ B^{\mu}_{\ \rho}\, B^{\nu}_{\ \sigma}\left(
M^{-1}\right)^{\mu \nu}\left(\omega^{-m} - \bar{\omega}^{-m}\right)
\left( \omega^{\prime m} - \bar{\omega^{\prime}}^{m}\right)\right.\right. \nonumber \\
&&\left.\left. - \left(
M^{-1}B\right)^{\mu \nu}\left(\omega^{-m} + \bar{\omega}^{-m}\right)
\left( \omega^{\prime m} - \bar{\omega^{\prime}}^{m}\right)
\right.\right. \nonumber \\
&&\left.\left. + \left(
M^{-1}B\right)^{\mu \nu}\left(\omega^{-m} + \bar{\omega}^{-m}\right)
\left( \omega^{\prime m} + \bar{\omega^{\prime}}^{m}\right)
\right]\right)\nonumber \\
\left[\Pi^{\mu}(\omega, \bar{\omega}), \Pi^{\nu}(\omega^{\prime},
\bar{\omega^{\prime}})\right] &=& 0 \nonumber.
\end{eqnarray}
Now proceeding as before, we can rewrite the above
commutation relation in $\sigma^1, \sigma^2$ coordinates to obtain the following,
\begin{eqnarray}
\label{4c25}
\left[X^{\mu}(\sigma^1, \sigma^2), \Pi^{\nu}(\sigma^{1\prime}, \sigma^{2\prime})\right]
&=& i \eta^{\mu \nu}\left(
\frac{1}{\pi}
+ \frac{1}{\pi} \sum_{m \neq 0} \exp^{-m\left(\sigma^2 -
\sigma^{2\,\prime}\right)}\left[ \mathrm{cos}\left(m \sigma^1\right)
\mathrm{cos}\left(m \sigma^{\prime\,1}\right)
\right.\right. \nonumber \\
&& \left.\left. + B^{\mu \nu}
\mathrm{sin}\left(m \sigma^1\right)
\mathrm{cos}\left(m \sigma^{\prime\,1}\right)\right]\right) \\
\left[X^{\mu}(\sigma^1, \sigma^2),
X^{\nu}(\sigma^{1\prime}, \sigma^{2\prime})\right] &=& 2\alpha^{\prime}
\left(M^{-1}\right)^{\mu \nu}\left(\sigma^{2\prime} - \sigma^{2}\right)
+2i \alpha^{\prime}\left(M^{-1} B
\right)^{\mu \nu}\left(\sigma^{1 \prime} + \sigma^{1} - \pi\right) \nonumber \\
&& + 2\alpha^{\prime}\left(\sum_{m \neq 0}\frac{1}{m}e^{-m\left(\sigma^{2}
- \sigma^{2\prime}\right)}\left[
\left(M^{-1}\right)^{\mu \nu}
\mathrm{cos}(m\sigma^1)\mathrm{cos}(m\sigma^{1\prime})\right.\right. \nonumber \\
&&\left.\left.+ B^{\mu}_{\ \rho}\, B^{\nu}_{\ \sigma}\,
\mathrm{sin}(m\sigma^1)\, \mathrm{sin}(m\sigma^{1\prime})
\right.\right. \nonumber \\
&&\left.\left. +i \left(M^{-1}B\right)^{\mu \nu}
\mathrm{sin} \left(m(\sigma^1 + \sigma^{1\prime})\right)\right]\right)\nonumber .
\end{eqnarray}
Finally we obtain the equal time commutation relations by
identifying $\tau = \tau^{\prime}$, i.e. $\sigma^2 = \sigma^{2\,\prime}$
and setting $\sigma^1 = \sigma$,
\begin{eqnarray}
\label{4c26}
\left[X^{\mu}(\tau, \sigma),
\Pi^{\nu}(\tau, \sigma^{\prime})\right] &=& i\eta^{\mu \nu}
\Delta_{+}(\sigma, \sigma^{\prime})\nonumber \\
\left[X^{\mu}(\tau, \sigma),
X^{\nu}(\tau, \sigma^{\prime})\right] &=& 2i \alpha^{\prime}
\left(M^{-1}B\right)^{\mu \nu}\left[\sigma + \sigma^{\prime} - \pi + \sum_{n \neq 0}
\frac{1}{n}\mathrm{sin}\, \left(n(\sigma + \sigma^{\prime})\right)\right].
\end{eqnarray}
One can explicitly check that these commutators are compatible
with BCs and also reproduces the result of 2nd chapter and
(\cite{chu, chu1, rb, br, jing}).

\section{Summary}
In this chapter we have discussed an open bosonic string moving in
the presence of a background Neveu-Schwarz two-form field $B_{\mu
\nu}$ in a conformal field theory approach. We find the
noncommutativity at the end point of the string. In contrast to
several discussions, in which boundary conditions are taken as
Dirac constraints, we have first obtained the mode algebra by
using the contour argument and the newly proposed normal ordering,
which satisfies both equations of motion and boundary conditions.
Using these the commutator among the string coordinates is
obtained. Interestingly, this new normal ordering  yields the same
algebra between the modes as the one satisfying only  the
equations of motion. In this approach, we find that
noncommutativity originates more transparently and our results
match with the noncommutative structure obtained in previous
chapters and existing results in the literature.

\chapter{Superstring in a constant antisymmetric background field}
We have already seen that open bosonic string, in presence of a
back-ground Neveu-Schwarz two form field, leads to a
noncommutative structure. In different approaches we have shown
the same noncommuyativity appears in the space-time coordinates of
$D$-branes, where the end points of open string are attached.

\noindent The particular string theory described in this chapter
is based on the introduction of a world-sheet supersymmetry that
relates the space-time coordinates $X^\mu(\tau, \sigma)$, taken to
be bosonic, to their fermionic counterparts  $\psi^\mu(\tau,
\sigma)$. The later are two component world-sheet spinors. Then we
study how non(anti) commutativity appears in the free super string
and in a constant B-field background. This non(anti)commutativity
is a direct consequence of the non trivial boundary conditions
which, contrary to several approaches, are not treated as
constraints. In this sense we have extended our methodology of
bosonic theory to the superstring theory. We start with RNS
superstring action in the conformal gauge. This also helps to fix
the notations. Then we discuss the boundary conditions of the
fermionic sector of the superstring and the non-anticommutativity
of the theory.

\section {Free superstring}
Let us consider the action for the free superstring, in conformal
gauge \cite{green, green13},
\begin{eqnarray}
\label{5c4} S \,=\, {i \over 4 } \int_{\Sigma} d^2\sigma \,d^2
\theta \Big( {\overline D} Y^\mu D Y_\mu  \Big)\,,
\end{eqnarray}
where the superfield
 \begin{eqnarray}
Y^\mu (\sigma, \theta) \,=\,
 X^\mu (\sigma)
+ {\overline \theta}  \psi^\mu (\sigma) + \frac{1}{2} {\overline
\theta} \theta B^\mu  (\sigma)
\label{5c4ka}
\end{eqnarray}
unites the bosonic ($X{^\mu}(\sigma)$) and fermionic
 ($\psi^{\mu}(\sigma)$) space-time string coordinates with
a new auxiliary bosonic field $B{^\mu}(\sigma)$ whose utility may not be
apparent at first. Note that both the bosonic ($X^\mu$) and fermionic
($\psi^\mu$) variables transform as vectors under target-space
Lorentz transformations $SO(D-1,1)$ and scalars under
arbitary world-sheet diffeomorphism, but as
scalars and spinors, respectively, under
world-sheet (local) Lorentz transformations $SO(1,1)$ i.e. orthonormal
transformation of tangent space at each point of the world-sheet.

Our signature of the induced world-sheet metric and target
space-time metric are $\eta^{ab} = \{-, +\}$,
 $\eta^{\mu\nu} =\{-, +, +, ...., +\}$ respectively and
  $\bar{\theta}$ is defined as
$\bar{\theta} = \theta^T\rho^0$. The derivative $$  D_A =
\frac{\partial}{\partial\bar{\theta}^A}
 - i(\rho^a\theta)_A\partial_a$$ is known as the
superspace covariant derivative. Here the symbol $\rho^{a}$
represents two-dimensional Dirac matrices. A convenient basis  is
\begin{equation}
\label{5c212} \rho^0 \,=\, \sigma^2 \, = \, \pmatrix{0&-i\cr
i&0\cr}\,\,\,,\,\,\, \rho^1 \,=\, i\sigma^1 \, = \,
\pmatrix{0&i\cr i&0\cr}\,,
\end{equation}
they obey the Clifford algebra
\begin{equation}
\{\rho^a , \rho^b\} = - 2 \eta^{ab}.
\label{5c38}
\end{equation}
The advantage of writing down the action in the superspace formalism,
which includes the $B^\mu$ field,
is that it is now manifestly invariant under infinitesimal
supersymmetric transformations (with $\epsilon$ being infinitesimal
Majorana spinor parameter)
\begin{eqnarray}
\label{5c17}
\delta X^{\mu} &=& \bar{\epsilon}\psi^{\mu}\,, \nonumber \\
\delta \psi^{\mu} &=& -i\rho^a\partial_aX^{\mu}\epsilon + B^{\mu}\, \epsilon\,, \\
\delta B^{\mu} &=& -i\bar{\epsilon} \rho^a\partial_a \psi^{\mu}\nonumber
\end{eqnarray}
even without going on-shell. In absence of the $B^\mu$ field, however,
one has to necessarily implement the on-shell condition.

\noindent In order to compute the $\theta$ integrals explicitly,
we first note that
\begin{eqnarray}
\label{5c9} DY^{\mu} &=& \psi^{\mu} + \theta B^{\mu} - i \rho^a
\theta \partial_{a}X^{\mu}
+ \frac{i}{2} \bar{\theta}\theta \rho^a \partial_ a \psi^\mu \,,\nonumber \\
\bar{D}Y^{\mu} &=& \bar{\psi^{\mu}} + B^{\mu}\bar{\theta} + i \partial_{a}X^{\mu}
\bar{\theta}\rho^a - \frac{i}{2} \bar{\theta}\theta  \partial_ a \bar{\psi^{\mu}}\rho^a .
\end{eqnarray}
It can easily be checked that both $D\Phi^{\mu}$  and
$\bar{D}\Phi^{\mu}$ are invariant under SUSY transformations
(\ref{5c17}), so that SUSY invariance of the action (\ref{5c4}) is
manifest.  In component form the action reads
\begin{eqnarray}
\label{5c1}
S &=& -\frac{1}{2} \int_{\Sigma} d^2\sigma \Big(
\eta_{\mu \nu } \partial_a X^\mu \partial^a X^\nu
\,- \,  i {\overline \psi}^\mu \rho^a \partial_ a \psi_\mu \Big)\\
 &=& S_B + S_F \,,\nonumber
\end{eqnarray}
where
\begin{eqnarray}
\label{5c1aa}
 S_B &=& -\frac{1}{2}\int_{\Sigma} d^2\sigma \eta_{\mu \nu }
 \partial_a X^\mu \partial^a X^\nu\,,  \nonumber \\
S_F&=& \frac{1}{2}\int_{\Sigma} d^2\sigma
 i {\overline \psi}^\mu \rho^a \partial_ a \psi_\mu
\end{eqnarray}
represent the decoupled bosonic and fermionic actions,
respectively. The fermions are taken to be
Majorana and we refer to the component of $\psi$ in the
 basis (\ref{5c212}) as $\psi_{\pm}$
\begin{equation}
\psi^\mu \,=\, \pmatrix{\psi^\mu_{-}\cr \psi^\mu_{+}\cr}\,,
\label{5c3}
\end{equation}
\noindent which in the representation (\ref{5c212}) are real (see
appendix A).
The structure of the fermionic part $S_F$ (\ref{5c1aa}) of the
action shows that the kinetic term is first order in the time
derivative, consequently one can either employ the Dirac bracket
formalism \cite{di,hrt} or the Faddeev-Jackiw method \cite{fj}
 to write down the following anti-bracket\footnote{In this chapter,
we have to clearly deal with a graded Lie-algebraic structure for the
Poisson/Dirac brackets, but we shall make no distinction here on the
notation.} :
\begin{equation}
\{\psi^{\mu}_{A}(\tau, \sigma), \psi^{\nu}_{B}(\tau,
\sigma^{\prime})\} _{D.B}=-i\eta^{\mu\nu}\delta_{AB}\delta(\sigma-
\sigma^{\prime})\,.
\label{5c3e}
\end{equation}
\noindent The above antibrackets read, in terms of the components of $\psi$,
\begin{eqnarray}
\{ \psi^\mu_{+} (\sigma) \,,\,\psi^\nu_{+}  (\sigma^\prime) \}_{D.B} &=&
\{ \psi^\mu_{-} (\sigma) \,,\,\psi^\nu_{-} (\sigma^\prime ) \}_{D.B}
\,\,=\,\,
 - i \eta^{\mu\nu} \delta (\sigma - \sigma^\prime )\,,
\nonumber\\
\{ \psi^\mu_{+}(\sigma)\,,\,\psi^\nu_{-}(\sigma^\prime) \}_{D.B}
&=& 0\,\,.
\label{5c5}
\end{eqnarray}
This, along with the brackets
\begin{eqnarray}
\label{5c18a}
\{X^\mu(\sigma) , \Pi^\nu(\sigma^{\prime})\} =
 \eta^{\mu \nu}\delta(\sigma - \sigma^{\prime})
\end{eqnarray}
from the bosonic sector, defines the preliminary symplectic
structure of the theory ($\Pi^\mu$ is the cannonically conjugate momentum
to $X^\mu$, defined in the usual way).

\noindent Confining our attention to $S_F$, we vary the action
(\ref{5c1aa})
\begin{eqnarray}
\label{5c18}
\delta S_F = i \int_{\Sigma} d^2\sigma \left[ \rho^a \, \partial_a
\psi^{\mu} \, \delta \bar{\psi}_{\mu} - \partial_{\sigma}
\left(\psi^\mu_{-}\, \delta \psi_{\mu -} -
\psi^\mu_{+}\, \delta \psi_{\mu +}\right)\right]
\end{eqnarray}
to obtain the Euler-Lagrange equation for fermionic field
\begin{eqnarray}
\label{5c10}
i \rho^a \partial_a \psi^{\mu} =0\,,
\end{eqnarray}
which further reduces to
\begin{eqnarray}
\label{5c11}
\left(\frac{\partial}{\partial\tau} +
\frac{\partial}{\partial\sigma}\right)\psi^\mu_{-} = 0 \ \ ; \ \
\left(\frac{\partial}{\partial\tau} -
\frac{\partial}{\partial\sigma}\right)\psi^\mu_{+} = 0\,.
\end{eqnarray}
This indicates that the functional dependence of fermi-fields are
given by $\psi_{\mp}(\tau \mp \sigma)$. The total divergence term
yields the necessary BC. We shall consider its consequences in the
following sections where the preliminary (anti) brackets will be
modified.
The action $S$ of (\ref{5c1}) is invariant (using equation of
motion) under the infinitesimal transformations
\begin{eqnarray}
\label{5c14}
\delta X^{\mu} &=& \bar{\epsilon}\psi^{\mu} \nonumber \\
\delta \psi^{\mu} &=& -i\rho^a\partial_aX^{\mu}\epsilon,
\end{eqnarray}
with $\epsilon$ a constant anticommuting spinor. One can generate
these transformations through the generator
\begin{eqnarray}
\label{5c15} Q_A = \frac{\partial}{\partial\bar{\theta}^A} +
i(\rho^a \theta)_A\partial_a.
\end{eqnarray}
The super charge $Q$ is used to define transformations of the
coordinates
\begin{eqnarray}
\label{5c16} \delta Y^{\mu} = \bar{\epsilon} Q Y^{\mu}
\end{eqnarray}
Since $\{D_A , Q_B\} = 0$ i.e. derivative operator is invariant
under supersymmetry, the action (\ref{5c1}) is also invariant
under supersymmetry.
 Using the standard Noether procedure\footnote{We now use
the supersymmetry transformations on-shell and hence we drop the
auxiliary field $B^{\mu}$ henceforth.}, the forms of the
supercurrent and the energy-momentum tensor (which are constraints
themselves \cite{green}) can be derived. The expressions are:
\begin{eqnarray}
J_{a}=-\frac{1}{2}\rho^{b}\rho_{a}\psi^{\mu}\partial_{b}X_{\mu}=
0\,, \label{5c17a}
\end{eqnarray}
\begin{eqnarray}
T_{ab}=\partial_{a}X^{\mu}\partial_{b}X_{\mu}
-\frac{i}{4}\bar{\psi^{\mu}}\rho_{a}\partial_{b}\psi_{\mu}
-\frac{i}{4}\bar{\psi^{\mu}}\rho_{b}\partial_{a}\psi_{\mu}
-\frac{1}{2}\eta_{ab}(\partial^{c}X^{\mu}\partial_{c}X_{\mu}+
\frac{i}{2}\bar{\psi^{\mu}}\rho^{a}\partial_{a}\psi_{\mu}) = 0\,.
\label{5c17b}
\end{eqnarray}
All the components of $T_{ab}$ are, however, not independent as the energy-
momentum tensor is traceless
\begin{eqnarray}
T^{a}_{\ a} = \eta^{ab}T_{ab} = 0\,,
\label{5c17c}
\end{eqnarray}
leaving us with only two independent components of $T_{ab}$.
These components, which are the constraints of the theory,
are given by
\begin{eqnarray}
\chi_1(\sigma) = 2T_{00} = 2T_{11} &=& \Omega_1(\sigma) +
\lambda_1(\sigma) = 0 \nonumber \\
\chi_2(\sigma) = T_{0 1} &=& \Omega_2(\sigma) + \lambda_2(\sigma)
= 0 \label{5c27}.
\end{eqnarray}
where,
\begin{eqnarray}
\Omega_1(\sigma) &=&
\left(\Pi^2(\sigma) + (\partial_{\sigma}X(\sigma))^2\right) \nonumber \\
\Omega_2(\sigma) &=&
\left(\Pi(\sigma)\partial_{\sigma}X(\sigma)\right) \nonumber \\
\lambda_1(\sigma)&=&
- i \bar{\psi^{\mu}}(\sigma)\rho_{1} \partial_{\sigma} \psi_{\mu}(\sigma)
=  - i \left(\psi^{\mu}_{-}(\sigma)
\partial_{\sigma} \psi_{\mu -}(\sigma) - \psi^{\mu}_{+}(\sigma)
\partial_{\sigma} \psi_{\mu +}(\sigma)\right) \nonumber \\
\lambda_2(\sigma)&=&
- \frac{i}{2}  \bar{\psi^{\mu}}(\sigma)\rho_{0}
\partial_{\sigma} \psi_{\mu}(\sigma)
= \frac{i}{2}\left(\psi^{\mu}_{-}(\sigma)\partial_{\sigma}
\psi_{\mu -}(\sigma) + \psi^{\mu}_{+}(\sigma)\partial_{\sigma}
\psi_{\mu +}(\sigma)\right)
\label{5c27q}.
\end{eqnarray}

\noindent The role of these constraints in generating
infinitesimal diffeomorphisms  is well known \cite{green, green13}
and we are not going to elaborate on this. Note that the
constraints that we obtain in this chapter are on-shell, i.e. we
have used the equation of motion (\ref{5c10})
 for the fermionic field $\psi$.
This allows us to write them down in terms of the phase-space
variables\footnote{This is in the true spirit of Dirac's classic
analysis of constrained hamiltonian dynamics \cite{di}.} and hence
they look quite different from the standard results found in the
literature \cite{green, green13} where they are written down in
the light-cone coordinates which involves time derivatives.

\noindent
 From the basic brackets (\ref{5c5}),
 it is easy to generate a first
class (involutive) algebra,
\begin{eqnarray}
\{\chi_1(\sigma) , \chi_1(\sigma^{\prime})\} &=& 4 \left(
\chi_2(\sigma) + \chi_2(\sigma^{\prime})\right)\partial_{\sigma}
\delta(\sigma - \sigma^{\prime})\,\nonumber \\
\{\chi_2(\sigma) , \chi_2(\sigma^{\prime})\} &=& \left(
\chi_2(\sigma) + \chi_2(\sigma^{\prime})\right)\partial_{\sigma}
\delta(\sigma - \sigma^{\prime})\,\nonumber\\
\{\chi_2(\sigma) , \chi_1(\sigma^{\prime})\} &=& \left(
\chi_1(\sigma) + \chi_1(\sigma^{\prime})\right)\partial_{\sigma}
\delta(\sigma - \sigma^{\prime})\,.
\label{5c4200}
\end{eqnarray}
It is interesting to observe that the structure of the super first
class constraint algebra is exactly similar to that of the
constraint algebra (\ref{alg}) of Bosonic theory.

\noindent
 Coming to the super current $J_{aA}$ \footnote{$A = 1,2$
being the spinor index},
 note that it is a two component spinor. Further, since $J_{a}$
obeys the relation $\rho^a J_a = 0$, the components of $J_{0A}$ and
 $J_{1A}$ are related to each other. Hence we only deal with the
components of $J_{0A}$ or simply $J_1$ and $ J_2$. These $J_1$, $
J_2$ along with $\chi_1(\sigma)$ and $\chi_2(\sigma)$ constitutes
the full set of super-Virasoro constraints. For convenience we
write:
\begin{eqnarray}
\tilde{J}_{1}(\sigma) &=& 2J_{1}(\sigma) =
(\psi^{\mu}_{-}(\sigma)\Pi_{\mu}(\sigma)
-\psi^{\mu}_{-}(\sigma)\partial_{\sigma}X_{\mu}) = 0\,, \nonumber\\
\tilde{J}_{2}(\sigma) &=& 2J_{2}(\sigma) =
(\psi^{\mu}_{+}(\sigma)\Pi_{\mu}(\sigma)
+\psi^{\mu}_{+}(\sigma)\partial_{\sigma}X_{\mu}) = 0\,.
\label{5c280}
\end{eqnarray}
The algebra between the above constraints read:
\begin{eqnarray}
\{\tilde{J}_{1}(\sigma) , \tilde{J}_{1}(\sigma^{\prime})\}
&=&-i(\chi_{1}(\sigma)-2\chi_{2}(\sigma))
\delta(\sigma-\sigma^{\prime})\,,\nonumber\\
\{\tilde{J}_{2}(\sigma) , \tilde{J}_{2}(\sigma^{\prime})\}
&=&-i(\chi_{1}(\sigma)+2\chi_{2}(\sigma))
\delta(\sigma-\sigma^{\prime})\,,\nonumber\\
\{\tilde{J}_{1}(\sigma) , \tilde{J}_{2}(\sigma^{\prime})\}&=&0\,.
\label{5c281}
\end{eqnarray}
The algebra between $\tilde J(\sigma)$ and $\chi(\sigma)$ is also given by
\begin{eqnarray}
\{\chi_{1}(\sigma) , \tilde{J}_{1}(\sigma^{\prime})\}
&=& - \left(2 \tilde{J}_{1}(\sigma) + \tilde{J}_{1}(\sigma^{\prime})\right)
\partial_{\sigma}\delta(\sigma-\sigma^{\prime})\,,\nonumber\\
\{\chi_{1}(\sigma) , \tilde{J}_{2}(\sigma^{\prime})\}
&=& \left(2 \tilde{J}_{2}(\sigma) + \tilde{J}_{2}(\sigma^{\prime})\right)
\partial_{\sigma}\delta(\sigma-\sigma^{\prime})\,,\nonumber\\
\{\chi_{2}(\sigma) , \tilde{J}_{1}(\sigma^{\prime})\}
&=& \left(\tilde{J}_{1}(\sigma) + \frac{1}{2}\tilde{J}_{1}(\sigma^{\prime})\right)\partial_{\sigma}\delta(\sigma-\sigma^{\prime})\,,\nonumber\\
\{\chi_{2}(\sigma) , \tilde{J}_{2}(\sigma^{\prime})\} &=&
\left(\tilde{J}_{2}(\sigma) +
\frac{1}{2}\tilde{J}_{2}(\sigma^{\prime})\right)\partial_{\sigma}\delta(\sigma-\sigma^{\prime})\,.
\label{5c282}
\end{eqnarray}

\section {Boundary conditions and super-Virasoro algebra for superstring}
As in the case of bosonic variables, Fermionic coordinates also
require careful consideration of the surface terms arising in the
variation of the action (\ref{5c18}). Vanishing of these surface
terms requires that ($\psi_{+}\delta\psi_{+} -
\psi_{-}\delta\psi_{-}$) should vanish at each end point of the
open string. This is satisfied by making $ \psi_{+} = \pm
\psi_{-}$ at each end. Without loss of generality we set
\begin{eqnarray}
\label{5c12}
\psi^{\mu}_{+}(0 , \tau) = \psi^{\mu}_{-}(0 , \tau).
\end{eqnarray}
The relative sign at the other end now becomes meaningful
and there are two cases to be considered. In the first case
(Ramond(R) boundary conditions)
\begin{eqnarray}
\label{5c6}
\psi^{\mu}_{+}(\pi , \tau) = \psi^{\mu}_{-}(\pi ,\tau)
\end{eqnarray}
and in the second case (Neveu-Schwarz (NS) boundary conditions)
\begin{eqnarray}
\label{5c13}
\psi^{\mu}_{+}(\pi , \tau) = - \psi^{\mu}_{-}(\pi ,
\tau).
\end{eqnarray}
Here we will work with Ramond boundary conditions and in the last
chapter we shall discuss the Neveu-Schwarz boundary conditions in
detail. Combining (\ref{5c12}) and (\ref{5c6}) we can write
\begin{eqnarray}
 \Big(\psi^\mu_{+} (\tau , \sigma)  - \psi^\mu_{-}
 (\tau , \sigma)\Big)\mid_{\sigma =0,\pi} = 0\,.
\label{5c7}
\end{eqnarray}

As discussed in the appendix, we have $\psi_L(\tau - \sigma) = -\
i \psi_R(\tau + \sigma)$ in the chiral representation (see
(\ref{chiral}) in appendix), which translates into
\begin{eqnarray}
\psi^{\mu}_{-}(-\sigma , \tau) &=& \psi^{\mu}_{+}(\sigma , \tau)
\label{5c31a}
\end{eqnarray}
in the representation (\ref{5c212}). Using the BC (\ref{5c6}), we
can therefore write
\begin{eqnarray}
\psi^{\mu}_{\pm}(\sigma = \pi , \tau) = \psi^{\mu}_{\pm}(\sigma =
-\pi , \tau) \label{5c30h}
\end{eqnarray}
in R-sector. We can then extend the range of definition of
$\psi^{\mu}_{\pm}$ from $[0 , \pi]$ to $[-\pi , \pi]$ with
periodic BC imposed on $\psi^{\mu}_{\pm}$ of period $2\pi$.
Consequently, the mode expansion of the components of Majorana
fermion takes the form
\begin{eqnarray}
\psi^{\mu}_{-}(\sigma , \tau) &=& \frac{1}{\sqrt{2}}\sum d^{\mu}_{n}
e^{-in(\tau - \sigma )}\,, \nonumber \\
\psi^{\mu}_{+}(\sigma , \tau) &=& \frac{1}{\sqrt{2}}\sum
d^{\mu}_{n} e^{-in(\tau + \sigma )}\,. \label{5c31}
\end{eqnarray}

\noindent On the other hand, in chapter 2 we have already enlarged
the domain of definition of the bosonic field $X^{\mu}$ as
\begin{eqnarray}
X^{\mu}(\tau , -\sigma) &=& X^{\mu}(\tau , \sigma) \label{5c37}
\end{eqnarray}
so that it is an even function and satisfies Neumann boundary
condition. This is in contrast to (\ref{5c31a}). Consistent with
this, we also have
\begin{eqnarray}
\Pi^{\mu}(\tau , -\sigma) &=& \Pi^{\mu}(\tau , \sigma)\nonumber \\
{X^{\mu}}^{\prime}(\tau , -\sigma) &=& -{X^{\mu}}^{\prime}(\tau ,
\sigma) \label{5c37aa}.
\end{eqnarray}

Now from  (\ref{5c31a}), we note that the constraints
$\chi_1(\sigma) = 0$ and $\chi_2(\sigma) = 0$ are even and odd
respectively under $\sigma \rightarrow -\sigma$. This also enables
us to increase the domain of definition of the length of the
string from ($0\leq \sigma \leq \pi $) to ($- \pi \leq \sigma \leq
\pi $). We may then write the generator of all $\tau$ and $\sigma$
reparametrization as the functional \cite{hrt}
\begin{eqnarray}
L[f] = \frac{1}{2}\int_{0}^{\pi} d\sigma \{f_+(\sigma) \chi_1(\sigma)
+ 2 f_-(\sigma) \chi_2(\sigma)\}\,,
\label{5c32}
\end{eqnarray}
where $f_{\pm}(\sigma)=\frac{1}{2}(f(\sigma)\pm f(-\sigma))$ are by
construction even/odd function and  $f(\sigma)$ is an arbitary
differentiable function defined in the extended interval $[-\pi, \pi]$.
The above expression can be simplified to
\begin{eqnarray}
L[f] = \frac{1}{4}\int_{-\pi}^{\pi} d\sigma f(\sigma)[\{\Pi(\sigma)+
\partial_{\sigma}X(\sigma)\}^2 + 2i\psi^{\mu}_{+}\partial_{\sigma}
\psi_{\mu+}]
\label{5c33}.
\end{eqnarray}
 Coming to the generators $J_{1}$ and $J_{2}$, note that $J_{1}(-\sigma)
= J_{2}(\sigma)$ (\ref{5c280}). This enables us to write
 down the functional $G[g]$
\begin{eqnarray}
G[g] &=& \int_{0}^{\pi} d\sigma(g(\sigma)J_{1}(\sigma)+
g(-\sigma)J_{2}(\sigma)) \nonumber \\
&=&\int_{-\pi}^{\pi} d\sigma g(\sigma)J_{1}(\sigma) =
\int_{-\pi}^{\pi} d\sigma g(-\sigma)J_{2}(\sigma) \label{5c34}
\end{eqnarray}
for any differentiable function $g(\sigma)$, defined again in the
extended interval $[-\pi,\pi]$.  These functionals (\ref{5c33}),
(\ref{5c34}) generate the following super Virasoro algebra
\begin{eqnarray}
\{L[f(\sigma)] , L[g(\sigma)]\} &=& L[f(\sigma)g^{\prime}(\sigma)
- f^{\prime}(\sigma)g(\sigma)]\,,\nonumber \\
\{G[g(\sigma)] , G[h(\sigma)]\} &=& -i L[g(-\sigma)h(-\sigma)]\,, \nonumber \\
\{L[f(\sigma)] , G[g(\sigma)]\} &=& G[f(\sigma)g^{\prime}(-\sigma)
-\frac{1}{2}f^{\prime}(\sigma)g(-\sigma)]\,.
\label{5c34a}
\end{eqnarray}
Defining
\begin{eqnarray}
L_m = L[e^{-im\sigma}]\ \ \mathrm{and} \ \ G_n = G[e^{in\sigma}]\,,
\label{5c34y}.
\end{eqnarray}
one can write down an equivalent form of the super-Virasoro algebra
\begin{eqnarray}
\{L_m , L_n\} &=& i (m-n) L_{m + n}\,,\nonumber \\
\{G_m , G_n\} &=& -i L_{m + n}\,,\nonumber \\
\{L_m , G_n\} &=& i \left(\frac{m}{2} - n\right)G_{m + n}\,.
\label{5c34b}
\end{eqnarray}
Note that we do not have a central extension here, as
the analysis is entirely classical.

\section{Non(anti)commutativity for open superstrings}
Coming back to the preliminary symplectic structure, given in
(\ref{5c5}), we note that the boundary conditions (\ref{5c7}) are
not compatible with  the brackets, although one could get the
super-Virasoro algebra (\ref{5c34a}) or (\ref{5c34b}) just by
using (\ref{5c5}) and (\ref{5c18a}).
 Hence the last of the brackets in
(\ref{5c5}) should be altered suitably. A simple inspection
suggests that
\begin{eqnarray}
\{\psi^\mu_{+}(\sigma)\,,\,\psi^\nu_{-}(\sigma^\prime)\}
 = -i \eta^{\mu\nu} \delta (\sigma - \sigma^\prime )\,.
\label{5c8}
\end{eqnarray}
Although the bracket structures (\ref{5c5}) and (\ref{5c8}) agree
with \cite{godinho} (in the free case), they can, however, not be
regarded as the final ones. This is because the
 presence of the usual Dirac delta function $\delta(\sigma - \sigma^{\prime})$
implicitly implies that the finite physical range of $\sigma \in
[0 , \pi]$ for the string has not been taken into account.
Besides, it is also not compatible with (\ref{5c31a}). Further,
the anti-brackets (\ref{5c8}) are a bit naive simply because the
support of $\sigma$ in the above (usual) delta function, or more
precisely a distribution, is $[-\infty , +\infty]$ and is not
compatible with the compact support of $\sigma$ in the physical
range of the string which is $[0 , \pi]$. This motivates us to
modify the above anti-brackets suitably. However, in order to
combine the fermionic and the bosonic sectors, one need to modify
the above antibrackets. In the previous chapters, the equal time
commutators were given in terms of $\Delta_{+}(\sigma ,
\sigma^{\prime})$, a certain combinations of periodic delta
function
\begin{eqnarray}
\{X^{\mu}(\tau , \sigma) ,  \Pi_{\nu}(\tau , \sigma^{\prime}) =
\delta^{\mu}_{\nu} \Delta_{+}(\sigma, \sigma^{\prime})\,,
\label{5c00}
\end{eqnarray}
where
\begin{eqnarray}
\Delta_{\pm}\left(\sigma , \sigma^{\prime}) = \delta_{P}(\sigma -
\sigma^{\prime})\right) \pm\delta_{P}(\sigma + \sigma^{\prime})\,,
\label{5c00z}
\end{eqnarray}
 rather than an ordinary delta function to ensure compatibility with Neumann BC
 in the bosonic sector. Basically, there one has to identify
the appropriate `` delta function " for the physical range $[0 ,
\pi]$ of  $\sigma$ starting from the periodic delta function
$\delta_P(\sigma - \sigma^{\prime})$ for the extended (but finite)
range  $[-\pi , \pi]$ and make use of the even nature of the
bosonic variables $X^{\mu}$ in the extended interval.

We can essentially follow the same methodology here in the
fermionic sector as $\psi^{\mu}_{\pm}(\tau , \sigma)$ also satisfy
periodic BC of period $2 \pi$ (\ref{5c30h}). The only difference
with the bosonic case, apart from the Grassmanian nature of the
latter,  is that, instead of their even property (\ref{5c37}), the
components of Majorana fermions satisfy (\ref{5c31a}). As
 we shall show now this condition is quite adequate to identify
the appropriate delta-functions for the ``physical interval"
$[0 , \pi]$.

We start by noting that the usual properties of a delta function is also
satisfied by $\delta_{P}(x)$
\begin{eqnarray}
\int_{-\pi}^{\pi}dx^{\prime}\delta_{P}(x^{\prime}-x)f(x^{\prime})=f(x)
\label{5c40}
\end{eqnarray}
for any periodic function $f(x)=f(x+2\pi)$ defined in the interval
 $[-\pi, \pi]$. Hence one  can immediately write down the following
expressions for $\psi^{\mu}_{-}$ and $\psi^{\mu}_{+}$:
\begin{eqnarray}
\int_{0}^{\pi}d\sigma^{\prime}\left[\delta_{P}(\sigma^{\prime} +
\sigma)\psi^{\mu}_{+}(\sigma^{\prime}) +
\delta_{P}(\sigma^{\prime}-\sigma)\psi^{\mu}_{-}(\sigma^{\prime})
\right] =\psi^{\mu}_{-}(\sigma)\,
\label{5c41}
\end{eqnarray}
\begin{eqnarray}
\int_{0}^{\pi}d\sigma^{\prime}\left[\delta_{P}(\sigma^{\prime}
+ \sigma)\psi^{\mu}_{-}(\sigma^{\prime}) +
\delta_{P}(\sigma^{\prime}-\sigma)\psi^{\mu}_{+}(\sigma^{\prime})
\right] =\psi^{\mu}_{+}(\sigma)\,.
\label{5c42}
\end{eqnarray}
Combining the above equations and writing them in a matrix form, we get,
\begin{eqnarray}
\int_{0}^{\pi}d\sigma^{\prime}\Lambda_{AB}(\sigma, \sigma^{\prime})
\psi^{\mu}_{B}(\sigma^{\prime})=\psi^{\mu}_{A}(\sigma)\quad;\quad (A=-,+)\,,
\label{5c43}
\end{eqnarray}
where $\Lambda_{AB}(\sigma, \sigma^{\prime})$, defined by
\begin{equation}
 \Lambda_{AB}(\sigma, \sigma^{\prime})\,=\, \pmatrix{\delta_{P}(\sigma^{\prime}-\sigma)&\delta_{P}(\sigma^{\prime}+\sigma)\cr
\delta_{P}(\sigma^{\prime}+\sigma)&\delta_{P}(\sigma^{\prime}-\sigma)\cr}\,,
\label{5c44}
\end{equation}
acts like a matrix valued ``delta function"
for the two component Majorana spinor in the reduced physical interval
$[0 , \pi]$ of the string.
We therefore propose the following anti-brackets in the fermionic sector:
\begin{equation}
\{\psi^{\mu}_{A}(\sigma), \psi^{\nu}_{B}(\sigma^{\prime})\}
=-i\eta^{\mu\nu}\Lambda_{AB}(\sigma, \sigma^{\prime})\,,
\label{5c45}
\end{equation}
instead of (\ref{5c3e}) which, when written down explicitly in
terms of components, reads
\begin{eqnarray}
\{\psi^{\mu}_{+}(\sigma), \psi^{\nu}_{+}(\sigma^{\prime})\}
&=&\{\psi^{\mu}_{-}(\sigma), \psi^{\nu}_{-}(\sigma^{\prime})\}
=-i\eta^{\mu\nu}\delta_{P}(\sigma - \sigma^{\prime})\,,\nonumber\\
\{\psi^{\mu}_{-}(\sigma), \psi^{\nu}_{+}(\sigma^{\prime})\}
&=&-i\eta^{\mu\nu}\delta_{P}(\sigma + \sigma^{\prime})\,.
\label{5c46}
\end{eqnarray}
We shall now investigate the consistency of this structure.
Firstly, this structure of the antibracket relations is completely
consistent with the boundary condition (\ref{5c7}). To see this
explicitly, we compute the anticommutator of
$\psi_{+}(\sigma^{\prime})$ with (\ref{5c7}), the left hand side
of which gives
\begin{eqnarray}
-i\left(\delta_{P}(\sigma - \sigma^{\prime})- \delta_{P}(\sigma +
\sigma^{\prime})\right)|_{\sigma = 0, \pi} &=& - i
\Delta_{-}\left(\sigma , \sigma^{\prime})\right)|_{\sigma = 0,
\pi} \nonumber \\
&=& \frac{1}{\pi}\sum_{n\neq 0}
\mathrm{sin}(n\sigma^{\prime})\mathrm{sin}(n\sigma)|_{\sigma = 0,
\pi}=0\,, \label{5c47}
\end{eqnarray}
where the form of the periodic delta function has been used. Not
only that, as a bonus, we reproduce the modified form of
(\ref{5c8}). Observe the occurrence of $\delta_{P}(\sigma +
\sigma^{\prime})$ rather than $\delta_{P}(\sigma -
\sigma^{\prime})$ in the mixed bracket $\{\psi_+ , \psi_-\}$,
which plays a crucial role in obtaining the following involutive
algebra in the fermionic sector\footnote{Since there is no
translational symmetry, the presence of $\delta_{P}(\sigma +
\sigma^{\prime})$does not give rise to any inconsistency here.}.
Indeed, using (\ref{5c45}), one can show that
\begin{eqnarray}
\{\lambda_1(\sigma) , \lambda_1(\sigma^{\prime})\} &=& 4 \left(
\lambda_2(\sigma)\partial_{\sigma}\Delta_{+}
\left(\sigma , \sigma^{\prime}\right) + \lambda_2(\sigma^{\prime})
\partial_{\sigma}\Delta_{-}\left(\sigma , \sigma^{\prime}\right)\right)\,,
\nonumber \\
\{\lambda_2(\sigma) , \lambda_2(\sigma^{\prime})\} &=&  \left(
\lambda_2(\sigma^{\prime})\partial_{\sigma}\Delta_{+}
\left(\sigma , \sigma^{\prime}\right) + \lambda_2(\sigma)
\partial_{\sigma}\Delta_{-}\left(\sigma , \sigma^{\prime}\right)\right) \,,
\nonumber \\
\{\lambda_2(\sigma) , \lambda_1(\sigma^{\prime})\} &=&  \left(
\lambda_1(\sigma) + \lambda_1(\sigma^{\prime})\right)
\partial_{\sigma}\Delta_{+}\left(\sigma , \sigma^{\prime}\right)
\label{5c49}
\end{eqnarray}
hold for the fermionic sector.

\noindent Remarkably the above constraint algebra is exactly
similar to the constraint algebra of the bosonic sector
(\ref{modbr}). This helps us to write down the complete algebra of
the super Virasoro constraints $\chi_{1}(\sigma)$ and
$\chi_{2}(\sigma)$:
\begin{eqnarray}
\{\chi_1(\sigma) , \chi_1(\sigma^{\prime})\} &=& 4 \left(
\chi_2(\sigma)\partial_{\sigma}\Delta_{+}
\left(\sigma , \sigma^{\prime}\right) + \chi_2(\sigma^{\prime})
\partial_{\sigma}\Delta_{-}\left(\sigma , \sigma^{\prime}\right)\right)\,,
\nonumber \\
\{\chi_2(\sigma) , \chi_2(\sigma^{\prime})\} &=&  \left(
\chi_2(\sigma^{\prime})\partial_{\sigma}\Delta_{+}
\left(\sigma , \sigma^{\prime}\right) + \chi_2(\sigma)
\partial_{\sigma}\Delta_{-}\left(\sigma , \sigma^{\prime}\right)\right) \,,
\nonumber \\
\{\chi_2(\sigma) , \chi_1(\sigma^{\prime})\} &=&  \left(
\chi_1(\sigma) + \chi_1(\sigma^{\prime})\right)
\partial_{\sigma}\Delta_{+}\left(\sigma , \sigma^{\prime}\right)
\label{5c49q}
\end{eqnarray}

\noindent The algebra between the constraints (\ref{5c280}) now
gets modified to
\begin{eqnarray}
\{\tilde{J}_{1}(\sigma) , \tilde{J}_{1}(\sigma^{\prime})\}
&=&-i(\chi_{1}(\sigma)-2\chi_{2}(\sigma))
\delta_P(\sigma-\sigma^{\prime})\,,\nonumber\\
\{\tilde{J}_{2}(\sigma) , \tilde{J}_{2}(\sigma^{\prime})\}
&=&-i(\chi_{1}(\sigma)+2\chi_{2}(\sigma))
\delta_P(\sigma-\sigma^{\prime})\,,\nonumber\\
\{\tilde{J}_{1}(\sigma) , \tilde{J}_{2}(\sigma^{\prime})\}&=&
-i(\chi_{1}(\sigma)-2\chi_{2}(\sigma))\delta_P(\sigma +
\sigma^{\prime})\,. \label{5c82a}
\end{eqnarray}
The algebra between $\tilde J(\sigma)$ and $\chi(\sigma)$ can now be computed
 by using the modified bracket (\ref{5c00}) to get,
\begin{eqnarray}
\{\chi_{1}(\sigma) , \tilde{J}_{1}(\sigma^{\prime})\}
&=& - \left(2 \tilde{J}_{1}(\sigma) + \tilde{J}_{1}(\sigma^{\prime})\right)
\partial_{\sigma}\delta_P(\sigma-\sigma^{\prime}) +
\left(2 \tilde{J}_{2}(\sigma) + \tilde{J}_{1}(\sigma^{\prime})\right)
\partial_{\sigma}\delta_P(\sigma + \sigma^{\prime})\,,
\nonumber\\
\{\chi_{1}(\sigma) , \tilde{J}_{2}(\sigma^{\prime})\}
&=& \left(2 \tilde{J}_{2}(\sigma) + \tilde{J}_{2}(\sigma^{\prime})\right)
\partial_{\sigma}\delta_P(\sigma-\sigma^{\prime}) -
\left(2 \tilde{J}_{1}(\sigma) + \tilde{J}_{2}(\sigma^{\prime})\right)
\partial_{\sigma}\delta_P(\sigma + \sigma^{\prime})\,,
\nonumber\\
\{\chi_{2}(\sigma) , \tilde{J}_{1}(\sigma^{\prime})\}
&=& \left(\tilde{J}_{1}(\sigma) + \frac{1}{2}\tilde{J}_{1}(\sigma^{\prime})\right)
\partial_{\sigma}\delta_P(\sigma-\sigma^{\prime}) +
\left(\tilde{J}_{2}(\sigma) + \frac{1}{2}\tilde{J}_{1}(\sigma^{\prime})\right)
\partial_{\sigma}\delta_P(\sigma + \sigma^{\prime})\,,
\nonumber\\
\{\chi_{2}(\sigma) , \tilde{J}_{2}(\sigma^{\prime})\}
&=& \left(\tilde{J}_{2}(\sigma) + \frac{1}{2}\tilde{J}_{2}(\sigma^{\prime})\right)
\partial_{\sigma}\delta_P(\sigma-\sigma^{\prime}) +
\left(\tilde{J}_{1}(\sigma) + \frac{1}{2}\tilde{J}_{2}(\sigma^{\prime})\right)
\partial_{\sigma}\delta_P(\sigma + \sigma^{\prime})\,,
\label{5c82b}
\end{eqnarray}
which clearly displays a new structure for the super-Virasoro algebra.

As a matter of consistency, we write down the hamiltonian of the
superstring and then study the time evolution of the
$\psi_{\pm}$ modes. This follows easily from the Virasoro functional
$L[f]$ (\ref{5c33}) by setting $f(\sigma)=e^{im\sigma}$, which gives
\begin{eqnarray}
L_{m} = \frac{1}{4}\int_{-\pi}^{\pi} d\sigma e^{-im\sigma}[\{\Pi(\sigma)+
\partial_{\sigma}X(\sigma)\}^2+2i\psi^{\mu}_{+}\partial_{\sigma}\,.
\psi_{\mu+}] \label{5c51}
\end{eqnarray}
Setting $m=0$, gives the hamiltonian
\begin{eqnarray}
H=L_{0} &=& \frac{1}{4}\int_{-\pi}^{\pi} d\sigma [\{\Pi(\sigma)+
\partial_{\sigma}X(\sigma)\}^2+2i\psi^{\mu}_{+}\partial_{\sigma}
\psi_{\mu+}]\nonumber\\
&=& \frac{1}{2}\int_{0}^{\pi} d\sigma[\Pi^2(\sigma)+
\partial_{\sigma}X(\sigma)^2+i(\psi^{\mu}_{+}(\sigma)\partial_{\sigma}
\psi_{\mu+}(\sigma)-\psi^{\mu}_{-}(\sigma)\partial_{\sigma}\psi_{\mu-}
(\sigma))]\,. \label{5c52}
\end{eqnarray}
This immediately leads to
\begin{eqnarray}
\dot{\psi_{-}}(\sigma)=\{\psi_{-}(\sigma), H\}
=-\partial_{\sigma}\psi_{-}(\sigma)\quad;\quad
\dot{\psi_{+}}(\sigma)=\{\psi_{+}(\sigma), H\}
=\partial_{\sigma}\psi_{+}(\sigma)\,, \label{5c52a}
\end{eqnarray}
which are precisely the equations of motion for the fermionic
fields. One can therefore regard (\ref{5c00}) and (\ref{5c46}) as
the final symplectic structure of the free superstring theory.

\section{The interacting theory :}
The action for a super string moving in the presence of
a constant background Neveu-Schwarz two form field
${\cal F}_{\mu \nu}$ is given by,
\begin{eqnarray}
\label{5c2} S &=& -{ 1 \over 2} \int_{\Sigma} d^2\sigma \Big(
\eta_{\mu \nu } \partial_a X^\mu \partial^a X^\nu
\,+ \, \epsilon^{ab}{\cal F}_{\mu \nu} \partial_a X^\mu \partial_b X^\nu  \nonumber\\
&-& i {\overline \psi}^\mu \rho^a \partial_ a \psi_\mu
+ i {\cal F}_{\mu\nu} {\overline \psi}^\mu \rho_b \epsilon^{ab}  \partial_ a \psi^\nu
 \Big)\,.
\label{5c53}
\end{eqnarray}
The bosonic and fermionic sectors decouple. We consider just the
fermionic sector since the bosonic sector has already been
discussed in chapter 2. In component the fermionic sector reads
\begin{equation}
S_F \,=\, {i \over 2} \int_{\Sigma} d\tau d\sigma \Big(
\psi^\mu_{-}  \partial_+\, \psi_{-\,\mu }\,+\, \psi^\mu_{+}
\partial_- \, \psi_{+\,\mu}\, \,-\, {\cal F}_{\mu\nu} \psi^\mu_{-}
\partial_+ \, \psi^\nu_{-} + {\cal F}_{\mu\nu} \psi^\mu_{+}
\partial_- \, \psi^\nu_{+} \,\Big)\,. \label{5c54}
\end{equation}
The minimum action principle $\delta S = 0$ leads to a volume term
that vanishes when the equations of motion hold, and also to a surface term
\begin{equation}
\label{5ca3} \Big( \psi^\mu_{-} (\eta_{\mu\nu} - {\cal
F}_{\mu\nu}) \delta \psi^\nu_{-}\,-\, \psi^\mu_{+} ( \eta_{\mu\nu}
+  {\cal F}_{\mu\nu}) \delta \psi^\nu_{+}
\Big)\vert_{0}^\pi\,\,=\,\,0\,\,.
\end{equation}
It is not possible to find non trivial boundary conditions
involving $\psi^{\mu}_{-}$ and $\psi^{\mu}_{+}$ that makes the above surface
term vanish.  However, the addition of a boundary term
\cite{haggi}, \cite{NL}
\begin{eqnarray}
S_{bound}=\frac{i}{2\pi\alpha^{\prime}}\int_{\Sigma}d\tau d\sigma
 \left(\cal F_{\mu\nu}\psi^{\mu}_{+}\partial_{-}\psi^{\nu}_{+}\right)
\label{5c550}
\end{eqnarray}
makes it possible to find a solution to the boundary condition.
Addition of this term to $S_{F}$ leads to the total action:
\begin{eqnarray}
S=\frac{-i}{4\pi\alpha^{\prime}}\int_{\Sigma}d\tau d\sigma
 \left( \psi^{\mu}_{-}E_{\nu\mu}\partial_{+}\psi^{\nu}_{-}+
\psi^{\mu}_{+}E_{\nu\mu}\partial_{-}\psi^{\nu}_{+}\right)\,,
\label{5c551}
\end{eqnarray}
where $E^{\mu\nu} \,=\, \eta^{\mu\nu} \,
+ {\cal F}^{\mu\nu}$.
The corresponding boundary term coming from $\delta S=0$ is given by
\begin{eqnarray}
\left(\psi^{\mu}_{-}E_{\nu\mu}\delta\psi^{\nu}_{-}-
\psi^{\mu}_{+}E_{\nu\mu}\delta\psi^{\nu}_{+}\right)|_{0}^{\pi}=0.
\label{5c552}
\end{eqnarray}

The above condition is satisfied by the following conditions that
preserve supersymmetry \cite{pi} at the string endpoints
 $\sigma=0$ and $\sigma=\pi$:
\begin{eqnarray}
 E_{\nu\mu}\,  \psi^\nu_{+} (0,\tau)\,&=&\,
 E_{\mu\nu} \, \psi^\nu_{-} (0,\tau) \,,\nonumber \\
 E_{\nu\mu}\,  \psi^\nu_{+} (\pi,\tau ) \,&=&\,\lambda
 E_{\mu\nu} \, \psi^\nu_{-} (\pi, \tau )\,\,,
\label{5c55}
\end{eqnarray}
 where $\lambda = \pm 1 \,$ with
the plus sign corresponding to Ramond
boundary condition and the minus corresponding to the Neveu-Schwarz case.
Here too we work with Ramond boundary conditions.
Now the BCs are recast as
\begin{eqnarray}
 \left(E_{\nu\mu}\,  \psi^\nu_{(+)} (\sigma,\tau)\,
-\, E_{\mu\nu} \, \psi^\nu_{(-)} (\sigma,\tau)\right)
\vert_{\sigma = 0, \pi} \, = \, 0\,. \label{5c56}
\end{eqnarray}
This nontrivial BC leads to a modification in the original (naive)
(\ref{5c5}) DBs. The \\  $\{\psi^\mu_{(+)} (\sigma,\tau) ,
\psi^\nu_{(+)} (\sigma^{\prime},\tau)\}_{DB}$  is the same as that
of the free string (\ref{5c5}). We therefore make an ansatz
\begin{eqnarray}
 \{\psi^\mu_{+} (\sigma,\tau) ,
\psi^\nu_{-} (\sigma^{\prime},\tau)\}_{DB}\, = \,
C^{\mu\nu}\delta_{P}\left(\sigma + \sigma^{\prime}\right)\,.
\label{5c57}
\end{eqnarray}

Taking brackets between the BCs (\ref{5c56}) and $\psi^\gamma_{-}
(\sigma^{\prime})$  we get
\begin{eqnarray}
 E_{\nu\mu}\,  C^{\nu \gamma}
\, = \, -i \, E_{\mu\gamma}. \label{5c58}
\end{eqnarray}
Solving this, we find
\begin{eqnarray}
 C^{\mu \nu}\, = \, -i \,
\left[\left(1 - {\cal F}^2\right)^{-1}\right]^{\mu \rho}\,
 E_{\rho \gamma}\, E^{\gamma \nu}.
\label{5c58a}
\end{eqnarray}
One can also take brackets between the BCs (\ref{5c56}) and
$\psi^\gamma_{+} (\sigma^{\prime})$, which yields
\begin{eqnarray}
 C^{\nu \mu}\, = \, -i \,
\left[\left(1 - {\cal F}^2\right)^{-1}\right]^{\mu \rho}\,
 E_{\gamma \rho}\, E^{\nu \gamma}.
\label{5c58b}
\end{eqnarray}
Although the expressions (\ref{5c58a}) and (\ref{5c58b}) look
different, they are actually the same as one can see easily by
taking transpose of (\ref{5c58b}) and using the fact that, for any
matrix $M$ we have the commuting property for the product :
 $f(M) g(M) = g(M) f(M)$, holding for any two polynomials
$f$ and $g$ of $M$. In other words, $f$ and $g$ can be regarded as
functions which map matrices to matrices of same dimension and are
constructed out of the same matrix $M$. Finally we can write the matrix
$C = \{C^{\mu \nu}\}$ more compactly as
\begin{eqnarray}
 C \, = \, -i \,
\left[\left(1 - {\cal F}^2\right)^{-1}\left(1 + {\cal
F}\right)^{2}\right]. \label{5c58y}
\end{eqnarray}

We therefore get the following modification:
\begin{eqnarray}
\{\psi^\mu_{+} (\sigma,\tau) , \psi^\nu_{-}
(\sigma^{\prime},\tau)\}_{DB}\, = \, -i\, \left[\left(1 - {\cal
F}^2\right)^{-1}\right]^{\mu \rho} \,  E_{\rho \gamma}\, E^{\gamma
\nu} \delta_{P}\left(\sigma + \sigma^{\prime}\right)\,,
\label{5c60}
\end{eqnarray}
which also reduces to those of \cite{godinho}, upto the
$\delta_{P}\left(\sigma + \sigma^{\prime}\right)$ factor. Finally,
note that in the limit ${\cal F}_{\mu \nu} \rightarrow 0$
(\ref{5c60}), the last of  (\ref{5c46}) is reproduced.

\section{Summary}
In this chapter we have extended the methodology of chapter 2 and
chapter 3 to an open fermionic string propagating freely and one
moving in a constant antisymmetric background field. Here also we
have observed that boundary conditions are incompatible with the
basic brackets. So the modification of the cannonical bracket was
necessary. Eventually we have constructed the appropriate delta
function for the physical interval $[0, \pi]$ of the string and in
the process we have obtained the non(anti)commutative structure of
the super string. Finally the above non(anti)commutative structure
led to new results in the algebra of superconstraints which still
remain involutive, indicating the internal consistency of our
analysis.

\chapter{ Normal ordering and non(anti)commutativity
in open super strings}
In this chapter we study non(anti)commutativity in an open
super string moving in the presence of a background antisymmetric tensor
field $\mathcal{B}_{\mu \nu}$ in a conformal
field theoretic approach.

\noindent In chapter 4, noncommutativity in an open bosonic string
moving in the presence of a background Neveu-Schwarz two-form
field $\mathcal{B}_{\mu \nu}$ is investigated in a conformal field
theory approach. The mode algebra is first obtained using the
newly proposed normal ordering, which satisfies both equations of
motion and BC(s). Using these the commutator among the string
coordinates is obtained. Interestingly, this new normal ordering
yields the same algebra between the modes as the one satisfying
only  the equations of motion. In this approach, we find that
noncommutativity originates more transparently and our results
match with the existing results in the literature.
In this chapter, we extend the same methodology  to analyse an
open super string propagating freely and one moving in a constant
antisymmetric background field. To start with we discuss the
recent results involving new normal ordered products (of fermionic
operators) in \cite{6cbr}. Then we study the symplectic structure
of the fermionic sector of both free and interacting super string.
The computational details of some of the key results in the
chapter are given in the appendix B.

\section{New Normal ordering for fermionic string coordinates}
The action for a super string moving in the presence of
a constant background antisymmetric tensor field
${\cal B}_{\mu \nu}$ is given by:
\begin{eqnarray}
S &=&  \frac{- 1}{4 \pi \alpha^\prime} \int_{\Sigma} d\tau
 d{\sigma}\,\Big[ \, \partial_a X^\mu \partial^a X_\mu
\,+\, \epsilon^{ab} B_{\mu\nu} \partial_a X^\mu \partial_b X^\nu
\nonumber\\
 & & + i \psi_{\mu (-)} E^{\nu \mu} \partial_{+} \psi_{\nu (-)} +
i \psi_{\mu (+)} E^{\nu \mu} \partial_{-} \psi_{\nu (+)} \,\Big]
\label{6c1}
\end{eqnarray}
\noindent where, $\partial_{+} =  \partial_{\tau} + \partial_{\sigma},\; \;
\partial_{-} =  \partial_{\tau} - \partial_{\sigma} $
and $ E^{\mu\nu} \,=\, \eta^{\mu\nu} \, + {\cal B}^{\mu\nu}\,$.

\noindent Now since the bosonic and fermionic sectors decouple, we
can consider the fermionic sector seperately\footnote{The bosonic
sector was already discussed in chapter 4.}. The variation of the
fermionic part of the action (\ref{6c1}) gives the classical
equations of motion:
\begin{equation}
\partial_{+} \psi_{\nu (-)} \,=\, 0\quad,\quad
\partial_{-} \psi_{\nu (+)} \,=\, 0
\label{6c2}
\end{equation}
and a boundary term that yields the following BCs:
\begin{eqnarray}
 E_{\nu\mu}\,  \psi^\nu_{(+)} (0,\tau)\, &=&\,
  E_{\mu\nu} \, \psi^\nu_{(-)} (0,\tau)\, \nonumber\\
\label{6c4}
E_{\nu\mu}\,  \psi^\nu_{(+)} (\pi,\tau ) \, &=& \,\lambda
E_{\mu\nu} \, \psi^\nu_{(-)} (\pi, \tau )
\label{6c3}
\end{eqnarray}
\noindent at the endpoints $\sigma \,=\,0$ and $\sigma = \pi\,$ of the string,
where $\lambda = \pm 1 \,$ corresponds to Ramond and Neveu-Schwarz BC(s)
respectively.

It is convenient now to change to complex world-sheet coordinates
and therefore we first make a Wick rotation by defining $\sigma^2
= i\tau$. Then we introduce the complex world sheet coordinates
\cite{pol}: $ z \,=\,  \sigma^1 +i \sigma^2 \, ; \,  \bar{z} \,=\,
\sigma^1 - i \sigma^2$ and $\partial_z = \frac{1}{2}(\partial_1 -
i\partial_2), \, \partial_{\bar{z}} = \frac{1}{2}(\partial_1 +
i\partial_2)$. In this notation the fermionic part of the action
(\ref{6c1}) reads:
\begin{eqnarray}
S_{F} =  \frac{-i}{4 \pi \alpha^\prime} \int_{\Sigma} dz
d \bar{z}[\psi_{\mu (-)} E^{\nu \mu} \partial_{\bar{z}} \psi_{\nu (-)} +
\psi_{\mu (+)} E^{\nu \mu} \partial_{z} \psi_{\nu (+)}]\,
\label{6c5}
\end{eqnarray}
while the classical equations of motion (\ref{6c2}) and the Ramond
BCs (\ref{6c3}) take the form:
\begin{eqnarray}
\label{6c6}
\partial_{\bar{z}} \psi_{\nu (-)} \,=\, 0\,\,\,, \,\,
\partial_{z} \psi_{\nu (+)} \,=\, 0 \\
\left( E_{\nu\mu}\,  \psi^\nu_{(+)} (z,\bar{z})\, -\,
  E_{\mu\nu} \, \psi^\nu_{(-)} (z,\bar{z})\right)\vert_{z\,=\,
-\bar{z}\, , \, 2\pi - \bar{z}}\, =\, 0 \, .
\label{6c7}
\end{eqnarray}
We now study the properties of quantum operators corresponding to
the classical variables by considering the expectation values
\cite{pol}. Using the fact that the path integral of a total
functional derivative vanishes and considering the insertion of
one fermionic operator one finds:
\begin{equation}
\label{6c8}
\int [ d  \psi ] \left[  \frac{\delta}{\delta \psi^{\mu} _{(a) }(z,\bar{z})}
[e^{-S_{F}} \psi^{\nu}_{(b)} (z', \bar{z}')]
  \right] \,=\,0
\end{equation}
\noindent where, $a,b \,=\,+,-\,$.
Considering first the case of
$\psi^{\nu}_{(b)} (z', \bar{z}')\,$ inside the world-sheet
and not at the boundary,
this equation yields the following expectation values:
\begin{eqnarray}
\label{6ceqmov}
\langle \partial_{z} \psi^{\mu}_{(+)} (z,\bar{z}) \psi^{\nu}_{(+)}
(z',\bar{z}') \rangle &=& 2\,\pi \,i \, \alpha'\, \langle \eta^{\mu \nu}
\delta^2(z-z',\bar{z}-\bar{z}') \rangle \nonumber \\
\langle \partial_{\bar{z}} \psi^{\mu}_{(-)} (z,\bar{z})
\psi^{\nu}_{(-)}(z',\bar{z}') \rangle &=& 2\,\pi \,i \, \alpha '\, \langle
 \eta^{\mu\nu}  \delta^2(z-z',\bar{z}-\bar{z}') \rangle \nonumber \\
\langle \partial_{\bar{z}} \psi^{\mu}_{(-)} (z,\bar{z})
\psi^{\nu}_{(+)}(z',\bar{z}') \rangle &=&  \langle
 \partial_{z} \psi^{\mu}_{(+)} (z,\bar{z})
\psi^{\nu}_{(-)}(z',\bar{z}') \rangle =  0\,\,.
\end{eqnarray}
Using these results one finds the appropriate way to define normal
ordered products that satisfy the equations of motion for
fermionic operators that are not at the world-sheet boundary
\cite{6cbr, pol1}:
\begin{eqnarray}
\label{6c9}
: \psi^{\mu}_{(+)}(z,\bar{z})\,\,\, \psi^{\nu}_{(+)}(z',\bar{z}')  : &=&
  \psi^{\mu}_{(+)}(z,\bar{z})\,\, \,\psi^{\nu}_{(+)}(z',\bar{z}')
\,-\, \frac{i\, \alpha'\,}{{\bar z} - {\bar z}'}\,\eta^{\mu \nu}
\nonumber\\
: \psi^{\mu}_{(-)}(z,\bar{z})\,\,\, \psi^{\nu}_{(-)}(z',\bar{z}')  : &=&
  \psi^{\mu}_{(-)}(z,\bar{z})\,\,\, \psi^{\nu}_{(-)}(z',\bar{z}')
\,-\, \frac{i\, \alpha'\,}{ z -  z'}\,\eta^{\mu \nu}
\nonumber\\
: \psi^{\mu}_{(+)}(z,\bar{z})\,\,\, \psi^{\nu}_{(-)}(z',\bar{z}')  : &=& 0
\nonumber\\
: \psi^{\mu}_{(-)}(z,\bar{z})\,\,\, \psi^{\nu}_{(+)}(z',\bar{z}')  : &=& 0\,.
\end{eqnarray}
The above products satisfy the equations of motion (\ref{6c6})
at the quantum level,
but fails to satisfy the BC(s) (\ref{6c7}).

\noindent At this point it is more convenient to choose world sheet
coordinates, related to these $z$ coordinates by conformal
transformation, that simplify the representation of the boundary,
\begin{eqnarray}
\omega\, = \,\mathrm{exp}\left(-iz\right)\, =\, e^{- i \sigma^1 +
\sigma^2} \,\,\,;\,\,\,{\bar \omega} = e^{i \sigma^1 + \sigma^2}.
\label{6c10}
\end{eqnarray}
Besides replacing
$\mathrm{exp}(-iz) \to \omega$, we must transform the fields \cite{pol1},
\begin{eqnarray}
\label{6c13}
\psi^{\mu}_{\omega^{\frac{1}{2}}}(\omega) \,=\, \left(\partial_\omega z\right)^{\frac{1}{2}}
\psi^{\mu}_{z^{\frac{1}{2}}}(z)
= \, i^{\frac{1}{2}} \omega^{-\frac{1}{2}}\, \psi^{\mu}_{z^{\frac{1}{2}}}(z).
\end{eqnarray}
The subscripts are a reminder that these transform with half the weight
of a vector.
In this present coordinates the complete boundary corresponds
just to the region $\omega = {\bar \omega}$.
Further, the action (\ref{6c5})  along with equations of motion (\ref{6c6})
in terms of $\omega, {\bar \omega}$ has still the same form, while the form of BC(s)
(\ref{6c7}) change to the following:
\begin{eqnarray}
\left( E_{\nu\mu}\,  \psi^\nu_{(+)} (\omega,\bar{\omega})\, + i\,
  E_{\mu\nu} \, \psi^\nu_{(-)} (\omega,\bar{\omega})\right)\vert_{\omega\,
=\,\bar{\omega}}\, =\, 0 \, .
\label{6c28}
\end{eqnarray}

\noindent Let us now consider the case of an insertion of a
fermionic string coordinate $\psi^{\nu}_{(\pm )} (\omega')\,$
located at the world-sheet boundary. Note that since $\omega' =
\bar{\omega}'$ at the boundary, the fermionic coordinate insertion
at the boundary depends only on $\omega'$. Working out equation
(\ref{6c8}), but now subject to constraint (\ref{6c28}) (with
$\omega$ replaced by $\omega'$ in (\ref{6c28})), we
find\footnote{Note that the fields $\psi$'s with unprimed
arguements are not located at the boundary.} (see appendix B for
the computational details):
\begin{eqnarray}
\label{6ceqmov2}
\langle \partial_{\omega} \psi^{\mu}_{(+)} (\omega,\bar{\omega}) \psi^{\nu}_{(+)}
(\omega') \rangle &=& \, 2\pi i\alpha^{\prime}\langle \eta^{\mu \nu}
\delta^2(\omega-\omega',\bar{\omega}-\omega') \rangle \nonumber \\
\langle \partial_{\omega} \psi^{\mu}_{(-)} (\omega,\bar{\omega}) \psi^{\nu}_{(-)}
(\omega') \rangle &=& \, 2\pi i\alpha^{\prime}\langle \eta^{\mu \nu}
\delta^2(\omega-\omega',\bar{\omega} - \omega') \rangle \nonumber \\
\langle \partial_{\omega} \psi^{\mu}_{(+)} (\omega,\bar{\omega}) \psi^{\nu}_{(-)}
(\omega') \rangle &=& \, -2\pi \alpha^{\prime}
\langle\Big[ ( \eta+{\cal B})^{^{-1}}\,
( \eta - {\cal B}) \Big]^{\nu \mu}
\delta^2(\omega-\omega',\bar{\omega}-\omega') \rangle \nonumber \\
\langle \partial_{\omega} \psi^{\mu}_{(-)} (\omega,\bar{\omega}) \psi^{\nu}_{(+)}
(\omega') \rangle &=& \, 2\pi \alpha^{\prime}\langle
\Big[ ( \eta+{\cal B})^{^{-1}}\,
( \eta - {\cal B}) \Big]^{\nu \mu}
\delta^2(\omega-\omega',\bar{\omega}-\omega') \rangle \,\,.
\end{eqnarray}
So the appropriate normal ordering for fermionic string
coordinates at the boundary reads:
\begin{eqnarray}
:\psi^{\mu}_{(+)} (\omega,\bar{\omega}) \psi^{\nu}_{(+)}
(\omega') : &=&  \psi^{\mu}_{(+)} (\omega,\bar{\omega}) \psi^{\nu}_{(+)}
(\omega')  \,-\, \frac{i\alpha^{\prime}}{\left(\bar{\omega} - \omega'
\right)} \eta^{\mu\nu}
\nonumber\\
: \psi^{\mu}_{(-)} (\omega,\bar{\omega}) \psi^{\nu}_{(-)}
(\omega') : &=& \psi^{\mu}_{(-)} (\omega,\bar{\omega}) \psi^{\nu}_{(-)}
(\omega')  \,-\, \frac{i\alpha^{\prime}}{\left(\omega - \omega'\right)}
\eta^{\mu\nu}
\nonumber\\
:  \psi^{\mu}_{(+)} (\omega,\bar{\omega}) \psi^{\nu}_{(-)}
(\omega')  : &=&   \psi^{\mu}_{(+)} (\omega,\bar{\omega}) \psi^{\nu}_{(-)}
(\omega') \,+\,\frac{ \alpha^{\prime}
\, \Big[ ( \eta + {\cal B})^{^{-1}}\,( \eta  -
{\cal B}) \Big]^{\nu \mu}}{\left(\bar{\omega} - \omega'\right)}
\nonumber\\
:  \psi^{\mu}_{(-)} (\omega,\bar{\omega}) \psi^{\nu}_{(+)}
(\omega')  : &=&  \psi^{\mu}_{(-)} (\omega,\bar{\omega}) \psi^{\nu}_{(+)}
(\omega') \,-\,\frac{ \alpha^{\prime}
\, \Big[ ( \eta + {\cal B})^{^{-1}}\,(\eta  - {\cal B}) \Big]^{\nu
\mu} }{\left(\omega - \omega'\right)}
\label{6c11}
\end{eqnarray}
The above results of normal ordering of fermionic operators are new and
incorporates the effect of BC(s).

\noindent Now for the functional ${\cal{F}}[X]$ (representing the
combinations occurring in the left hand side of the above
equation), the new normal ordering (in absence of the
$\mathcal{B}$ field) can be compactly written as:
\begin{eqnarray}
\label{6ccompact}
:{\cal{F}}:= \exp\left(
\frac{i\alpha^{\prime}}{2}
\int d^2 \omega^{\prime\prime}
d^2 \omega^{\prime\prime\prime}\left[\frac{1}{(\omega^{\prime\prime}-
\omega^{\prime\prime\prime})}\frac{\delta}{\delta \psi^{\mu}_{(-)}
(\omega^{\prime\prime}, \bar{\omega}^{\prime\prime})}\,
\frac{\delta}{\delta\psi_{\mu(-)}
(\omega^{\prime\prime\prime}, \bar{\omega}^{\prime\prime\prime})}+\left(\omega\leftrightarrow\bar\omega,
-\leftrightarrow +\right)\right]\right)\cal{F}\nonumber\\
\end{eqnarray}
Note that the fields $\psi$'s with double prime and triple
prime arguements in (\ref{6ccompact}) are not located at the boundary.

\noindent
We shall see now that normal ordered products are important to
compute the central charge which gives us the critical dimension.
\noindent The energy-momentum tensor
(in the absence of the $\mathcal{B}$ field) for the fermionic sector
for points inside the world-sheet (in the $z$-frame) is given by:
\begin{eqnarray}
\label{6cenergy}
T^{zz}=-\frac{1}{2}\psi_{\mu(+)}\partial_{\bar{z}}\psi^{\mu}_{(+)}
\equiv\bar{T}\nonumber\\
T^{\bar{z}\bar{z}}=-\frac{1}{2}\psi_{\mu(-)}\partial_{z}\psi^{\mu}_{(-)}
\equiv T
\end{eqnarray}
while at the boundary, the BC(s) (\ref{6c7}) (with
${\mathcal{B}}=0$) relating $\psi_{\nu(-)}$ to $\psi_{\nu(+)}$
lead to :
\begin{eqnarray}
\label{6cenergy1}
\bar{T}=-\frac{1}{2}\psi_{\mu(+)}\partial_{\bar{z}}\psi^{\mu}_{(+)}=-T
\end{eqnarray}
where we have used $\partial_{\bar{z}}=-\partial_{z}$ (since
$dz=-d\bar{z}$ at the boundary).
The central charge can now be computed from the most singular term
in the normal ordered product of energy-momentum tensor.
This involves two contractions of the fermionic coordinate operator
products and is proportional to \cite{6cbr}:
\begin{eqnarray}
\label{6cenergy2}
\int dz^{\prime}...dz^{\prime\prime\prime\prime}\frac{1}{2}
\left[\frac{i\alpha^{\prime}}{(z^{\prime}-z^{\prime\prime})}
\frac{\delta^{2}}{\delta\psi_{\mu(-)}(z^{\prime})
\delta\psi^{\mu}_{(-)}(z^{\prime\prime})}\right]
\left[\frac{i\alpha^{\prime}}{(z^{\prime\prime\prime}-
z^{\prime\prime\prime\prime})}
\frac{\delta^{2}}{\delta\psi_{\mu(-)}(z^{\prime\prime\prime})
\delta\psi^{\mu}_{(-)}(z^{\prime\prime\prime\prime})}\right] \nonumber \\
\times \left[T(z_{1})T(z_{2})\right]\nonumber \\
\sim \frac{D\alpha^{\prime 2}}{4(z_{1}- z_{2})^{4}}\quad \quad \quad
\quad \quad \quad \quad \quad \quad \quad \quad \quad \quad \quad \quad \quad
\quad \quad \quad \quad \quad \quad \quad \quad \quad \quad
\quad \quad \quad \quad
\end{eqnarray}
where $\sim$ mean ``equal up to nonsingular terms"\footnote{The
other less singular terms are not given explicitly.}. The above
computation gives the well known result $D/2$ as the central
charge where $D$ is the dimension of space-time \cite{kaku},
\cite{pol1}. The results are also in conformity with \cite{6cbr}.

\noindent We shall make use of the results discussed here in the next section
where we study both free and interacting open super strings.
\section{Mode expansions and Non(anti)Commutativity for super strings}
\subsection{Free open strings}
In this section, we consider the mode expansions of free (${\cal
B}_{\mu \nu} = 0$) open super strings. We first expand
$\psi^{\mu}_{(-)}(z)$ and $\psi^{\mu}_{(+)}(\bar{z})$ in Fourier
modes in ($z, \bar{z}$) coordinates \cite{pol1}:
\begin{eqnarray}
\label{6c12}
\psi^{\mu}_{(-)}(z) = \frac{1}{\sqrt{2\pi}}\sum_{m \in {\bf{Z}}}
d^{\mu}_{m}\,\mathrm{exp}(imz)\ \ \ \ ;\ \ \ \
\psi^{\mu}_{(+)}(\bar{z}) = \frac{1}{\sqrt{2\pi}} \sum_{m \in {\bf{Z}}}
\tilde{d}^{\mu}_{m}\,\mathrm{exp}(-im\bar{z}).
\end{eqnarray}
Let us also write these as Laurent expansions in ($\omega, \bar{\omega}$) coordinates:
\begin{eqnarray}
\label{6c14}
\psi^{\mu}_{(-)}(\omega) = \frac{i^{\frac{1}{2}}}{\sqrt{2\pi}}
\,\sum_{m \in {\bf{Z}}}
\frac{d^{\mu}_{m}}{\omega^{m + \frac{1}{2}}}
\ \ \ \ ;\ \ \ \
\psi^{\mu}_{(+)}(\bar{\omega}) = \frac{i^{-\frac{1}{2}}}{\sqrt{2\pi}}
\,\sum_{m \in {\bf{Z}}}
\frac{\tilde{d}^{\mu}_{m}}{\bar{\omega}^{m + \frac{1}{2}}}.
\end{eqnarray}
Now the BC(s) (\ref{6c28}) in case of free open super strings
(${\mathcal{B}}_{\mu\nu} = 0$) requires $d = \tilde{d}$ in the
expansions (\ref{6c14}). The expressions (\ref{6c14}) can be
equivalently written as:
\begin{eqnarray}
\label{6c15}
d^{\mu}_{m} &=& \frac{\sqrt{2\pi}}{\sqrt{i}}\oint \frac{d\omega}{2\pi i}\,
\omega^{m - \frac{1}{2}}\, \psi^{\mu}_{(-)}(\omega)
 = \,-\sqrt{2\pi}\sqrt{i}\oint \frac{d\bar{\omega}}{2\pi i}\,
\bar{\omega}^{m - \frac{1}{2}}\, \psi^{\mu}_{(+)}(\bar{\omega}).
\end{eqnarray}
The anticommutation relation between $d$'s can be worked out from
the contour argument \cite{pol} and the operator product expansion
(OPE) (\ref{6c11}) (with ${\mathcal{B}}_{\mu\nu}=0$):
\begin{eqnarray}
\label{6c16}
\left\{d^{\mu}_{m}, d^{\nu}_{n}\right\} &=& \frac{1}{i}
\oint \frac{d\omega_2}{2\pi i} \mathrm{Res}_{\omega_1 \to \omega_2}\left(
\omega^{m - \frac{1}{2}}_1\,  \psi^{\mu}_{(-)}(\omega_1)\,
\omega^{n - \frac{1}{2}}_2\,  \psi^{\nu}_{(-)}(\omega_2)\right) \nonumber \\
&=& 2\pi\alpha^{\prime}\,\eta^{\mu \nu}\, \delta_{m+n,0}
=\eta^{\mu \nu}\, \delta_{m+n,0}
\end{eqnarray}
where we have set $2\pi\alpha^{\prime}=1$.
The anti-commutation relations between $\psi^{\mu}_{(-)}(\omega, \bar{\omega})$ and
$\psi^{\nu}_{(+)}(\omega^{\prime}, \bar{\omega^{\prime}})$ are then obtained by using
(\ref{6c16}):
\begin{eqnarray}
\left\{\psi^{\mu}_{(-)}(\omega, \bar{\omega}),
\psi^{\nu}_{(-)}(\omega^{\prime},
\bar{\omega^{\prime}})\right\} &=& \frac{i\eta^{\mu \nu}}{2 \pi}
 \sum_{m \in {\bf{Z}}}\left(\omega^{-m - \frac{1}{2}}\,
 \omega^{\prime m - \frac{1}{2}} \right) \nonumber \\
\left\{\psi^{\mu}_{(+)}(\omega, \bar{\omega}),
\psi^{\nu}_{(+)}(\omega^{\prime},
\bar{\omega^{\prime}})\right\} &=& -\frac{i\eta^{\mu \nu}}{2 \pi}
 \sum_{m \in {\bf{Z}} }\left(\bar{\omega}^{-m - \frac{1}{2}}\,
 \bar{\omega}^{\prime m - \frac{1}{2}} \right)\nonumber \\
\left\{\psi^{\mu}_{(-)}(\omega, \bar{\omega}),
\psi^{\nu}_{(+)}(\omega^{\prime},
\bar{\omega^{\prime}})\right\} &=& \frac{\eta^{\mu \nu}}{2 \pi}
 \sum_{m \in {\bf{Z}}}\left(\omega^{-m - \frac{1}{2}}\,
 \bar{\omega}^{\prime m - \frac{1}{2}} \right).
\label{6c17}
\end{eqnarray}
To obtain the usual equal time ($\tau = \tau^{\prime}$)
anticommutation relation we first rewrite (\ref{6c17}) in ``$z$
frame'' using (\ref{6c10}, \ref{6c13}) and then in terms of
$\sigma^1, \ \sigma^2$ to find:
\begin{eqnarray}
\left\{\psi^{\mu}_{(-)}(\sigma^1, \sigma^2), \psi^{\nu}_{(-)}(\sigma^{1 \prime},
\sigma^{\prime 2})\right\} &=& \frac{\eta^{\mu \nu}}{2 \pi}
 \sum_{m \in {\bf{Z}}}\left[\exp\left({im(\sigma^1 + i \sigma^2
- \sigma^{\prime 1} - i \sigma^{\prime 2})} \right)\right]\nonumber \\
\left\{\psi^{\mu}_{(+)}(\sigma^1, \sigma^2), \psi^{\nu}_{(+)}(\sigma^{1 \prime},
\sigma^{\prime 2})\right\} &=& \frac{\eta^{\mu \nu}}{2 \pi}
 \sum_{m \in {\bf{Z}}}\left[\exp\left({im(\sigma^1 - i \sigma^2
- \sigma^{\prime 1} + i \sigma^{\prime 2})} \right)\right]\nonumber \\
\left\{\psi^{\mu}_{(-)}(\sigma^1, \sigma^2), \psi^{\nu}_{(+)}(\sigma^{1 \prime},
\sigma^{\prime 2})\right\} &=& \frac{\eta^{\mu \nu}}{2 \pi}
 \sum_{m \in {\bf{Z}}}\left[\exp\left({im(\sigma^1 + i \sigma^2
+ \sigma^{\prime 1} - i \sigma^{\prime 2})} \right)\right].
\label{6c18}
\end{eqnarray}
Finally substituting $\tau = \tau^{\prime}$ (i.e. $\sigma^2 = \sigma^{2\,\prime}$)
and $\sigma^1 = \sigma$ we get
back the equal time anti-commutation relations:
\begin{eqnarray}
\left\{\psi^{\mu}_{(-)}(\sigma, \tau), \psi^{\nu}_{(-)}(\sigma^{\prime},
\tau)\right\} &=&\eta^{\mu \nu}
 \delta_{P}(\sigma - \sigma^\prime)\nonumber \\
\left\{\psi^{\mu}_{(+)}(\sigma, \tau), \psi^{\nu}_{(+)}(\sigma^{\prime},
\tau)\right\} &=& \eta^{\mu \nu}
 \delta_{P}(\sigma - \sigma^\prime)\nonumber \\
\left\{\psi^{\mu}_{(-)}(\sigma, \tau), \psi^{\nu}_{(+)}(\sigma^{\prime},
\tau)\right\} &=& \eta^{\mu \nu}
 \delta_{P}(\sigma + \sigma^\prime).
\label{6c19}
\end{eqnarray}
where, $\delta_{P}(\sigma - \sigma^\prime)$ is the so called periodic delta function
which is defined as:
\begin{eqnarray}
\delta_{P}(\sigma - \sigma^\prime) = \frac{1}{2 \pi}\sum_{m \in {\bf{Z}}}
\exp\left(im(\sigma - \sigma^{\prime})\right).
\label{6c20}
\end{eqnarray}
This structure of anticommutator is completely consistent with the
BCs (\ref{6c28}) for ${\mathcal{B}}_{\mu\nu} = 0$. Note that not
only the usual Dirac delta function is replaced by the periodic
delta function but also the anticommutator among $\psi_{(-)},
\psi_{(+)}$ are non-vanishing even in case of the free open
fermionic string as we have already seen in previous chapter
\cite{jing1, agh1}. \noindent The most important feature of the
above analysis is that unlike the bosonic case, the new normal
ordering of the fermionic operators (that incorporates the BC(s))
(\ref{6c11}) leads to the nonanticommutative structures
(\ref{6c19}) among the fermionic string coordinates.

\subsection{Open superstring in the constant ${\mathcal{B}}$-field background}
We now analyse the open superstring moving in presence of a
background antisymmetric tensor field ${\cal{B}}_{\mu \nu}$. To
begin with, let us again consider the Laurent expansion of
$\psi_{(-)}^{\mu}(\omega)$ and $\psi_{(+)}^{\mu}(\bar{\omega})$
(\ref{6c14}). Now due to the BC(s) (\ref{6c28}) (with
${\mathcal{B}}_{\mu \nu} \neq 0$), the modes $d$ and $\tilde{d}$
are no longer independent but satisfy the following relation:
\begin{eqnarray}
E_{\mu \nu}\, d^{\nu}_{m}\, = \, E_{\nu \mu}\, \tilde{d}^{\nu}_{m}.
\label{6c21}
\end{eqnarray}
Hence there exists only one set of independent modes
$\alpha^{\mu}_{m}$, which can be thought of as the modes of free open
strings and is related to $d^{\mu}_{m}$ and $\tilde{d}^{\mu}_{m}$ by:
\begin{eqnarray}
\label{6c22}
d^{\mu}_{m} &=& \left(\delta^{\mu}_{\ \nu} - {\cal{B}}^{\mu}_{\ \nu}
\right)\alpha^{\nu}_{m} := \left[({1\!\mbox{l}} - {\cal{B}})\alpha
\right]^{\mu}_{m} \nonumber \\
\tilde{d}^{\mu}_{m} &=& \left(\delta^{\mu}_{\ \nu} + {\cal{B}}^{\mu}_{\ \nu}
\right)\alpha^{\nu}_{m} := \left[({1\!\mbox{l}} +
{\cal{B}})\alpha\right]^{\mu}_{m} .
\end{eqnarray}
Note that under world-sheet parity transformation (i.e.
$\sigma\leftrightarrow-\sigma$), $d^{\mu}_{m} \leftrightarrow
\tilde{d}^{\mu}_{m}$, since ${\cal{B}}_{\mu \nu}$ is a world-sheet
pseudo-scalar (similar to bosonic part). Substituting (\ref{6c22})
in (\ref{6c14}), we obtain  the following Laurent expansions for
$\psi^{\mu}_{-}$ and $\psi^{\mu}_{+}$:
\begin{eqnarray}
\label{6c23}
\psi^{\mu}_{(-)}(\omega) &=& \frac{i^{\frac{1}{2}}}{\sqrt{2\pi}}
\sum_{m \in {\bf{Z}}}
\frac{\left[({1\!\mbox{l}} -
{\cal{B}}) \alpha\right]^{\mu}_{m}}{\omega^{m + \frac{1}{2}}} \\
\psi^{\mu}_{(+)}(\bar{\omega}) &=& \frac{i^{- \frac{1}{2}}}{\sqrt{2\pi}}
\sum_{m \in {\bf{Z}}}
\frac{\left[({1\!\mbox{l}} + {\cal{B}})
\alpha\right]^{\mu}_{m}}{\bar{\omega}^{m + \frac{1}{2}}}\nonumber.
\end{eqnarray}
These are the appropriate mode expansions for the fermionic part of the
interacting superstring,
that satisfy both the equations of motion (\ref{6c6}) and the BC(s) (\ref{6c28}).

\noindent Now the expressions (\ref{6c23}) for interacting superstrings can
also be written as:
\begin{eqnarray}
\left[({1\!\mbox{l}} - {\cal{B}})\alpha\right]^{\mu}_{m} &=&
\frac{\sqrt{2\pi}}{i}\oint \frac{d\omega}{2\pi i}\,
\omega^{m - \frac{1}{2}}\, \psi^{\mu}_{(-)}(\omega) \nonumber \\
\left[({1\!\mbox{l}} + {\cal{B}})\alpha\right]^{\mu}_{m} &=&
\,\frac{\sqrt{2\pi}}{i}\oint \frac{d\bar{\omega}}{2\pi i}\,
\bar{\omega}^{m - \frac{1}{2}}\, \psi^{\mu}_{(+)}(\bar{\omega}) .
\label{6c24}
\end{eqnarray}
The anticommutation relation between $\alpha$'s can be obtained
once again from the contour argument (using (\ref{6c24})) and the
$\psi\, \psi$ OPE (\ref{6c11}):
\begin{eqnarray}
\label{6c25} \left\{\alpha^{\mu}_{m}, \alpha^{\nu}_{n}\right\} =
\left[\left({1\!\mbox{l}} - {\cal{B}}^2 \right)^{-1}\right]^{\mu
\nu} \delta_{m, -n} = \left({\mathcal{M}}^{-1}\right)^{\mu \nu}
\delta_{m, -n}
\end{eqnarray}
where, ${\mathcal{M}} = ({1\!\mbox{l}} - {\cal{B}}^2)$ ;
$({\cal{B}}^2)^{\mu \nu} = {\cal{B}}^{\mu}_{\ \rho}{\cal{B}}^{\rho
\nu}$\footnote{Here we should note that
$\left({1\!\mbox{l}}\right)^{\mu \nu} = \eta^{\mu \nu}$.}. Now the
anticommutator between the fermionic string coordinates can be
computed using (\ref{6c23}), (\ref{6c25}). The antibrackets
between $\left\{\psi^{\mu}_{(-)}, \psi^{\nu}_{(-)}\right\}$ and
$\left\{\psi^{\mu}_{(+)}, \psi^{\nu}_{(+)}\right\}$ are the same
as that of free case but the anticommutator between
$\psi^{\mu}_{(-)}$ and $\psi^{\mu}_{(+)}$ gets modified to the
following form:
\begin{eqnarray}
\left\{\psi^{\mu}_{-}(\omega, \bar{\omega}), \psi^{\nu}_{+}(\omega^{\prime},
\bar{\omega^{\prime}})\right\} &=& \frac{1}{2 \pi} \,
 \sum_{m \in {\bf{Z}}} \left[\frac{\left({1\!\mbox{l}} - {\cal{B}}
\right)^{\mu}_{\ \rho}\, \left[\left({1\!\mbox{l}} - {\cal{B}}^2\right)^{-1}
\right]^{\rho \sigma}\,
\left({1\!\mbox{l}} - {\cal{B}}\right)_{\sigma}^{\ \nu}
}{\omega^{m + \frac{1}{2}}\, \bar{\omega}^{\prime -m + \frac{1}{2}}}\right].
\label{6c26}
\end{eqnarray}
Now proceeding as before, we can write the above anticommutation relation
in $(\tau, \sigma)$ coordinates to obtain the usual equal time
(i.e. $\tau = \tau^{\prime}$) anticommutation relation:
\begin{eqnarray}
\left\{\psi^{\mu}_{(-)}(\sigma, \tau), \psi^{\nu}_{(+)}(\sigma^{\prime},
\tau)\right\} &=& E^{\rho\, \mu}\,
\left[\left({1\!\mbox{l}} - {\cal{B}}^2\right)^{-1}
\right]_{\rho\, \sigma}\, E^{\nu \,\sigma}\,
 \delta_{P}(\sigma + \sigma^\prime).
\label{6c27}
\end{eqnarray}
The above result reduces to the the free case result in the
${\mathcal{B}}_{\mu\nu} = 0$ limit and also agrees with the
existing results in the previous chapter and literature
\cite{jing1}.


\section{Summary}
In this chapter, we have used conformal field theoretic techniques
to compute the anticommutator among Fourier components of
fermionic sector of super strings. Using this the anticommutator
between the basic fermionic fields is obtained. This is the
extension of our earlier work on bosonic strings discussed in
chapter 4. The method is also different from (\cite{jing1}), where
the algebra among the Fourier components have been computed using
the Faddeev-Jackiw symplectic formalism. The advantage of this
approach is that the results one obtains takes into account the
quantum effects right from the beginning, in contrary to the
previous investigations, which were made essentially at the
classical level \cite{chu, chu1, rb, br, jing1}. Interestingly,
the new normal ordering that takes into account the effect of the
BC(s) plays a crucial role in obtaining the nonanticommutative
symplectic structure among the fermionic string coordinates
(\ref{6c19}). This is in contrast to the analysis in case of the
bosonic strings where the new normal ordering has no bearing on
the symplectic structure.
\noindent Finally, we also computed the oscillator algebra in
presence of the $\mathcal{B}$ field which is a parity-odd field on
the string world-sheet. As in the bosonic case, in presence of
this $\mathcal{B}$ field, the fourier modes appearing in the
Laurent series expansions of the fermionic fields
$\psi^{\mu}_{(-)}$ and $\psi^\mu_{(+)}$ of the closed string are
no longer equal when open string BCs are imposed. These rather get
related to the free oscillator modes $d^{\mu}_{m}$. Using these
expressions of the modes, we rewrite the fermionic fields
$\psi^{\mu}_{(-)} $ and $\psi^{\mu}_{(+)}$ entirely in terms of
the free oscillator modes $\alpha^{\mu}_{m}$ (\ref{6c22}). Then a
straight forward calculation, involving $\psi\psi$ OPE  and
contour argument yields the NC anticommutator, thereby reproducing
the results of previous chapter.

\chapter{String non(anti)commutativity for Neveu-Schwarz boundary conditions}
We have already seen in the case of fermionic string there is a
choice between Ramond boundary conditions and Neveu Schwarz (NS)
boundary conditions. Surprisingly a common point of all the
studies in the superstring theory is that all the literatures are
solely confined for the Ramond (R) BC(s) only and the second type
of BC is less studied in the research area. Here in this chapter,
we extend our methodology (which has already been discussed in
chapter 5) to the superstring satisfying the NS boundary
conditions. A nontrivial result we have found from the whole
analysis is that, contrary to the R case, bosonic sector of the
superstring satisfies Dirichlet BC at one end and Neumann BC at
the other end provided the bosonic variable $X^{\mu}$ is allowed
to be antiperiodic. This observation is completely new and has not
been discussed elsewhere. Further, the symplectic structure of the
bosonic sector also keeps the superconstraint algebra involutive.
The bracket structures have also been computed using the mode
expansions of the bosonic and the fermionic coordinates.

\noindent In the next section, the R-Neveu Schwarz (RNS)
superstring action in the conformal gauge is briefly discussed to
fix the notations. The section is then subdivided into two parts.
In the first subsection, the BC(s) and the mode expansions of the
fermionic sector of the superstring is given and the
nonanticommutativity of the theory is revealed in the conventional
Hamiltonian framework. In the next subsection, the PB structure
and the BC(s) of the bosonic sector is discussed. Then in next
section we compute the super constraint algebra with the modified
symplectic structure obtained in the previous section. The results
obtained in the subsections 7.1.1 and 7.1.2 are further confirmed
in section 7.3 by the mode expansion method. This consistency
check is performed separately for the bosonic and the fermionic
sector. Finally in Section 7.4 we discuss the
non(anti)commutativity in the interacting superstring theory in
the RNS formulation.

\section {RNS free superstring}
In the first part of this section we briefly mention the canonical
algebra of the basic fields of a free open superstring (we have
already discussed the superstring in chapter 5). Later we shall
show how these algebraic structures get modified as a result of
the boundary conditions of the theory. The action we take for our
analysis is given by \cite{green, green13}\footnote{We follow the
same conventions as in chapter 5, $\rho^0 \,=\, \sigma^2 \, = \,
\pmatrix{0&-i\cr i&0\cr}\,\,\,,\,\,\, \rho^1 \,=\, i\sigma^1 \, =
\, \pmatrix{0&i\cr i&0\cr}\ $ and take the induced world-sheet
metric and target space-time metric as $\eta^{ab} = \{-, +\}$,
 $\eta^{\mu\nu} =\{-, +, +, ...., +\}$ respectively.}
\begin{eqnarray}
\label{7c1} S &=& -\frac{1}{2} \int_{\Sigma} d^2\sigma \Big(
\eta_{\mu \nu } \partial_a X^\mu \partial^a X^\nu \,- \,  i
{\overline \psi}^\mu \rho^a \partial_ a \psi_\mu \Big).
\end{eqnarray}
The bosonic and the fermionic part of the above action can be
separated out as
\begin{eqnarray}
S &=& S_B + S_F
\end{eqnarray}
where,
\begin{eqnarray}
\label{7c1aa} S_B = -\frac{1}{2}\int_{\Sigma} d^2\sigma \eta_{\mu
\nu }
\partial_a X^\mu \partial^a X^\nu \  \ {\textrm{and}} \
\ S_F = \frac{1}{2}\int_{\Sigma} d^2\sigma i {\overline \psi}^\mu
\rho^a \partial_ a \psi_\mu.
\end{eqnarray}
The components of the Majorana spinor $\psi$ are denoted as
$\psi_{\pm}$
\begin{equation}
\psi^\mu \,=\, \pmatrix{\psi^\mu_{-}\cr \psi^\mu_{+}\cr}\, .
\label{7c3}
\end{equation}
\noindent The Dirac antibracket of the first order action $S_F$ is
easily read off
\begin{eqnarray}
\{ \psi^\mu_{+} (\sigma) \,,\,\psi^\nu_{+}  (\sigma^\prime)
\}_{D.B} &=& \{ \psi^\mu_{-} (\sigma) \,,\,\psi^\nu_{-}
(\sigma^\prime ) \}_{D.B} \,\,=\,\,
 - i \eta^{\mu\nu} \delta (\sigma - \sigma^\prime )\,
\nonumber\\
\{ \psi^\mu_{+}(\sigma)\,,\,\psi^\nu_{-}(\sigma^\prime) \}_{D.B}
&=& 0\,\,. \label{7c5}
\end{eqnarray}
On the other hand the action $S_B$ gives the following brackets
among the bosonic variables
\begin{eqnarray}
&&\{X^\mu(\sigma) , \Pi^\nu(\sigma^{\prime})\} =
\eta^{\mu \nu}\delta(\sigma - \sigma^{\prime})\nonumber\\
&&\{X^\mu(\sigma) , X^\nu(\sigma^{\prime})\} =0=\{\Pi^\mu(\sigma)
, \Pi^\nu(\sigma^{\prime})\} \label{7c18a}
\end{eqnarray}
where $\Pi_\mu$ is the canonically conjugate momentum to $X^\mu$,
defined in the usual way. Eqs. (\ref{7c5}) and (\ref{7c18a})
defines the preliminary symplectic structure of the theory. We
shall now discuss the effects of BC(s) on these symplectic algebra
for the fermionic and the bosonic sectors separately.
\subsection{Fermionic sector}
Varying the fermionic part of the action (\ref{7c1aa})
\begin{eqnarray}
\label{7c18} \delta S_F = i \int_{\Sigma} d^2\sigma \left[ \delta
\bar{\psi}_{\mu}\rho^a \, \partial_a \psi^{\mu} \, -
\partial_{\sigma} \left(\psi^\mu_{-}\, \delta \psi_{\mu -} -
\psi^\mu_{+}\, \delta \psi_{\mu +}\right)\right]
\end{eqnarray}
we obtain the Euler-Lagrange equation for the fermionic field
\begin{eqnarray}
\label{7c10} i \rho^a \partial_a \psi^{\mu} =0.
\end{eqnarray}
together with the following BC(s):
\begin{eqnarray}
\psi^{\mu}_{+}(0 , \tau) &=& \psi^{\mu}_{-}(0 , \tau) \nonumber \\
\psi^{\mu}_{+}(\pi , \tau) &=& \lambda \psi^{\mu}_{-}(\pi , \tau)
\label{7c12}
\end{eqnarray}
where $\lambda = \pm 1$ corresponds to the R BC(s) and the NS
BC(s), respectively. In this chapter, we shall work with the NS
BC(s) which we write in the following manner
\begin{eqnarray}
\label{7c6aaa} (\psi^{\mu}_{+}(\sigma , \tau) -
\psi^{\mu}_{-}(\sigma , \tau))|_{\sigma=0}=0
\end{eqnarray}
\begin{eqnarray}
\label{7c13aaa} (\psi^{\mu}_{+}(\sigma , \tau) +
\psi^{\mu}_{-}(\sigma , \tau))|_{\sigma=\pi}=0.
\end{eqnarray}
Now the mode expansion of the components of Majorana fermion,
satisfying the above set of BC(s) is given by \cite{green,
green13}:
\begin{eqnarray}
\psi^{\mu}_{-}(\sigma , \tau) &=& \frac{1}{\sqrt{2}}\sum_{n\in Z +
\frac{1}{2}} d^{\mu}_{n}
e^{-i\, n(\tau - \sigma )} \nonumber \\
\psi^{\mu}_{+}(\sigma , \tau) &=& \frac{1}{\sqrt{2}}\sum_{n\in Z +
\frac{1}{2}} d^{\mu}_{n} e^{-i\,n(\tau + \sigma )}. \label{7c7}
\end{eqnarray}
From the above mode expansions it follows automatically that
\begin{eqnarray}
\psi^{\mu}_{-}(-\sigma , \tau) \,=\, \psi^{\mu}_{+}(\sigma ,
\tau)\,. \label{7c8}
\end{eqnarray}
Furthermore, making use of eq. (\ref{7c12}), we obtain
\begin{eqnarray}
\psi^{\mu}_{\pm}(\sigma = -  \pi , \tau)
&=& - \psi^{\mu}_{\pm}(\sigma=\pi , \tau) \nonumber\\
\psi^{\mu}_{\pm}(\sigma = -  2\pi , \tau) &=&
\psi^{\mu}_{\pm}(\sigma = 2\pi , \tau) \label{7c9}
\end{eqnarray}
in the NS-sector. Hence $\psi^{\mu}_{\pm}(\sigma , \tau)$ is an
antiperiodic function of antiperiodicity $2\pi$ which naturally
implies that it is a periodic function of periodicity $4\pi$.
\noindent We now essentially follow the methodology discussed in
chapter 5 for the present case. First, we introduce the
antiperiodic delta function $\delta_{(a)P}(x)$ of antiperiodicity
$2\pi$ and periodicity $4\pi$
\begin{eqnarray}
\delta_{(a)P}(x)= - \delta_{(a)P}(x + 2\pi)=\frac{1}{4\pi}
\sum_{n\in Z + \frac{1}{2}}e^{i\,n x} \label{7cap}
\end{eqnarray}
which satisfies the defining property of a periodic
$\delta$-function i.e.
\begin{eqnarray}
\int_{-2\pi}^{2\pi}dx^{\prime}\delta_{(a)P}(x^{\prime}-x)f(x^{\prime})=f(x)
\label{7c40}
\end{eqnarray}
where $f(x)$ is an arbitrary periodic function with periodicity
$4\pi$. Using this we write the following expression for
$\psi^{\mu}_{-}$ and $\psi^{\mu}_{+}$ in the physical interval
$[0,\pi]$ of the string
\begin{eqnarray}
2\int_{0}^{\pi}d\sigma^{\prime}\left[\delta_{(a)P}(\sigma^{\prime}
+ \sigma)\psi^{\mu}_{+}(\sigma^{\prime}) +
\delta_{(a)P}(\sigma^{\prime}-\sigma)\psi^{\mu}_{-}(\sigma^{\prime})
\right] &=& \psi^{\mu}_{-}(\sigma)\,
\label{7c41} \\
2\int_{0}^{\pi}d\sigma^{\prime}\left[\delta_{(a)P}(\sigma^{\prime}
+ \sigma)\psi^{\mu}_{-}(\sigma^{\prime}) +
\delta_{(a)P}(\sigma^{\prime}-\sigma)\psi^{\mu}_{+}(\sigma^{\prime})
\right] &=& \psi^{\mu}_{+}(\sigma)\,. \label{7c42}
\end{eqnarray}
We define a matrix $\Lambda_{AB}(\sigma, \sigma^{\prime})$
\begin{equation}
\Lambda_{AB}(\sigma, \sigma^{\prime})\,=\,
\pmatrix{\delta_{(a)P}(\sigma^{\prime}-\sigma)&
\delta_{(a)P}(\sigma^{\prime}+\sigma)\cr
\delta_{(a)P}(\sigma^{\prime}+\sigma)&
\delta_{(a)P}(\sigma^{\prime}-\sigma)\cr}\, \label{7c44}
\end{equation}
to write the equations (\ref{7c41}) and (\ref{7c42}) in a compact
form
\begin{eqnarray}
2\int_{0}^{\pi}d\sigma^{\prime}\Lambda_{AB}(\sigma,
\sigma^{\prime})
\psi^{\mu}_{B}(\sigma^{\prime})=\psi^{\mu}_{A}(\sigma)\quad;\quad
(A, \ B=-,+). \label{7c43}
\end{eqnarray}
From the above equation $\Lambda$ can be interpreted as a matrix
valued ``delta function" which acts on the two component Majorana
spinor. Instead of (\ref{7c5}) we therefore propose the following
antibrackets in the fermionic sector
\begin{equation}
\{\psi^{\mu}_{A}(\sigma), \psi^{\nu}_{B}(\sigma^{\prime})\}
=-2i\eta^{\mu\nu}\Lambda_{AB}(\sigma, \sigma^{\prime}).
\label{7c45}
\end{equation}
Making use of eq. (\ref{7c44}) we write this in its component form
\begin{eqnarray}
\{\psi^{\mu}_{+}(\sigma), \psi^{\nu}_{+}(\sigma^{\prime})\}
&=&\{\psi^{\mu}_{-}(\sigma), \psi^{\nu}_{-}(\sigma^{\prime})\}
=-2i\eta^{\mu\nu}\delta_{(a)P}(\sigma - \sigma^{\prime})\,\nonumber\\
\{\psi^{\mu}_{-}(\sigma), \psi^{\nu}_{+}(\sigma^{\prime})\}
&=&-2i\eta^{\mu\nu}\delta_{(a)P}(\sigma + \sigma^{\prime})\,.
\label{7c46}
\end{eqnarray}
Remarkably the above set of antibracket algebra is now completely
consistent with the BC(s). To see this explicitly, we compute the
anticommutator of $\psi_{+}^{\nu}(\sigma^{\prime})$ with
(\ref{7c6aaa}) and (\ref{7c13aaa}), the left hand side of which
gives:
\begin{eqnarray}
-2i\left(\delta_{(a)P}(\sigma - \sigma^{\prime})-
\delta_{(a)P}(\sigma + \sigma^{\prime})\right)|_{\sigma = 0}&=& -
2i \Delta_{-(a)}\left(\sigma , \sigma^{\prime}\right)|_{\sigma =
0}
\nonumber\\
&=& \frac{i}{\pi}\sum_{n \in Z +
\frac{1}{2}}{\textrm{sin}}(n\sigma) \ {\textrm{sin}} (n
\sigma')|_{\sigma = 0}=0
\label{7c2237}\\
-2i\left(\delta_{(a)P}(\sigma - \sigma^{\prime})+
\delta_{(a)P}(\sigma + \sigma^{\prime})\right)|_{\sigma = \pi}&=&
- 2i \Delta_{+(a)}\left(\sigma , \sigma^{\prime}\right)|_{\sigma =
\pi}
\nonumber\\
&=& -\frac{i}{\pi}\sum_{n \in Z + \frac{1}{2}}{\textrm{cos}}(n
\sigma) \ {\textrm{cos}} (n \sigma')|_{\sigma = \pi}= 0
\label{7c47}
\end{eqnarray}
where the form of the antiperiodic delta function (\ref{7cap}) has
been used. This completes the analysis of the fermionic algebra
for the NS BC(s). In the next section we shall use these relations
(\ref{7c46}) to compute the super constraint algebra.
\subsection{Bosonic sector}
Let us now study the bosonic sector of the superstring action
(\ref{7c1}). Varying the bosonic part of the action (\ref{7c1}),
we obtain the equation of motion for the bosonic field
\begin{eqnarray}
(\partial_{\sigma}^2-\partial_{\tau}^2)X^{\mu} = 0 \label{7ceqb}
\end{eqnarray}
together with Dirichlet and Neumann BC(s)
\begin{eqnarray}
\delta X^{\mu}|_{\sigma=0,\pi} = 0 \nonumber \\
X'^{\mu}|_{\sigma=0,\pi}=0. \label{7cbcb}
\end{eqnarray}
Now there are two cases depending on the periodicity of the
bosonic variable $X^{\mu}$. Usually, one is interested in theories
with maximum Poincar\'{e} invariance and hence $X^{\mu}$ must be
periodic (with a periodicity of $2\pi$). This case has already
been discussed in previous chapters. On the other hand
antiperiodicity of $X^{\mu}$ is interesting because one encounters
it for twisted strings on an orbifold \cite{pol1}. In this chapter
we shall discuss this case in details.

\noindent We let the bosonic string coordinates $X^{\mu} (\sigma)$
to have a periodicity of $4 \pi$ (antiperiodicity of
$2\pi$)\footnote{Note that this is also in accord with the
fermionic sector.}:
\begin{eqnarray}
X^{\mu}(\sigma+4\pi)=X^{\mu}(\sigma). \label{7cperiodic}
\end{eqnarray}
Hence the integral (\ref{7c40}) once again holds for the bosonic
coordinate $X^{\mu}(\sigma)$. Restricting to the case of even(odd)
functions $X^{\mu}_{\pm}(- \sigma) = \pm X^{\mu}_{\pm}(\sigma)$,
it can be easily seen that (\ref{7c40}) reduces to:
\begin{eqnarray}
2\int^{\pi}_{0} d\sigma^{\prime}\,
\Delta_{\pm(a)}(\sigma,\sigma')\, X^{\mu}_{\pm}(\sigma^{\prime})
\,=\, X^{\mu}_{\pm}(\sigma) \label{7capb}
\end{eqnarray}
where $\Delta_{\pm(a)}$ were defined in the eqs. (\ref{7c2237})
and (\ref{7c47}). We therefore propose the following equal time
PB:
\begin{eqnarray}
\left\{X^{\mu}(\tau , \sigma) ,  \Pi_{\nu}(\tau ,
\sigma^{\prime})\right\} = 2\,\delta^{\mu}_{\nu}\,
\Delta_{\pm(a)}(\sigma, \sigma^{\prime}). \label{7cetpb}
\end{eqnarray}
It is now easy to observe that for $\Delta_{+
(a)}(\sigma,\sigma')$ to appear in the above PB the end points
must satisfy following BC(s)
\begin{eqnarray}
&& X^{' \mu}(0) = 0 \nonumber\\
&& X^{\mu}(\pi) = 0 \label{7capbcb}
\end{eqnarray}
and for $\Delta_{- (a)}(\sigma,\sigma')$, the appropriate BC(s)
that the end points must satisfy, reads
\begin{eqnarray}
&& X^{\mu}(0) = 0 \nonumber\\
&& X^{' \mu}(\pi) = 0. \label{7capbcb1}
\end{eqnarray}
We shall find in the next section that the symplectic structure of
the bosonic sector also plays a crucial role in the closure of the
super constraint algebra.

\section {Super constraint algebra }
In this section we shall compute the algebra of the super-Virasoro
constraints using the modified symplectic structures derived in the first section 2.\\
\noindent The complete set of super constraints are given by
\cite{ agh1,green}:
\begin{eqnarray}
\chi_1(\sigma) &=& \Phi_1(\sigma) + \lambda_1(\sigma)
= 0 \nonumber \\
\chi_2(\sigma) &=& \Phi_2(\sigma) + \lambda_2(\sigma) = 0
\label{7cchi}
\end{eqnarray}
where,
\begin{eqnarray}
\Phi_1(\sigma) &=&
\left(\Pi^2(\sigma) + (\partial_{\sigma}X(\sigma))^2\right) \nonumber \\
\Phi_2(\sigma) &=&
\left(\Pi(\sigma)\partial_{\sigma}X(\sigma)\right) \nonumber \\
\lambda_1(\sigma)&=& - i \bar{\psi^{\mu}}(\sigma)\rho_{1}
\partial_{\sigma} \psi_{\mu}(\sigma) =  - i
\left(\psi^{\mu}_{-}(\sigma)
\partial_{\sigma} \psi_{\mu -}(\sigma) - \psi^{\mu}_{+}(\sigma)
\partial_{\sigma} \psi_{\mu +}(\sigma)\right) \nonumber \\
\lambda_2(\sigma)&=& - \frac{i}{2}
\bar{\psi^{\mu}}(\sigma)\rho_{0}
\partial_{\sigma} \psi_{\mu}(\sigma)
= \frac{i}{2}\left(\psi^{\mu}_{-}(\sigma)\partial_{\sigma}
\psi_{\mu -}(\sigma) + \psi^{\mu}_{+}(\sigma)\partial_{\sigma}
\psi_{\mu +}(\sigma)\right) \label{7clambda}
\end{eqnarray}
and using the basic algebra of fermionic and bosonic variables
(\ref{7c46}, \ref{7cetpb}), we get the following algebra for
super-Virasoro constraints:
\begin{eqnarray}
\{\chi_1(\sigma) , \chi_1(\sigma^{\prime})\} &=& 8 \left(
\chi_2(\sigma)\partial_{\sigma}\Delta_{+ (a)} \left(\sigma ,
\sigma^{\prime}\right) + \chi_2(\sigma^{\prime})
\partial_{\sigma}\Delta_{- (a)}\left(\sigma , \sigma^{\prime}\right)\right)\,
\nonumber \\
\{\chi_2(\sigma) , \chi_2(\sigma^{\prime})\} &=&  2\left(
\chi_2(\sigma^{\prime})\partial_{\sigma}\Delta_{+ (a)}
\left(\sigma , \sigma^{\prime}\right) + \chi_2(\sigma)
\partial_{\sigma}\Delta_{- (a)}\left(\sigma , \sigma^{\prime}\right)\right) \,
\nonumber \\
\{\chi_2(\sigma) , \chi_1(\sigma^{\prime})\} &=&  2\left(
\chi_1(\sigma) + \chi_1(\sigma^{\prime})\right)
\partial_{\sigma}\Delta_{+ (a)}\left(\sigma , \sigma^{\prime}\right)\,.
\label{7cchi-chi}
\end{eqnarray}
Apart from a numerical factor the above algebra has the same
structure as in 5th chapter with the only difference that
$\delta_{P}(\sigma)$ occurring in chapter 5 has been replaced by
$\delta_{(a)P}(\sigma)$. Similarly one can show that the algebra
of super currents
\begin{eqnarray}
&&\tilde{J_1}(\sigma)=2J_{01}(\sigma)=\psi_-^{\mu}(\sigma)\Pi_{\mu}(\sigma)-\psi_-^{\mu}(\sigma)\partial_{\sigma}X_{\mu}\nonumber\\
&&\tilde{J_2}(\sigma)=2J_{02}(\sigma)=\psi_+^{\mu}(\sigma)\Pi_{\mu}(\sigma)+\psi_+^{\mu}(\sigma)\partial_{\sigma}X_{\mu}
\label{7cJtilde}
\end{eqnarray}
among themselves and also with the super constraints (\ref{7cchi})
close. It is also interesting to note that both $\Delta_{+(a)}$
and $\Delta_{-(a)}$ appearing in the PB of the bosonic variables
(\ref{7cetpb}) gives the same constraint algebra
(\ref{7cchi-chi}). Furthermore, the closure of the algebra also
indicates the internal consistency of our analysis.
\section{Mode expansions and symplectic algebra}
In this section, we shall derive the fermionic algebra
(\ref{7c46}) and the bosonic algebra (\ref{7cetpb}) from a mode
expansion of the constituting fields. To do that we consider the
mode expansions of the fermionic field (\ref{7c7}). Here
$d^{\mu}_{n}$ are Fourier modes and they satisfy the algebra
\begin{eqnarray}
\left\{d^{\mu}_{m}, d^\nu_n\right\} = -\frac{i}{\pi}\, \eta^{\mu
\nu}\, \delta_{m+n,0}. \label{7cs0}
\end{eqnarray}
This algebra can be obtained just by  following the procedure of
\cite{jing1}, in which they have computed the anti brackets among
Fourier components of fermionic sector of superstrings (R sector)
using Faddeev-Jackiw symplectic formalism \cite{fj}. This relation
(\ref{7cs0}) between $d$'s can also be worked out from the contour
argument (discussed in chapter 6) \cite{pol1} and the operator
product expansion. \noindent The antibracket relations between
$\psi^{\mu}_{A}(\sigma), \psi^{\nu}_{B}(\sigma^{\prime})$ are then
obtained by using (\ref{7c7}) and (\ref{7cs0})
\begin{eqnarray}
\left\{\psi^{\mu}_{-}(\sigma),
\psi^{\nu}_{+}(\sigma^{\prime})\right\} \, &=& \, \frac{1}{2}
\sum_{r, s\in{Z + \frac{1}{2}}} e^{- i r (\tau - \sigma)}\, e^{- i
s (\tau +\sigma)} \left\{d^{\mu}_{r} , d^{\nu}_{s}\right\} \\
\nonumber &=& - \frac{i}{2\pi}\, \eta^{\mu \nu}\sum_{r\in{Z +
\frac{1}{2}}}e^{- i r (\tau - \sigma)}\, e^{i r(\tau +\sigma)}\\
\nonumber &=& -2 i \eta^{\mu \nu} \delta_{(a)P}(\sigma +
\sigma^\prime). \label{7cs1}
\end{eqnarray}
Proceeding exactly in the similar manner one can get back the
other anti-brackets of (\ref{7c46}).

\noindent In order to study the bosonic sector, we first need the
expressions
 of the mode expansion for the two different types of BC(s) (\ref{7capbcb})
 and (\ref{7capbcb1}). \\
\noindent For the first case (BC (\ref{7capbcb})) it is given by:
\begin{eqnarray}
X^{\mu}(\tau, \sigma) = \sum_{n\in{Z + \frac{1}{2}}}
\frac{\alpha^{\mu}_{n}}{n}\, e^{in\tau}\, \mathrm{sin}\,n\sigma
\label{7cs2}
\end{eqnarray}
and for the other case (BC (\ref{7capbcb1})) the mode expansion is
\begin{eqnarray}
X^{\mu}(\tau, \sigma) = \sum_{n\in{Z + \frac{1}{2}}}
\frac{\alpha^{\mu}_{n}}{n}\, e^{in\tau}\, \mathrm{cos}\,n\sigma.
\label{7cs3}
\end{eqnarray}
The canonical momenta corresponding to (\ref{7cs2}) and
(\ref{7cs3}) are given by
\begin{eqnarray}
\Pi_{\mu}(\tau, \sigma) &=& \eta_{\mu \nu}
\partial_{\tau}X^{\nu}(\tau,
\sigma)  \nonumber\\
&=& i\eta_{\mu \nu}\sum_{n\in{Z + \frac{1}{2}}} \alpha^{\nu}_{n}\,
e^{in\tau}\, \mathrm{sin}\,n\sigma, \quad  i\eta_{\mu
\nu}\sum_{n\in{Z + \frac{1}{2}}} \alpha^{\nu}_{n}\, e^{in\tau}\,
\mathrm{cos}\,n\sigma. \label{7cs5}
\end{eqnarray}
Here also the algebra between the modes can be computed by
following the methodology of \cite{jing, fj}:
\begin{eqnarray}
\left\{\alpha^{\mu}_{m}, \alpha^\nu_n\right\} = - \frac{i}{\pi}\,
\eta^{\mu \nu}\, m\delta_{m+n,0}. \label{7cs4}
\end{eqnarray}
Using (\ref{7cs4}) we obtain the same equal time PB given in
(\ref{7cetpb}).
\section{The interacting theory}
After finishing the analysis for the free theory, we shall now
study the interacting case where a superstring moves in the
presence of a constant antisymmetric tensor field ${\cal B}_{\mu
\nu}$. The action given by \cite{haggi, NL}:
\begin{eqnarray}
S &=&  \frac{- 1}{2} \int_{\Sigma} d\tau
 d{\sigma}\,\Big[ \, \partial_a X^\mu \partial^a X_\mu
\,+\, \epsilon^{ab} {\cal B}_{\mu\nu} \partial_a X^\mu \partial_b
X^\nu
\nonumber\\
 & & + i \psi_{\mu -} E^{\nu \mu} \partial_{+} \psi_{\nu -} +
i \psi_{\mu +} E^{\nu \mu} \partial_{-} \psi_{\nu +} \,\Big]
\label{7c1.1}
\end{eqnarray}
\noindent where, $\partial_{+} =  \partial_{\tau} +
\partial_{\sigma},\; \;
\partial_{-} =  \partial_{\tau} - \partial_{\sigma} $
and $ E^{\mu\nu} \,=\, \eta^{\mu\nu} \, + {\cal B}^{\mu\nu}\,$.
\noindent Now since the bosonic and fermionic sectors decouple, we
can study them separately. \noindent

Here we concentrate on the fermionic sector. The variation of the
fermionic part of the action (\ref{7c1.1}) gives the classical
equations of motion:
\begin{equation}
\partial_{+} \psi_{\nu -} \,=\, 0\quad,\quad
\partial_{-} \psi_{\nu +} \,=\, 0
\label{7c1.2}
\end{equation}
and a boundary term that yields the following NS BC(s)\footnote{
The boundary term also leads to R BC(s). Detailed investigations
involving R BC(s) has already been carried out in chapter 5 and
6.}:
\begin{eqnarray}
 E_{\nu\mu}\,  \psi^\nu_{+} (0,\tau)\, &=&\,
  E_{\mu\nu} \, \psi^\nu_{-} (0,\tau)\, \nonumber\\
\label{7c4} E_{\nu\mu}\,  \psi^\nu_{+} (\pi,\tau ) \, &=& \, -
E_{\mu\nu} \, \psi^\nu_{-} (\pi, \tau ) \label{7c1.3}
\end{eqnarray}
\noindent at the endpoints $\sigma \,=\,0$ and $\sigma = \pi\,$ of
the string.

\noindent As in the free case, the above non-trivial BC(s) leads
to a modification in the symplectic structure (\ref{7c5}). The
$\{\psi^\mu_{(\pm)} (\sigma,\tau) , \psi^\nu_{\pm}
(\sigma^{\prime},\tau)\}$ is the same as (\ref{7c46}). In the case
of mixed bracket, we make the following ansatz:
\begin{eqnarray}
 \{\psi^\mu_{+} (\sigma,\tau) ,
\psi^\nu_{-} (\sigma^{\prime},\tau)\}\, = \,
C^{\mu\nu}\delta_{(a)P}\left(\sigma + \sigma^{\prime}\right)\,.
\label{7c57}
\end{eqnarray}
Brackets $\psi^\gamma_{-} (\sigma^{\prime})$ with the BC(s)
(\ref{7c1.3}) one obtains
\begin{eqnarray}
 E_{\nu\mu}\,  C^{\nu \gamma}
\, = \, -2i \, E_{\mu\gamma} \label{7c58}
\end{eqnarray}
which on solving gives
\begin{eqnarray}
 C^{\mu \nu}\, = \, -2i \,
\left[\left(1 - {\cal B}^2\right)^{-1}\right]^{\mu \rho}\,
 E_{\rho \gamma}\, E^{\gamma \nu}.
\label{7c58a}
\end{eqnarray}
Above solution is written in a matrix notation as,
\begin{eqnarray}
 C \, = \, -2i \,
\left[\left(1 - {\cal B}^2\right)^{-1}\left(1 + {\cal
B}\right)^{2}\right] \label{7c58y}
\end{eqnarray}
where $C = \{C^{\mu \nu}\}$. Thus we get the modified mixed
bracket in the form
\begin{eqnarray}
\{\psi^\mu_{+} (\sigma,\tau) , \psi^\nu_{-}
(\sigma^{\prime},\tau)\}\, = \, -2i\, \left[\left(1 - {\cal
B}^2\right)^{-1}\right]^{\mu \rho} \,  E_{\rho \gamma}\, E^{\gamma
\nu} \delta_{(a)P}\left(\sigma + \sigma^{\prime}\right)\,.
\label{7c60}
\end{eqnarray}
If we take the limit ${\cal B}_{\mu \nu} \rightarrow 0$ in the
above equation we get back the last relation of (\ref{7c46}).
\section{Summary}
In string theory the modification of Poisson algebra is a
consequence of the nontrivial BC(s). In this chapter, we have
studied this problem for an open superstring satisfying the NS
BC(s). Here also we have obtained non(anti)commutative structure
for the fermionic string coordinates following the approach discussed in 
chapter 5. So in that sense this is an extension of the chapter 5 and 6.

%
\chapter{Concluding remarks}
Noncommutativity of spacetime and its consequences for quantum
field theory have been one of the main objects of interest in the
last few years. An important source of noncommutativity in string
theory is the presence of an antisymmetric constant tensor field
along the D-brane world volumes (where the string end points are
located). The quantisation of strings attached to branes involves
mixed (combination of Dirichlet and Neumann) boundary conditions.
This makes the quantisation procedure more subtle  since the
quantum commutators must be consistent with these boundary
conditions. The aim of this thesis is to go further with these
investigations by making a thorough study on the role played by
boundary conditions in noncommutativity/non(anti)commutativity in
string theory.

We started, in chapter 1, with a brief introduction of how
noncommutativity appears in string theory. Different approaches
have been adopted to obtain this result. In some of the earlier
papers in the literature, the authors have regarded the boundary
conditions as constraints. The interpretation of the boundary
condition as primary constraints usually lead to an infinite tower
of second class constraints in contrast to the usual Dirac
formulation of constrained systems. Besides, in this approach,
where one tries to obtain non-commutativity through Dirac brackets
between coordinates, one encounters ambiguous factor like
$\delta(0)$. Furthermore, different results are obtained depending
on the interpretations of these factors. On the other hand Hanson,
Regge and Teitelboim \cite{hrt}, modified the cannonical Poisson
bracket structure, so that it is compatible with the boundary
conditions. The modified  Poisson brackets were obtained for the
free NG string, in the orthonormal gauge, which is  the
counterpart of the conformal gauge in the free Polyakov string. We
essentially followed the  same procedure in chapter 2, to modify
the basic brackets of Polyakov string so that it is compatible
with the boundary conditions.

In chapter 2, we first presented a review of noncommutativity in
an open string moving in a background Neveu-Schwarz field in a
gauge independent hamiltonian approach. The noncommutativity was
seen to be a direct consequence of the nontrivial boundary
conditions, which in contrary to several approaches, were not
treated as constraints. The origin of any modification in the
usual Poisson algebra was the presence of boundary conditions. In
a gauge independent formulation of a free Polyakov string, the
boundary conditions naturally led to a noncommutative structure
among the string coordinates. This noncommutativity vanished in
the conformal gauge, as expected. For the interacting string, a
more involved boundary condition led to a more general type of
noncommutativity. In contrary to the standard conformal gauge
expressions, this noncommutative structure survived at all points
of the string and not just at the boundaries. Also in contrast to the
free string theory, this noncommutativity could not be entirely removed in
any gauge. In the conformal gauge, noncommutativity survived only
at the string end points.

We then discussed a new form of the action that interpolates
between the Nambu-Goto and Polyakov form of interacting bosonic
string, without the need of any gauge fixing. The interpolating
Lagrangian introduced in this chapter contains as many fields as
there are independent degrees of freedom. Being already in the
first order form, this action is free from nonlinearity problems
associated with the Nambu-Goto action. It also does not contain
redundant fields as in the Polyakov forms. 
Here also it was seen that the basic brackets are not
compatible with the interpolating boundary conditions. So we
modified the basic Poission brackets in order to establish
consistency of the boundary condition with the basic Poission
brackets. A thorough analysis of the gauge symmetries of
interpolating actions was then performed in this noncommutative
set up using a general method based on Dirac's theory of
constrained Hamiltonian analysis. Specifically we demonstrated the
equivalence of the reparametrisation invariances of different
string actions with the gauge invariances generated by the first
class constraints. Indeed, the whole analysis of the interpolating
Lagrangian formalism was based on the local gauge symmetries only.
Finally, we feel that it would be interesting to investigate
whether non-critical strings can be discussed using the
interpolating action in a path-integral framework.

So far we basically discussed the appearances of noncommutativity
in bosonic string at classical level. So we extended our analysis
to the quantum level in chapter 4. We first discussed new normal
ordered products for open string position operators that satisfy
both the equations of motion and the boundary conditions. Using
the contour argument and the new $X$-$X$ operator product
expansion we calculated the commutator among the Fourier
components and then the commutation relations among string
coordinates. In this chapter, we used conformal field theory techniques
to compute the commutator among Fourier components which was
unlike the method in (\cite{jing}), where the algebra among the
Fourier components were computed using the Faddeev-Jackiw
symplectic formalism. This was then used to obtain the commutator
between the basic fields. The advantage of this approach was that
the results one obtained took into account the quantum effects
right from the beginning, in contrary to the previous
investigations, which were made essentially at the classical
level, so that the question of the existence of quantum effects,
if any, can be addressed immediately. For example, it was checked
that the new normal ordering, as proposed in \cite{4cbr}, which
took into account the boundary conditions had no bearing on the
central charge in case of free bosonic string. Finally, we also
computed the oscillator algebra in presence of the $B$ field which
is a parity-odd field on the string world-sheet. Consequently in
presence of this $B$ field, the left and right moving modes
appearing in the  Laurent series expansions of the
(anti)holomorphic fields $\partial X^{\mu}$ and $\bar{\partial}
X^\mu$ (\ref{4c12}) of the closed string were no longer equal when
open string BCs were imposed to obtain the corresponding Laurent
expansions. These rather got related to the free oscillator modes
$\gamma^{\mu}_{m}$ (\ref{4c35}) in a parity asymmetric way. Using
these expressions of left and right moving modes, we rewrote the
(anti)holomorphic fields $\partial X^{\mu}$ and $\bar{\partial}
X^\mu$ entirely in terms of the free oscillator modes
$\gamma^{\mu}_{m}$ (\ref{4c19}). Then a straight forward
calculation, involving $XX$ OPE and contour argument yield the NC
commutator given in (\ref{4c26}), thereby reproducing the previous
chapters results and that of \cite{chu, chu1, rb, br, jing}, even
though we had made use of newly proposed normal ordering
\cite{4cbr} which was compatible with boundary conditions.

We then extended our analysis from bosonic to the superstring case in the next
few chapters. As pointed out earlier the origin of any
modification in the usual canonical algebra is the presence of
boundary conditions. This phenomenon is quite well known for a
free scalar field subjected to periodic boundary conditions.
Besides this method was also used earlier by \cite{hrt} in the
context of Nambu-Goto formulation of the bosonic string. We show
that the same thing also held true in the fermionic sector of the
conformal gauge fixed free superstring. It should be mentioned that in the case 
of fermionic string there is a choice between two boundary conditions 
{\it{viz}} Ramond boundary conditions and Neveu Schwarz boundary conditions.
In chapter 5 and 6 we worked in detail with Ramond boundary conditions and
finally in chapter 7 we worked with  Neveu Schwarz boundary conditions.
Here also the boundary conditions became periodic once we extended the domain of
definition of the length of the string from $[0, \pi]$ to $[-\pi,
\pi]$. This mathematical trick led to a modification where the
usual Dirac delta function got replaced by a periodic delta
function. Eventually one constructs the appropriate ``delta
function" for the physical interval $[0, \pi]$ of the string to
write down the basic symplectic structure. Interestingly, there we
got a $2 \times 2$ matrix valued ``delta function" appropriate for
the two component Majorana spinor. This is in contrast to the
bosonic case, where one had a single component ``delta function"
$\Delta_{+}(\sigma , \sigma^{\prime})$ satisfying Neumann boundary
condition. This symplectic structure, interestingly, led to a new
involutive structure for the super-Virasoro algebra at the
classical level. The interesting thing to be noted is that, unlike
the bosonic case, we got an anticommutative structure in the
fermionic sector even for the free superstring. Our results differ
from those in \cite{godinho} and were mathematically consistent
which was reflected from the closure of the constraint algebras.
The analysis of this chapter is a direct generalisation of bosonic
string discussed in 2nd and 3rd chapter. 
The same technique was adopted for the interacting case
also where the boundary condition got more involved and led to a
more general type of non(anti)-commutativity that had been
observed before. However, our results were once again different
from the existing results since we got a  periodic delta function
instead of the usual delta function, apart from the relative sign
of $\sigma , \sigma^{\prime}$. This change of relative sign indeed
played a crucial role in the internal consistency of our analysis.
Further, the interacting results go over smoothly to the free case
once the interaction was switched off.

It is important to note that all these discussion in the context of superstrings
were at classical level. So in the next chapter we calculated the normal
ordered products for fermionic open string coordinates in the
presence of an antisymmetric tensor background taking the boundary
conditions into account. Then we again computed the anticommutator
between the basic fermionic fields using the conformal field
theoretic techniques. In that sense this was an extension of
chapter 4 (in which we discussed normal ordering for bosonic
string coordinates only).

Finally we extended our methodology to the
superstring satisfying the  Neveu Schwarz boundary conditions in
chapter 7. Following the approach of 5th chapter, here also the domain
of the string length was extended from $[0, \pi]$ to $[-\pi, \pi]$
to got the antiperiodic boundary conditions. That construction enables us to got
the $2 \times 2$ matrix valued $\delta$ function in the algebra of
the fermionic sector. Apart from a numerical factor the fermionic
algebra was identical to the result obtained in chapter 5.
However for the bosonic part of the superstring the
result was drastically different. We stress that
the symplectic algebra of the bosonic variables, in that chapter
contained both $\Delta_{+(a)}(\sigma , \sigma^{\prime})$ and
$\Delta_{-(a)}(\sigma , \sigma^{\prime})$ (certain combination of anti periodic 
delta function) which was completely
different from the Ramond case where only $\Delta_{+}(\sigma ,
\sigma^{\prime})$ was present. Interestingly that the symplectic
structure containing both $\Delta_{\pm(a)}(\sigma ,
\sigma^{\prime})$, kept the superconstraint algebra closed
provided one imposes Neumann boundary conditions at one end and Dirichlet 
 boundary conditions at the other end of the string in the bosonic sector. That
observation was completely new and had not been noticed before in
the literature. Finally to complete the analysis, we
calculated the non(anti)commutative structures for the
interacting case by employing the same procedure. As one expected,
without the background field term, the interacting results took the
limiting value of the free case

\begin{appendix}
\chapter{Reality condition for a Majorana spinor}
We have chosen a convenient basis of Dirac matrices as
\begin{equation}
\rho^1 \,=\, \pmatrix{0&i\cr i&0\cr}\,\,\,,\,\,\,\rho^0 \,=\,
\pmatrix{0&-i\cr i&0\cr} \label{app1}
\end{equation}
satisfying the Clifford algebra (\ref{5c38}). In this
representation the component of $\Psi_D$ is given by
$\Psi_{D{\pm}}$
\begin{equation}
\Psi_{D}^\mu \,=\, \pmatrix{\psi^\mu_{-}\cr \psi^\mu_{+}\cr}.
\label{app2}
\end{equation}
Now suppose it satisfies the Dirac equation
\begin{equation}
i\rho^{\mu}\partial_{\mu}\Psi_D = 0. \label{app3}
\end{equation}
In the presence of a background electromagnetic field, the
corresponding equation is obtained by replacing
\begin{eqnarray*}
\partial_{\mu} \longrightarrow D_{\mu} = \partial_{\mu} - i e A_{\mu}
\end{eqnarray*}
and (\ref{app3}) reduces to
\begin{eqnarray}
i \rho^{\mu} \left(\partial_{\mu} - i e A_{\mu}\right)\Psi_D =
0\,. \label{app4}
\end{eqnarray}
Hole theory interpretation ensures that there exists a
corresponding solution $\Psi^c_D$ for the antiparticle of charge
($-e$) satisfying
\begin{eqnarray}
-i \rho^{\mu} \left(\partial_{\mu} + i e A_{\mu}\right)\Psi^c_D =
0\,. \label{app5}
\end{eqnarray}
Defining $\bar{\Psi} = \Psi^{\dagger}\rho_0$ we find the conjugate
Dirac equation as
\begin{eqnarray}
-i {\rho^{\mu}}^T \left(\partial_{\mu} + i e
A_{\mu}\right)\bar{\Psi}^T_D = 0\,. \label{app6}
\end{eqnarray}
Now in order that the matrices $- \rho^{\mu T}$ also satisfy (the
Clifford algebra) (\ref{5c38}), and there must exist a nonsingular
matrix $C$ such that
\begin{eqnarray}
C^{-1} \rho^{\mu} C = - {\rho^{\mu}}^T \label{app7}
\end{eqnarray}
so that (\ref{app4}) matches with (\ref{app6}).  Thus, if we
define the `charge-conjugate spinor' $\Psi^c_D$ by putting
\begin{eqnarray}
\Psi^c_D = C \bar{\Psi}^T_D \ \ \ \ \ \ (\mathrm {upto\  a\
phase})\,, \label{app8}
\end{eqnarray}
we see that it satisfies (\ref{app5}).  It is easy to show that
$C$ is always antisymmetric and, in this representation
(\ref{app1}), we may choose $C$ to be (proportional to) $\sigma^3
\rho^1$, i.e.\footnote{where $\sigma^3 = \, \pmatrix{1&0\cr
0&-1\cr}.$},
\begin{eqnarray}
C = \, \pmatrix{0&i\cr -i&0\cr}. \label{app9}
\end{eqnarray}
A Majorana spinor $\Psi_M$ is  defined as one that equals its
charge-conjugate spinor, i.e., $\Psi^c_M = \Psi_M$. For a Majorana
spinor we therefore have
\begin{equation}
\pmatrix{\psi^\mu_{-}\cr \psi^\mu_{+}\cr} = \pmatrix{\psi^{\mu\
\star}_{-}\cr \psi^{\mu\ \star}_{+}\cr}\,, \label{app10}
\end{equation}
which is the reality condition.

Now we proceed to establish a relation between the chiral
representation and the representation in this chapter, of a
Majorana spinor.  In chiral representation
\begin{equation}
\{\Psi_{M}\}_{c}= \pmatrix{\psi_{R}\cr \psi_{L}\cr} \label{app11}
\end{equation}
and the $\gamma$ matrices satisfying the Clifford algebra
(\ref{5c38})
 in the chiral representation read
\begin{equation}
\gamma^0 \,=\, \pmatrix{0&-1\cr -1&0\cr}\,\,\,,\,\,\,\gamma^1
\,=\, \pmatrix{0&-1\cr 1&0\cr}. \label{app12}
\end{equation}
Let $S$ be a matrix such that
\begin{eqnarray}
\rho^a = S \gamma^a S^{-1}\,, \nonumber \\
\Psi = S  \{\Psi_{M}\}_{c}\,, \label{app13}
\end{eqnarray}
where $\Psi$ is given by (\ref{5c3}).  This immediately leads to the
following solution for the matrix $S$:
\begin{equation}
S \,=\, \pmatrix{0&1\cr -i&0\cr}. \label{app14}
\end{equation}
Hence, from (\ref{app13}), we have
\begin{equation}
\psi_L(\sigma) = \psi_{-}(\sigma)\,, \quad -i\psi_R(\sigma) =
\psi_{+}(\sigma)\,. \label{chiral}
\end{equation}
Clearly it follows that $\psi_L(\sigma)$ is real but
$\psi_R(\sigma)$ is purely imaginary. Also one can easily identify
$\psi_{+}(\sigma)$ and $\psi_{-}(\sigma)$ to be the real chiral
components themselves. Therefore from physical grounds one can
easily expect
\begin{equation}
\psi_{+}(- \sigma) = \psi_{-}(\sigma). \label{chiralq}
\end{equation}

\chapter{Computational details of some of the key result of the chapter 6}
Here we would like to give some of the computational details
involved in deriving (\ref{6ceqmov2}) from (\ref{6c8}) and
(\ref{6c28}) (for convenience we treat the free case, i.e.
${\mathcal{B}} = 0$). Eq.(\ref{6c8}) with $z$ replaced by $\omega$
yields:
\begin{eqnarray}
\label{a1} 0&=&\int [d\psi]\left[  \frac{\delta}{\delta \psi^{\mu}
_{(a)} (\omega,\bar{\omega})}[e^{-S_{F}} \psi^{\nu}_{(b)}
(\omega', \bar{\omega}')]
  \right] \nonumber\\
&=&\int[d\psi]e^{-S_{F}} \left[-\frac{\delta
S_{F}}{\delta\psi_{(a)}(\omega,\bar{\omega})}
\psi_{(b)}(\omega^{\prime},\bar{\omega}^{\prime})+
\frac{\delta\psi_{(b)}(\omega^{\prime},\bar{\omega}^{\prime})}
{\delta\psi_{(a)}(\omega,\bar{\omega})}\right]
\end{eqnarray}
Putting $a=+$, $b=-$; we obtain:
\begin{eqnarray}
\label{a2} 0&=&\int[d\psi]e^{-S_{F}}
\left[\frac{i}{2\pi\alpha^{\prime}}\partial_{\omega}\psi_{+}
(\omega,\bar{\omega})\psi_{-}(\omega^{\prime},
\bar{\omega}^{\prime})+
\frac{\delta\psi_{(-)}(\omega^{\prime},\bar{\omega}^{\prime})}
{\delta\psi_{(+)}(\omega,\bar{\omega})} \right. \nonumber \\
&&+ \left.\frac{i}{4\pi\alpha^{\prime}}
\oint_{\partial\Sigma}d\omega^{\prime\prime}\delta^{2}
(\omega^{\prime\prime}-\omega,
\bar{\omega}^{\prime\prime}-\bar{\omega})
\psi_{(-)}(\omega^{\prime}, \bar{\omega}^{\prime})\right. \nonumber \\
&& \quad \quad \quad \quad \quad \quad \quad \quad \quad \quad
\times \left. \left(\psi_{(-)}(\omega^{\prime \prime},
\bar{\omega}^{\prime \prime}) - i  \psi_{(+)}(\omega^{\prime
\prime}, \bar{\omega}^{\prime \prime})\right) \right] \
\end{eqnarray}
Now we discuss two distinct cases seperately.

\noindent $\bullet$ Case 1: The insertion
$\psi_{(-)}(\omega^{\prime},\bar{\omega}^{\prime})$ is not located
at the boundary:

\noindent In this case
\begin{eqnarray}
\label{a3}
\frac{\delta\psi_{(-)}(\omega^{\prime},\bar{\omega}^{\prime})}
{\delta\psi_{(+)}(\omega,\bar{\omega})}=0
\end{eqnarray}
and therefore one finds:
\begin{eqnarray}
\label{a4} \langle \partial_{\omega} \psi^{\mu}_{(+)}
(\omega,\bar{\omega}) \psi^{\nu}_{(-)} (\omega',\bar{\omega}')
\rangle &=& 0.
\end{eqnarray}

\noindent $\bullet$ Case 2: The insertion
$\psi_{(-)}(\omega^{\prime})$ is located at the boundary (since
$\omega' = \bar{\omega}'$ at the boundary, the insertion
$\psi_{(-)}(\omega^{\prime})$ depends only on the arguement $\omega'$): \\
In this case the computation of the second term in (\ref{a2})
needs to be done more carefully. One finds
\begin{eqnarray}
\label{a5}
\frac{\delta\psi_{(-)}(\omega^{\prime},\bar{\omega}^{\prime})}
{\delta\psi_{(+)}(\omega,\bar{\omega})}\Big{\vert}_{\omega' =
\bar{\omega}'} &=& i\,
\frac{\delta\psi_{(+)}(\omega^{\prime},\bar{\omega}^{\prime})}
{\delta\psi_{(+)}(\omega,\bar{\omega})}\Big{\vert}_{\omega' =
\bar{\omega}'}
\nonumber \\
&=& i\, \delta^2 \left(\omega-\omega', \bar{\omega}- \bar{\omega}'\right)\Big{\vert}_{\omega' = \bar{\omega}'} \nonumber \\
&=&  i\, \delta^2 \left(\omega-\omega', \bar{\omega}-
\omega'\right).
\end{eqnarray}
where we have used the BC (\ref{6c28}) (with $\omega$ repaced by
$\omega^{\prime}$) in the first line of (\ref{a5}).

\noindent Substituting (\ref{a5}) in (\ref{a2}) and equating the
volume term to zero, one finds the third of the equations in
(\ref{6ceqmov2}) (with ${\mathcal{B}} = 0$).

\noindent Similarly, for other choices of $a, b$ the rest of the
equations in (\ref{6ceqmov2}) can be derived (with ${\mathcal{B}}
= 0$).

\end{appendix}


\end{document}